\documentclass[aps,showpacs,pra,superscriptaddress,preprintnumbers,amsmath,amssymb,twocolumn,longbibliography]{revtex4-2}
\usepackage[utf8]{inputenc}
\usepackage[T1]{fontenc}
\usepackage{graphicx}
\usepackage{dcolumn}
\usepackage{appendix}
\usepackage{amsmath}
\usepackage{bm}
\usepackage[dvipsnames]{xcolor}
\usepackage{multirow}
\usepackage{pifont}
\newcommand{\relaxket}[1]{\lvert{#1}\rangle}
\newcommand{\relaxbra}[1]{\langle{#1}\rvert}

\usepackage{subfigure}
\usepackage{bbm}
\usepackage{enumerate}

\usepackage[
bookmarks=true,
colorlinks,
linkcolor=blue,
urlcolor=blue,
citecolor=black,
plainpages=false,
pdfpagelabels,
final,
breaklinks=true
]{hyperref}

\usepackage{titlesec}

\usepackage{array}

\usepackage{physics}

\begin{document}
\allowdisplaybreaks

\title{Erasing photons from bright squeezed vacuum light via above-threshold ionization}

\author{J.~Rivera-Dean}
\email{physics.jriveradean@proton.me}
\email{javier.dean@ucl.ac.uk}
\affiliation{Department of Physics and Astronomy, University College London, Gower Street, London WC1E 6BT, UK}

\author{T. Rook}
\affiliation{Clarendon Laboratory, University of Oxford, Parks Road, Oxford OX1 3PU, United Kingdom}

\author{G.~Singh}
\affiliation{Perimeter Institute for Theoretical Physics, Waterloo, Ontario, N2L 2Y5, Canada}
\affiliation{Department of Physics and Astronomy, University of Waterloo, Waterloo, Ontario, N2L 3G1, Canada}

\author{P.~Stammer}
\affiliation{ICFO -- Institut de Ciencies Fotoniques, The Barcelona Institute of Science and Technology, 08860 Castelldefels (Barcelona)}
\affiliation{Atominstitut, Technische Universität Wien, 1020 Vienna, Austria}

\author{M.~Khokhlova}
\affiliation{Attosecond Quantum Physics Laboratory, Department of Physics, King's College London, Strand Campus, London WC2R 2LS, UK}

\author{E.~Pisanty}
\affiliation{Attosecond Quantum Physics Laboratory, Department of Physics, King's College London, Strand Campus, London WC2R 2LS, UK}

\author{C. Figueira de Morisson Faria}
\affiliation{Department of Physics and Astronomy, University College London, Gower Street, London WC1E 6BT, UK}

\begin{abstract}
	While the interface between strong-field physics and quantum optics offers a unique regime for combining extreme nonlinearity with quantum optical resources, its potential for generating non-classical states of light remains largely unexplored.~Standard protocols for generating optical Schrödinger cat states, such as photon subtraction from squeezed light, are inherently limited in the achievable macroscopicity of the state and its scalability.~In this work, we bridge this gap by demonstrating that above-threshold ionization driven by bright squeezed light provides a strong-field analogue of photon subtraction, where photoelectron detection acts as a high-intensity heralding mechanism, enabling the generation of large amplitude optical Schrödinger cat states.~We characterize the resulting non-Gaussian features and show that they can be tuned via the detected photoelectron momentum, and study their robustness against the experimental imperfections arising from finite momentum resolution at the heralding step.~Despite the noise, we show that the generated states can be manipulated to violate a Bell inequality, thereby highlighting their potential for foundational and practical applications.~Our results establish strong-field processes as a scalable platform for macroscopic quantum state engineering, opening a route to quantum optics in previously inaccessible regimes.
\end{abstract}
\maketitle

\section{INTRODUCTION}
The distinction between classical and quantum technologies lies in our ability to generate, manipulate and exploit non-classical states and their correlations---properties of physical systems that require quantum mechanics for description.~Across applications ranging from computation and simulation~\cite{georgescu_quantum_2014}, to sensing~\cite{degen_quantum_2017} and communication~\cite{portmann_security_2022}, resources such as entanglement and quantum coherence enable capabilities beyond those achievable within classical physics~\cite{horodecki_quantum_2009}.~A central challenge, therefore, is to identify physical platforms that permit these resources to be generated, engineered, and measured with high levels of precision~\cite{acin_quantum_2018}.~In this pursuit, photonics has emerged over the past few decades as a preeminent platform~\cite{gisin_quantum_2007,kok_linear_2007,giovannetti_advances_2011,aspuru-guzik_photonic_2012,pirandola_advances_2018,usenko_continuous-variable_2026}.~Owing to its intrinsic robustness to decoherence and the sophisticated control afforded by both linear and nonlinear optical operations, photonic systems have demonstrated remarkable versatility in generating and manipulating a vast landscape of non-classical states of light, both in purely optical setups and within hybrid architectures~\cite{van_loock_optical_2011,andersen_hybrid_2015}.

In this regard, nonlinear optics stands as a cornerstone for the generation of non-classical states of light, enabling the production of single-photon states and photon pairs~\cite{burnham_observation_1970,hong_measurement_1987}, foundational to discrete-variable protocols; as well as continuous-variable resources such as squeezed states~\cite{slusher_observation_1985,wu_squeezed_1987,schneider_generation_1998,lam_optimization_1999}, where reduced noise in selected quadratures enhances measurement sensitivity~\cite{walls_squeezed_1983,xiao_precision_1987,abadie_gravitational_2011}.~Furthermore, when combined with non-Gaussian operations~\cite{dakna_generating_1997,ourjoumtsev_generating_2006,neergaard-nielsen_generation_2006,ourjoumtsev_generation_2007,hacker_deterministic_2019}, nonlinear optics allows for engineering increasingly complex, and more diverse, quantum light.~A paradigmatic example is the optical Schrödinger cat state:~a coherent superposition of two coherent states with equal amplitude but opposite phase.~These states are of central interest for a variety of applications in quantum information science~\cite{van_enk_entangled_2001,gilchrist_schrodinger_2004,lund_fault-tolerant_2008,lee_teleportation_2011} as well as from a more fundamental perspective for probing the quantum-to-classical transition~\cite{zavatta_quantum--classical_2004,paavola_finite-time_2011}, both of which frequently intersect in the study of nonlocal behaviors via Bell inequality violations~\cite{einstein_can_1935,bell_einstein_1964,brunner_bell_2014}. However, the feasibility of these demonstrations, and the broader utility of optical cat states, is crucially dictated by their size, defined by the separation of the coherent state components in phase-space.~As current generation techniques typically yield only small-amplitude superpositions~\cite{ourjoumtsev_generating_2006,neergaard-nielsen_generation_2006,ourjoumtsev_generation_2007,hacker_deterministic_2019}, scaling these states remains a major outstanding challenge~\cite{laghaout_amplification_2013,sychev_enlargement_2017}.

\begin{figure}
	\centering
	\includegraphics[width=1\columnwidth]{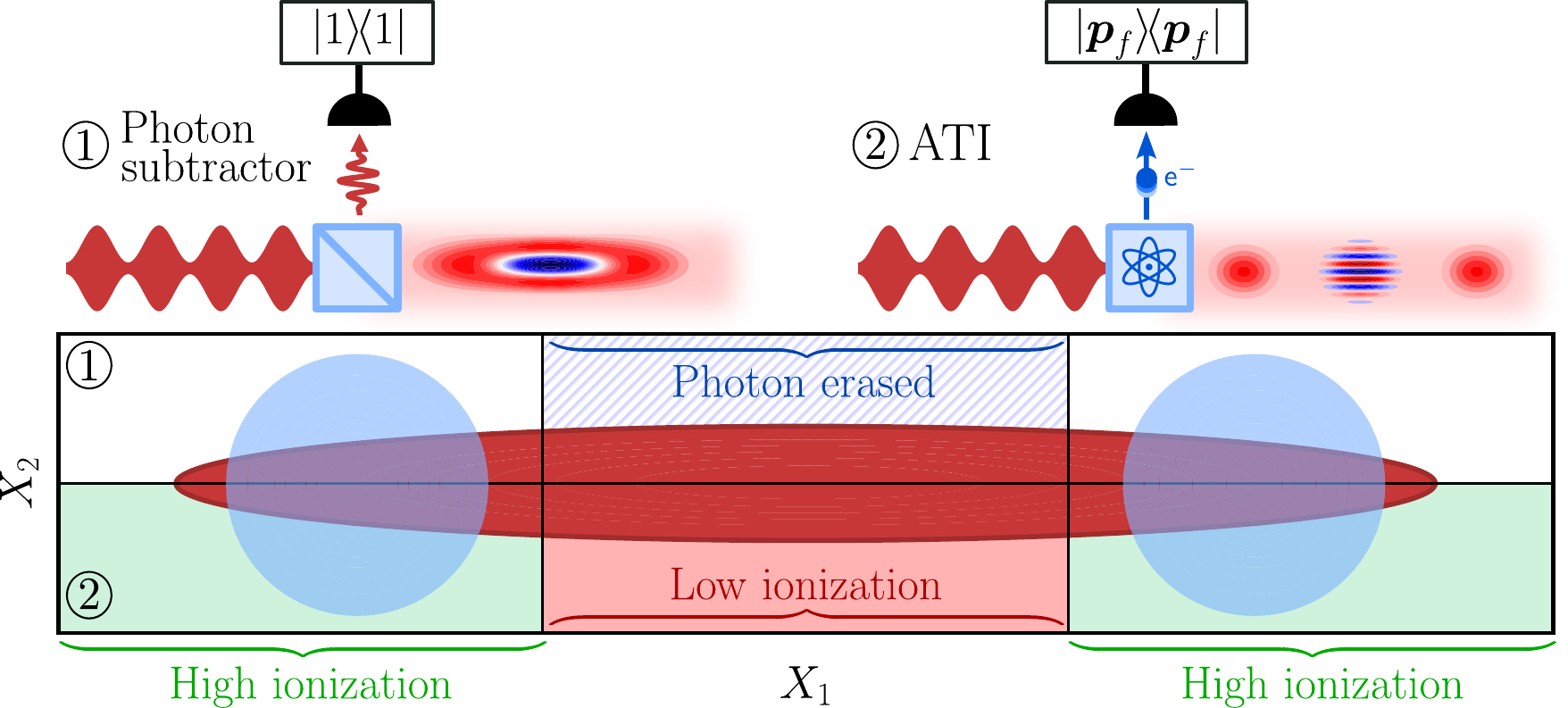}
	\caption{Pictorial representation highlighting the similarities between photon subtraction and photoelectron heralding. In the first case (\ding{172}), a low intensity squeezed vacuum state passes through a high transmissivity beam splitter; the detection of a single-photon state on the reflected mode heralds the presence of a small optical Schrödinger cat state in the transmitted one.~In the second case (\ding{173}), a bright squeezed vacuum state interacts with an atomic system; the generation of high-energy photoelectrons heralds the presence of a large optical Schrödinger cat state.~In both cases, the detection of either a single photon on the reflected mode or high-energy electrons effectively induces a phase-space filter (highlighted by the dashed blue and red regions respectively) which effectively removes part of the driving field, allowing the remaining components to coherently interfere and thus generate the desired superposition.}
	\label{Fig:Scheme}
\end{figure}

To identify mechanisms capable of scaling such states, it is natural to turn to physical regimes that inherently operate at large field amplitudes.~In this context, strong-field physics provides a particularly promising setting.~Traditionally, this field explores light-matter interactions driven by intense electromagnetic fields, giving rise to highly nonlinear phenomena, such as exotic ionization regimes~\cite{keldysh_ionization_1965,dimauro_50_2014,amini_symphony_2019} and the generation of ultrafast pulses in extreme spectral regimes~\cite{antoine_attosecond_1996,drescher_x-ray_2001,paul_observation_2001}, that form the foundation of attosecond science~\cite{corkum_attosecond_2007,krausz_attosecond_2009}.~Recently, the convergence of this field with quantum optics has birthed the interdisciplinary frontier of strong-field quantum optics~\cite{cruz-rodriguez_quantum_2024,stammer_colloquium_2025}, aiming to harness these extreme regimes for ultrafast quantum optical applications~\cite{stammer_metrological_2024,sennary_attosecond_2025,sennary_attosecond_2026}.~Within this framework, leveraging either intrinsic light-matter correlations or externally engineered resources, such as bright squeezed fields~\cite{spasibko_multiphoton_2017,manceau_indefinite-mean_2019,rasputnyi_high_2024,heimerl_multiphoton_2024,heimerl_driving_2025,lemieux_photon_2025,tzur_measuring_2025,kern_single-shot_2026,lyu_attosecond_2026}, has enabled the uncovering of novel physics~\cite{gorlach_high-harmonic_2023,even_tzur_photon-statistics_2023,wang_high-order_2023,lyu_effect_2025,liu_atomic_2025,rivera-dean_structured_2025,stammer_fluctuation-induced_2026}, the development of new characterization methods~\cite{tzur_measuring_2025,rivera-dean_attosecond_2026,singh_interferometrically_2026}, and the demonstration of non-classical states of light across a broad spectral range~\cite{lange_electron-correlation-induced_2024,stammer_squeezing_2023,theidel_evidence_2024,theidel_observation_2025,yi_generation_2025}. Critically, the integration of non-Gaussian operations has opened pathways toward more complex non-classical states of light~\cite{lewenstein_generation_2021,stammer_high_2022,stammer_quantum_2023,stammer_information-theoretic_2026,theidel_sub-poissonian_2026}, in some instances producing states that share features with high-photon number Schrödinger cat states, albeit with limited phase-space separation.

In this work, we investigate the strong-field process of above-threshold ionization (ATI) driven by a bright squeezed vacuum (BSV) field, casting it as a strong-field analogue of standard quantum optical protocols in which small Schrödinger cat states are generated via single-photon subtraction from weak squeezed states~\cite{dakna_generating_1997,ourjoumtsev_generating_2006} [Fig.~\ref{Fig:Scheme}].~By instead \emph{subtracting} the high-energy photoelectrons produced in the ATI process, we show that this mechanism enables the generation of large amplitude optical-like Schrödinger cat states with controllable quantum features.~We further analyze how resolution-limited photoelectron heralding impacts the purity and non-classicality of the resulting states. Finally, we assess the capability of these states to violate a Bell inequality across different momentum values, specifically examining the robustness of these nonlocal correlations against imperfect heralding resolution.~Taken together, our results identify strong-field processes as a viable route to overcoming the scaling limitations of conventional quantum optical approaches, opening a pathway toward the generation of large optical Schrödinger cat states in previously unexplored regimes.

The manuscript is organized as follows.~In Sec.~\ref{Sec:Theory}, we provide the mathematical foundation for our results by first introducing photon subtraction and establishing an analogy with the photoelectron heralding considered here, and then presenting a more technical discussion of the detailed form of the ATI-heralded state.~Section~\ref{Sec:Results} presents the main results of our analysis, including the non-Gaussian features of the generated state, their sensitivity to finite measurement resolution in the heralding step, and their use for the violation of a Bell inequality.~Finally, we conclude in Sec.~\ref{Sec:Dicussion}, where we summarize our results and provide an outlook for future research.

\section{THEORY BACKGROUND}\label{Sec:Theory}
To provide a clear foundation for our results, this section is organized in two parts.~In Sec.~\ref{Sec:Intution}, we revisit the photon subtraction protocol theoretically proposed in Ref.~\cite{dakna_generating_1997} and experimentally implemented in Ref.~\cite{ourjoumtsev_generating_2006}, highlighting both its operational mechanism and its scaling bottlenecks.~We then use these insights to provide the physical picture motivating how the use of ATI driven by bright squeezed light can provide a natural strong-field alternative, a conceptual bridge that we formalize in Sec.~\ref{Sec:maths}.

\subsection{From photon subtraction to photoelectron heralding}\label{Sec:Intution}
When discussing photon subtraction, we refer to the operation in which a photon is \emph{extracted} from a given quantum state of light after it passes through a beam splitter, typically of low reflectivity. Since an ideal beam splitter conserves the total number of photonic excitations, it therefore redistributes all input excitations between the reflected and transmitted modes according to its reflectivity-transmissivity ratio.~Conditioning on the detection of a photon in the reflected mode induces a transformation in the transmitted mode which, in the Fock basis, maps each $n$-photon component onto a $(n-1)$-photon component, that is, effectively subtracting a photon in the transmitted mode.~This is seen explicitly  by expanding the input state $\ket{\Phi}$ in the Fock basis
\begin{equation}
	\ket{\Phi} = \sum_{n=0}^\infty \dfrac{c_n}{\sqrt{n!}} \hat{a}_{\text{in}}^{\dagger n} \ket{0},
\end{equation}
where $c_n = \braket{n}{\Phi}$ and $\hat{a}^\dagger_{\text{in}}$ is the creation operator acting on the input mode. After passing through a beam splitter of transmissivity $t$, the state becomes
\begin{equation}
	\begin{aligned}
	\ket{\Phi}
		= \sum^\infty_{n=0} \dfrac{c_n}{\sqrt{n!}}
				\sum_{l=0}^n& \mqty(n \\ l) t^{(n-l)/2}(1-t)^{l/2}
				\\&\hspace{1cm} \times
					 (\hat{a}_t^{\dagger})^{n-l}(\hat{a}_r^{\dagger})^{l}
						\ket{0_t,0_r},
	\end{aligned}
\end{equation}
where $\hat{a}_t^\dagger$ and $\hat{a}_r^\dagger$ represent the creation operators acting on the transmitted and reflected output ports, respectively.~Projecting the reflected mode onto a single-photon state $\ket{1_r}$ yields, up to a normalization factor,
\begin{equation}
	\ket{\Phi_t}
		= \sum^{\infty}_{n=1}
				c_n
				 	t^{(n-1)/2} (1-t)(\hat{a}_t^{\dagger})^{n-1}
				 	\ket{0_t},
\end{equation}
confirming that each $n$-photon component of the input mode is mapped onto a $(n-1)$-photon component in the transmitted mode.
 
Interestingly, this operation can also be interpreted as an \emph{erasure} process acting on the input state.~This becomes more transparent when expanding $\ket{\Phi}$ in the overcomplete basis of coherent states
\begin{equation}\label{Eq:state:coh:exp}
	\ket{\Phi} 
		= 	\int \dd^2 \alpha \ c(\alpha)\ket{\alpha},
\end{equation}
where $c(\alpha) = \pi^{-1} \braket{\alpha}{\Phi}$.~After the beam splitter, the state is
\begin{equation}\label{Eq:BS:state}
	\ket{\Phi_{\text{BS}}} 
		= 	\int \dd^2 \alpha \ c(\alpha)\relaxket{\alpha\sqrt{t},-\alpha\sqrt{1-t}}.
\end{equation}
Upon projecting the reflected mode onto a single-photon state, we find, up to normalization,
\begin{equation}
	\begin{aligned}
	\ket{\Phi_{t}}\label{Eq:subtracted:state}
		&=\sqrt{1-t} \int \dd^2 \alpha 
			\ c(\alpha) \alpha e^{-(1-t)\abs{\alpha}^2/2} \relaxket{\alpha\sqrt{t}}
		\\& \equiv \int \dd^2 \alpha \ c_t(\alpha)\relaxket{\alpha\sqrt{t}}.
	\end{aligned}
\end{equation}
From this expression, it becomes clear that the coherent state amplitude at the origin is suppressed:~while initially the vacuum component $\ket{\alpha=0}$ contributes with amplitude $c(0)$, after conditioning one has $c_t(0) = 0$ [case \ding{172} in Fig.~\ref{Fig:Scheme}].~In this sense, photon subtraction acts as an \emph{erasure of the vacuum component} in phase-space. 

\begin{figure}
	\centering
	\includegraphics[width=1\columnwidth]{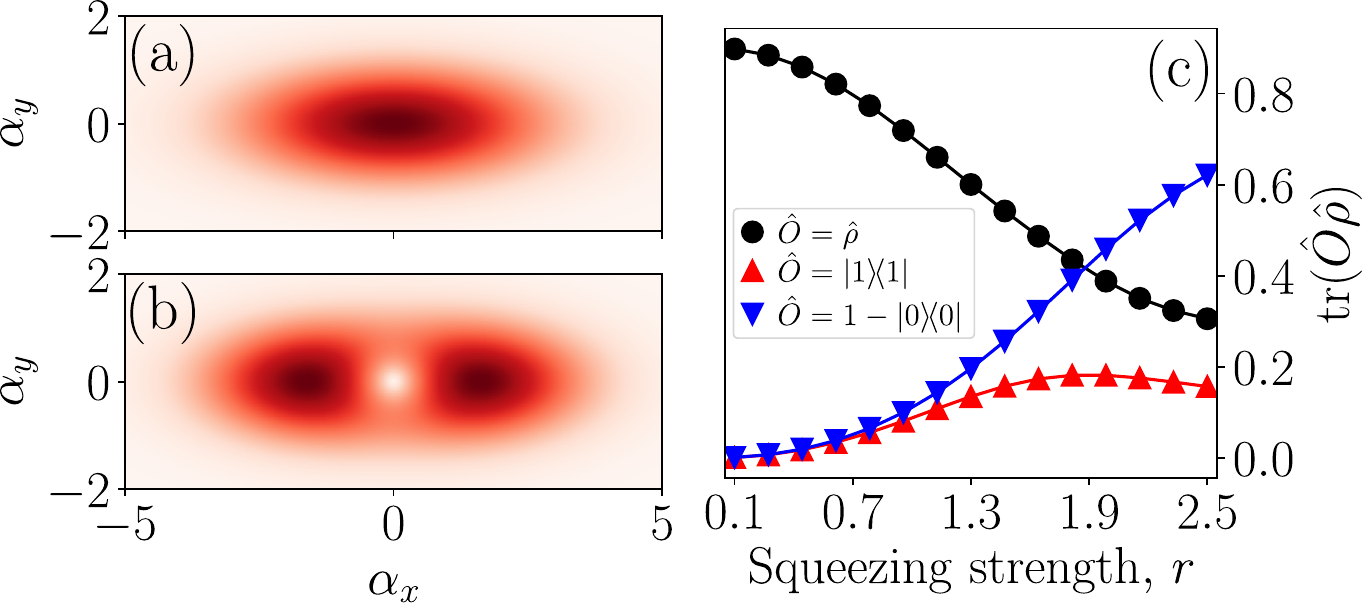}
	\caption{Norm of the coherent state probability amplitudes before [$\abs{c(\alpha)}$, panel (a)] and after [$\abs{c_t(\alpha)}$, panel (b)] photon subtraction for a squeezed state input.~Panel (c) shows the expectation value of different operators $\hat{O}$ with respect to $\hat{\rho} = \dyad{\Phi}$ in Eq.~\eqref{Eq:BS:state}.~Specifically, we display the purity ($\hat{O}=\hat{\rho}$, black curve with circular markers); the probability of detecting a single photon in the reflected mode ($\hat{O} =\dyad{1}$, red curve with upward triangles); and the probability of detecting any other state than a vacuum state in the reflected mode ($\hat{O} = \mathbbm{1}-\dyad{0}$, blue curve with downward triangles).}
	\label{Fig:Photon:subtractor:example}
\end{figure}

However, the extent to which operations performed in the reflected mode can modify the properties of the transmitted mode depends crucially on whether the two output modes of the beam splitter are entangled.~If, after the beam splitter, the state can be written as $\ket{\Phi_t}\otimes\ket{\Phi_r}$, measurements performed on the reflected mode do not alter the transmitted state.~A necessary condition for non-trivial backaction is therefore that the input state be non-classical~\cite{rivera-dean_condition_202X}.~Nevertheless, even if this condition is fulfilled, it does not guarantee that the post-selected state exhibits non-trivial features.~This is precisely why squeezed vacuum states are particularly well suited in this context.

For a squeezed vacuum state, the $c(\alpha)$ probability amplitude (related to the Husimi $Q$-function~\cite{SchleichBookCh12} via $Q(\alpha) = \pi \abs{c(\alpha)}^2$) has the form of an elongated Gaussian along one quadrature, while being squeezed along the orthogonal one [Fig.~\ref{Fig:Photon:subtractor:example}~(a)]. In this case, as seen from Eq.~\eqref{Eq:subtracted:state}, the heralding process breaks the Gaussian character through the additional linear factor $\alpha$, enforcing $c_t(\alpha) = 0$ at $\alpha = 0$ while leaving two symmetric regions around the origin where $\abs{c(\alpha)} > 0$ [Fig.~\ref{Fig:Photon:subtractor:example}~(b)] which coherently interfere. In this context, Eq.~\eqref{Eq:subtracted:state} can be written as
\begin{equation}\label{Eq:Ph:sub:state}
	\ket{\Phi_t}
		= \int_{\mathcal{C}} \dd \alpha \ c_t(\alpha)\relaxket{\alpha\sqrt{t}}
			-  \int_{\mathcal{C}}  \dd \alpha \ c_t(-\alpha) \relaxket{-\alpha\sqrt{t}},
\end{equation}
where $\mathcal{C} :=\{\alpha_x \in \mathbbm{R}^+\backslash \{0\}, \alpha_y\in \mathbbm{R}\backslash\{0\}\}$ with $\alpha = \alpha_x + i \alpha_y$.~The resulting state therefore bears a close resemblance to an odd optical Schrödinger cat state of the form $\ket{\alpha} - \ket{-\alpha}$~\cite{dakna_generating_1997,ourjoumtsev_generating_2006}.

Nonetheless, as with any heralding-based protocol, the generation of Eq.~\eqref{Eq:Ph:sub:state} is inherently probabilistic, as it depends on the success rate of detecting a single-photon excitation in the reflected mode.~This probabilistic nature is precisely what limits the scalability of the technique.~While, in principle, the method remains applicable to arbitrarily squeezed states, the probability of successful single-photon heralding decreases as the squeezing parameter becomes sufficiently large~[Fig.~\ref{Fig:C_matrix}~(c), red curve].~To circumvent this, one might consider a broader heralding condition by projecting onto any non-vacuum excitation, an operation described by the measurement operator $\mathbbm{1}-\dyad{0_r}$.~Although this approach successfully increases the heralding probability [Fig.~\ref{Fig:C_matrix}~(c), blue curve] and effectively mimics single-photon subtraction for small squeezing parameters, as squeezing increases the heralded state becomes increasingly mixed and rapidly loses its purity [Fig.~\ref{Fig:C_matrix}~(c), black curve].

It is within this context that we identify ATI as a strong-field analogue to the photon subtraction protocol.~In ATI, atoms are ionized through the absorption of more photons than necessary to overcome their ionization potential~\cite{agostini_free-free_1979,amini_symphony_2019}.~For low-frequency, intense driving fields, this can result in electrons with an energy excess of dozens or even hundreds of photons~\cite{chin_observation_1983,hansch_resonant_1997,milosevic_above-threshold_2006,becker_plateau_2018}.~Thus, generating such high-energy photoelectrons necessitates intense driving fields, a situation that becomes particularly relevant for our purposes when considering BSV fields as drivers~\cite{spasibko_multiphoton_2017,manceau_indefinite-mean_2019,rasputnyi_high_2024,heimerl_multiphoton_2024,heimerl_driving_2025,lemieux_photon_2025,tzur_measuring_2025,kern_single-shot_2026}.~From the perspective of Eq.~\eqref{Eq:state:coh:exp}, the ATI process is predominantly sensitive to the tails of the distribution $c(\alpha)$. In these regions, the coherent state amplitudes are sufficiently large to trigger ionization, whereas contributions from the central phase-space region remain negligible due to insufficient field strength [case \ding{173} in Fig.~\ref{Fig:Scheme}].~Consequently, ATI effectively selects phase-space components associated with large amplitudes.~This observation suggests a direct analogy with heralding in photon subtraction protocols:~the atom serves as the entangling medium, while the detection of photoelectrons acts as the heralding event.~As a result, the post-selected optical state is expected to resemble the structure of Eq.~\eqref{Eq:Ph:sub:state}, where the effective integration domain $\mathcal{C}$ is now defined by the regions of phase-space where the ionization probability is significant. In the following section, we formalize this connection.

\subsection{ATI driven by bright squeezed vacuum}\label{Sec:maths}
Having established the physical picture underlying our analysis, we now proceed to develop the mathematical framework.~In particular, we focus on the fundamental equations that establish the connection with the heralding mechanism, while a detailed, step-by-step derivation is provided in the Supplementary Material (Secs.~\ref{Sec:SM:Preliminaries} to \ref{Sec:SM:smcl:SPA}).~Atomic units are used throughout the text unless otherwise specified.

We begin with the light-matter interaction between an atomic system, initially in its ground state, and a BSV field. Working in the single-active electron and dipole approximations, and adopting the interaction picture with respect to the free-field Hamiltonian, the time evolution operator $\hat{U}(t,t_0)$ describing the light-matter interaction dynamics satisfies the Schrödinger equation
\begin{equation}\label{Eq:initial:HE:prop}
	i \pdv{\hat{U}(t)}{t}
		= \big[
				\hat{H}_{\text{at}}
				+ \hat{\boldsymbol{r}}\cdot \hat{\boldsymbol{E}}(t)
			\big] \hat{U}(t).
\end{equation}
Here, $\hat{H}_{\text{at}}$ denotes the atomic Hamiltonian, while $\hat{\boldsymbol{r}}\cdot \hat{\boldsymbol{E}}(t)$ describes the light-matter interaction in the length gauge. The electric field operator is given by
\begin{equation}
	\begin{aligned}
	\hat{\boldsymbol{E}}(t) 
		&= -i\sum_{\mu,q=1} \boldsymbol{\epsilon}_\mu g(\omega_q)[\hat{a}_{q,\mu}e^{-i\omega_q t} + \text{h.c.}] 
		\\&\equiv \sum_{\mu,q} \hat{\boldsymbol{E}}_{q,\mu}(t),
	\end{aligned}
\end{equation}
where $\hat{a}_{q,\mu}$ annihilates a photon in mode $(q,\mu)$, $\boldsymbol{\epsilon}_\mu$ denotes the polarization direction $(\mu = \perp,\parallel)$, and $g(\omega_q) = \sqrt{\hbar \omega_q/(2\epsilon_0V)}$ is the mode-dependent coupling constant, with $V$ the quantization volume.~Therefore, at any time $t \geq t_0$, the quantum state of the joint light-matter system is given by $\ket{\Psi(t)} = \hat{U}(t,t_0)\ket{\Psi(t_0)}$, with the initial state taken here as
\begin{equation}\label{Eq:init:state}
	\ket{\Psi(t_0)}
		= \ket{\text{g}}\otimes \hat{S}_L(r)\ket{\boldsymbol{0}},
\end{equation}
where $\ket{\text{g}}$ denotes the atomic ground state and $\hat{S}_{L}(r) = \exp[r(\hat{a}_{L}^2 - \text{h.c.})]$ the squeezing operator acting on the fundamental mode  $L\equiv \{q=1,\parallel \}$, while all other modes remain in a vacuum state, such that $\ket{\boldsymbol{0}} \equiv \bigotimes_{q,\mu}\ket{0_{q,\mu}}$. In what follows, we take $r \in \mathbbm{R}$, which is physically sufficient since, in the absence of an external phase reference (such as a coherent displacement or a second driving field), the characteristics of the interaction depend only on the magnitude of the squeezing.

Following the intuition developed in the previous subsection, the role of photoelectron heralding in ATI is best understood when expressing the initial state [Eq.~\eqref{Eq:init:state}] in the coherent state basis of the driving field mode,
\begin{equation}
	\ket{\Psi(t_0)}
		= \int \dd^2 \alpha \ 
				c(\alpha) \ket{\text{g}} \otimes \hat{D}_L(\alpha)\ket{\boldsymbol{0}},
\end{equation}
where $\hat{D}_{L}(\alpha) = \exp[-\alpha \hat{a}_L + \text{h.c.}]$ is the displacement operator acting on the fundamental mode.~Within this representation, the time-evolved joint state is
\begin{equation}\label{Eq:time:evolved:state}
	\ket{\Psi(t)}
		= \int \dd^2 \alpha \ c(\alpha)
				\hat{D}_L(\alpha)\hat{U}_{\alpha}(t,t_0)
					\ket{\text{g}}\otimes \ket{\boldsymbol{0}},
\end{equation}
where we define $\hat{U}_{\alpha}(t,t_0) \equiv \hat{D}_L(\alpha)^\dagger \hat{U}(t,t_0)\hat{D}_L(\alpha)$. From Eq.~\eqref{Eq:initial:HE:prop}, it follows that $\hat{U}_\alpha(t,t_0)$ satisfies
\begin{equation}
	i \pdv{\hat{U}_\alpha(t)}{t}
		= \Big[
				\hat{H}_{\text{at}}
				+\hat{\boldsymbol{r}}\cdot 
				\big(
					\hat{\boldsymbol{E}}(t)
					+ \boldsymbol{E}_{\text{cl}}(t;\alpha)
				\big)
			\Big]\hat{U}_{\alpha}(t),
\end{equation}
where $\boldsymbol{E}_{\text{cl}}(t;\alpha) = \mel{\alpha}{\hat{\boldsymbol{E}}_L(t)}{\alpha}$ denotes the classical field associated with the coherent state amplitude $\alpha$.

To probe the formation of the desired non-classical optical states, we condition the joint state in Eq.~\eqref{Eq:time:evolved:state} on the detection of a photoelectron with final momentum $\boldsymbol{p}_f$. The heralded optical state is then
\begin{equation}\label{Eq:final:ATI:state}
	\begin{aligned}
	\ket{\Phi(\boldsymbol{p}_f,t)}
		&= \braket{\boldsymbol{p}_f}{\Psi(t)}
		\\&
		= \int \dd^2\alpha	\
				c(\alpha)
				\hat{D}_L(\alpha)
				\mel{\boldsymbol{p}_f}{\hat{U}_{\alpha}(t,t_0)}{\text{g}} \ket{\boldsymbol{0}}.
	\end{aligned}
\end{equation}
We emphasize that, unlike in semiclassical approaches~\cite{faria_it_2020}, the matrix element $\mel{\boldsymbol{p}_f}{\hat{U}_{\alpha}(t,t_0)}{\text{g}}$ remains an operator acting on the Hilbert space of the optical modes.~To evaluate this term, we employ a combination of Dyson expansion and Feynman path-integral techniques~\cite{kleinert_path_2009}, as commonly used in semiclassical strong-field theories~\cite{milosevic_phase_2013,lai_influence_2015,maxwell_strong-field_2019,faria_it_2020}, and recently extended to the quantum optical domain~\cite{mao_benchmarking_2025}.~A detailed derivation is provided in the Supplementary Material (Secs.~\ref{Sec:SM:Dyson} and \ref{Sec:SM:Path:Int}).

Restricting our analysis to direct ATI events~\cite{amini_symphony_2019}, i.e., neglecting rescattering contributions, we obtain,
\begin{equation}\label{Eq:ATI:propagator}
	\begin{aligned}
	\mel{\boldsymbol{p}_f}{\hat{U}_{\alpha}(t,t_0)}{\text{g}}
		&\approx i\int \dd t'
			e^{
				i S(\boldsymbol{p}_f,t,t';\alpha)}
			\\&\quad\times	
				\langle \Tilde{\boldsymbol{p}}_f(\alpha)+\boldsymbol{A}_{\text{cl}}(t';\alpha)\rvert
				\hat{\boldsymbol{r}}\lvert \text{g}\rangle\cdot \hat{\boldsymbol{E}}(t'),
	\end{aligned}
\end{equation}
where the semiclassical action is given by
\begin{equation}
	\begin{aligned}
	S(\boldsymbol{p}_f,t,t';\alpha)
		&=I_p(t'-t_0)
			\\&\quad
			-\frac{i}{2}
					\int^{t}_{t'}\dd \tau
					\big[
						\Tilde{\boldsymbol{p}}_f(\alpha)
						+ \boldsymbol{A}_{\text{cl}}(\tau;\alpha)
					\big]^2.
	\end{aligned}
\end{equation}
Here, $\boldsymbol{A}_{\text{cl}}(t;\alpha) = -\int \dd t \boldsymbol{E}_{\text{cl}}(t;\alpha)$ is the classical vector potential and $\tilde{\boldsymbol{p}}_f(\alpha) = \boldsymbol{p}_f -\boldsymbol{A}_{\text{cl}}(t;\alpha)$ is the canonical momentum.~Notably, the canonical momentum depends explicitly on $\alpha$ and, unlike in the semiclassical case, is not globally conserved in general.~It remains, however, locally well-defined within each coherent state component of the superposition.
	
Equation~\ref{Eq:ATI:propagator} is derived under the assumption that the electronic backaction on the quantum optical modes is negligible within the support of $c(\alpha)$, a condition typically satisfied for strong-field interactions in the near-infrared regime~\cite{rivera-dean_light-matter_2022}.~We note, however, that strong squeezing can effectively enhance the light-matter coupling (with $g(\omega_L)\sim 10^{-8}$ a.u.)~and potentially lead to non-negligible backaction in displaced squeezed states with coherent state components intense enough to drive strong-field phenomena by themselves~\cite{rivera-dean_microscopic_2025}.~In contrast, for BSV drivers, this effect remains strongly suppressed due to the absence of a large coherent offset.

\begin{figure}
	\centering
	\includegraphics[width=1\columnwidth]{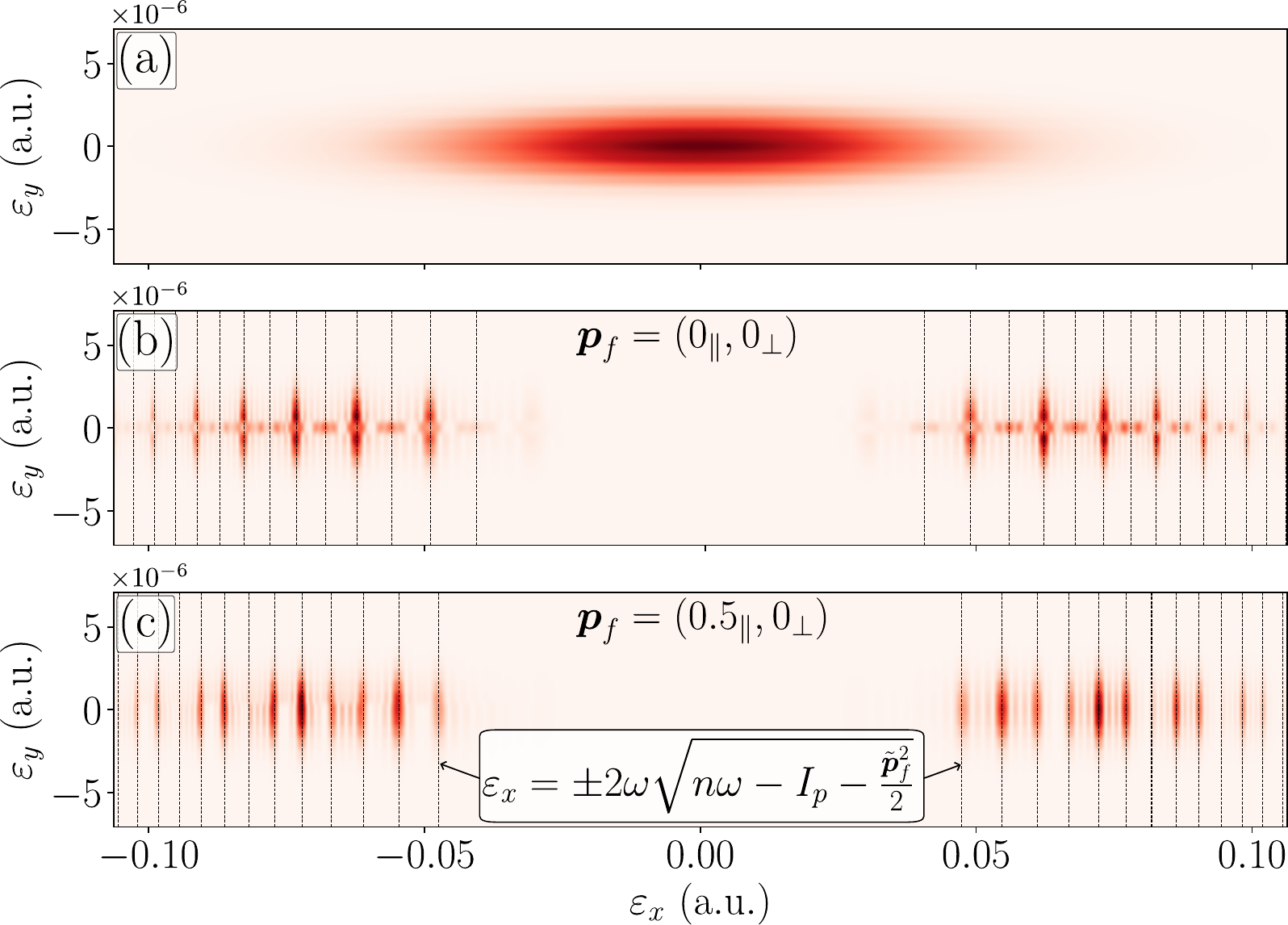}
	\caption{Norm of the coherent state probability amplitudes before [panel (a)] and after [panel (b), (c)] the ATI heralding. In panel (b), the heralding is performed for $\boldsymbol{p}_f=\boldsymbol{0}$, while in panel (c) for $\boldsymbol{p}_f = (0.5_{\parallel},0_\perp)$.~The vertical dashed lines highlight the location of the ATI rings, computed as indicated in the inset of panel (c). The numerical evaluation is carried out for $r=10$ and $g(\omega_L) = 10^{-6}$ a.u., considering a light-matter interaction lasting four optical cycles ($\omega_L = 0.057$ a.u.) and involving hydrogen atoms ($I_p \simeq 0.5$ a.u.).~A discrete coherent state expansion is employed on a rectangular lattice with spacing $k=0.2\sqrt{\pi}$ (see Supplementary Material~\ref{Sec:SM:Numerics}).}
	\label{Fig:C_matrix}
\end{figure}

Combining Eqs.~\eqref{Eq:final:ATI:state} and \eqref{Eq:ATI:propagator}, and neglecting vacuum excitation contributions proportional to $g(\omega_L)$, we arrive at
\begin{equation}\label{Eq:heralded:state}
	\ket{\Phi(\boldsymbol{p}_f,t)}
			= \int \dd^2\alpha
			\underbrace{
				\int \dd t'
				c(\alpha)
				b(\boldsymbol{p}_f,t';\alpha)}_{\tilde{c}(\alpha)}\ket{\alpha},
\end{equation}
where we have defined
\begin{equation}\label{Eq:direct:ATI:prob:amp}
	\begin{aligned}
	b(\boldsymbol{p}_f,t';\alpha)
		&= i e^{iS(\boldsymbol{p}_f,t,t';\alpha)}
			\\&\quad\times
			\langle \Tilde{\boldsymbol{p}}_f(\alpha)+\mathsf\boldsymbol{A}_{\text{cl}}(t';\alpha)\rvert
			\hat{\boldsymbol{r}}\lvert \text{g}\rangle\cdot \boldsymbol{E}_{\text{cl}}(t';\alpha),
	\end{aligned}
\end{equation}
corresponding to the semiclassical direct ATI probability amplitude when using the driving field $\boldsymbol{E}_{\text{cl}}(t;\alpha)$. Since these amplitudes become significant only for field strengths $\abs{E_{\text{cl}}(t)} \gtrsim 10^{-2}$ a.u., corresponding to $\abs{\alpha} \gtrsim 10^{6}$, while remaining negligible for smaller values, this demonstrates that ATI acts as a phase-space filter that selectively weights large amplitude coherent state components, providing a direct strong-field analogue of photon subtraction. This behavior is explicitly illustrated in Fig.~\ref{Fig:C_matrix}, where we consider a squeezed state with $r = 10$ and $g(\omega_L) = 10^{-6}$ a.u.~\footnote{We numerically restrict to smaller values of $g(\omega_L)$ than those typically estimated theoretically, as realistic values would lead to computational requirements beyond our current capabilities. This point is discussed in Sec.~\ref{Sec:Results} and in Sec.~\ref{Sec:SM:Numerics} of the Supplementary Material.}.

Prior to the interaction, the distribution $\abs{c(\alpha)}$ exhibits a Gaussian profile strongly squeezed along the $y$-axis [Fig.~\ref{Fig:C_matrix}~(a)]. Here, phase-space coordinates are expressed in terms of the electric field strength $\varepsilon(\alpha) = 2g(\omega_L)\alpha$. Following ATI heralding for different final electron momenta [panels (b) and (c)], the structure of the effective coefficient $\tilde{c}(\alpha)$ is fundamentally transformed.~Two pronounced lobes emerge symmetrically around the origin, marking the phase-space regions where the ATI events for $\boldsymbol{p}_f$ are most probable.~Crucially for the generation of optical Schrödinger cat-like states, the central region ($\abs{\varepsilon_x} \lesssim 0.01$ a.u.) is strongly suppressed, while each lobe possesses a rich internal structure.~Given the coherent superposition inherent to Eq.~\eqref{Eq:heralded:state}, one expects strong interference not only between the contributions from lobes placed at opposite sides of phase-space but also from the fine structure within each lobe.

Interestingly, the fine structure within each lobe depends sensitively on $\boldsymbol{p}_f$, and is rooted in the energy conservation of the ATI process, i.e., the formation of ATI rings~\cite{lewenstein_rings_1995}. Specifically, ATI events occur at energies satisfying
\begin{equation}
	\dfrac{\Tilde{\boldsymbol{p}}_f^2}{2}
		= n\omega_L -I_p - \dfrac{\varepsilon^2}{4\omega_L^2}, \ n\in \mathbbm{N},
\end{equation}
from which one obtains $\varepsilon = \pm 2\omega_L \sqrt{n\omega_L - I_p-\Tilde{\boldsymbol{p}}_f^2/2}$. These solutions, indicated by vertical dashed lines in Fig.~\ref{Fig:C_matrix}~(b) and (c), coincide with the maxima of $\abs{\tilde{c}(\alpha)}$. This alignment confirms that ATI peaks determine the dominant contributions to the coherent-state superposition, highlighting the tunability afforded by the detected electron momentum.~Finally, for a temporally symmetric ionization process, as for $\boldsymbol{p}_f = \boldsymbol{0}$ where ionization occurs at intensity maxima separated by half a cycle, the spacing between consecutive peaks in phase-space corresponds to twice the photon energy of the driving field [panel (b)].

\section{RESULTS}\label{Sec:Results}
Having established the capability of ATI to act as a massive photon subtraction mechanism, we now turn to the quantum optical and quantum information properties of the resulting heralding states.~In particular, we analyze their non-Gaussian features, how these are affected by imperfect heralding, and their potential for violating Bell inequalities, thereby highlighting both their foundational and practical relevance. However, we first discuss some details about the numerical analysis and provide some analytical insights regarding the large nature of the generated coherent state superpositions.

\subsection{Numerical analysis and macroscopic nature of the state}\label{Sec:SPA}
Before analyzing the quantum optical properties of the generated states, we briefly outline the numerical approach employed (see Supplementary Material~\ref{Sec:SM:Numerics} for more details).~To numerically implement Eq.~\eqref{Eq:heralded:state}, we use a discrete coherent state expansion on a lattice $\mathsf{L} = \{\ket{\alpha_{m,n}} : \alpha_{m,n} = k \sqrt{\pi}(m+in); m,n\in \mathbbm{Z}\}$, with $k=0.035$, ensuring completeness of the representation~\cite{bargmann_completeness_1971,perelomov_completeness_1971}.~In this framework, a key numerical challenge arises from the large phase-space support of highly squeezed states, which leads to prohibitively large coefficients matrices $c(\alpha)$.~Interestingly, however, is that the physically relevant parameter governing the electron dynamics is the field strength $\varepsilon(\alpha)$.~Therefore, to maintain the strong-field regime while keeping the computation tractable, we adopt a rescaling strategy~\cite{wang_high-order_2025}, increasing $g(\omega_L)$ to the order of $10^{-3}$ a.u., and correspondingly restricting the sampled coherent state amplitudes to $\abs{\alpha}\sim 10$, such that $\varepsilon \sim 10^{-2}$ a.u.~remains unchanged.~While this procedure effectively rescales the phase-space structure of the optical state, its qualitative features are preserved, which are the relevant aspects of our analysis.

Semiclassical-based methods may nevertheless provide analytical insight into the macroscopic character of the generated states under more realistic conditions.~Starting from Eq.~\eqref{Eq:heralded:state} and noting that strong squeezing localized $c(\alpha)$ in that direction, we approximate the state as
\begin{equation}
	\ket{\Phi(\boldsymbol{p}_f,t)}
		\simeq - i
			\int\dd \alpha_x
				c(\alpha_x)
					\int^{t}_{t_0} \dd t'
						b(\boldsymbol{p}_f,t;\alpha)
							\ket{\alpha}.
\end{equation}
From the definition of both $c(\alpha_x)$ and $b(\boldsymbol{p}_f,t;\alpha)$ [Eq.~\eqref{Eq:direct:ATI:prob:amp}], their product can be expressed as a slowly varying prefactor multiplied by a rapidly oscillating phase, the second governed by the effective action
\begin{equation}
	\begin{aligned}
		&S_{\text{QO}}(\tilde{\boldsymbol{p}}_f,t',\alpha_x)
		= I_p(t'-t_0)
		+
		\dfrac{i\varepsilon_x^2}{4g(\omega_L)^2e^{r}\cosh(r)},
		\\&\hspace{1cm}
		-\dfrac12
		\int_{t'}^{t} \dd \tau
		\Bigg\{
			\Big[
				\tilde{p}_{f,\parallel}
				+\dfrac{\varepsilon_x}{\omega_L}\cos(\omega_L \tau)
			\Big]^2
			+ \tilde{p}_{f,\perp}^2
		\Bigg\},
	\end{aligned}
\end{equation}
which contains both the semiclassical propagation phase of the electron and an additional contribution arising from the quantum statistics of the driver.~In analogy with stationary-phase arguments used in strong-field physics~\cite{lewenstein_theory_1994,olga_simpleman}, we analyze the regions of parameter space ($\boldsymbol{\mathsf{x}}\equiv (\alpha_x,t')$) where this phase becomes stationary, i.e., $\nabla_{\boldsymbol{\mathsf{x}}}S_{\text{QO}}(\boldsymbol{p}_f,\boldsymbol{\mathsf{x}}) = \boldsymbol{0}$. These stationary configurations provide qualitative insight into the regions of parameter space that may play a relevant role in shaping the coherent state superposition.~Such configurations are determined by
\begin{align}
	&\begin{aligned}
	&\pdv{S_{\text{QO}}}{t'} = 0
	\Rightarrow
	\\& \hspace{1cm}
	\dfrac12
	\bigg[ 
	\tilde{p}_{f,\parallel}
	+ \frac{\varepsilon_x}{\omega_L}\cos(\omega_L t')
	\bigg]^2
	+ \dfrac{\tilde{p}_{f,\perp}^2}{2}
	+ I_p = 0,\label{Eq:SP:time}
	\end{aligned}
	\\&
	\begin{aligned}\label{Eq:SP:varex}
		&\pdv{S_{\text{QO}}}{\varepsilon_x} = 0
		\Rightarrow
		\\&\hspace{1cm}
		\bigg\{\!
		-\dfrac{1}{\omega_L}
		\int^{t}_{t'} \dd \tau
			\Big[
				\tilde{p}_{f,\parallel}
				+\dfrac{\varepsilon_x}{\omega_L}
				\cos(\omega_L \tau)
			\Big]\cos(\omega_L \tau)
		\\&\hspace{1cm}+ 
			\dfrac{i\varepsilon_x}{2g(\omega_L)^2e^{r}\cosh(r)}
		\bigg\} = 0,
	\end{aligned}
\end{align}
where we have made the substitution $\alpha_x = \varepsilon_x/(2g(\omega_L))$.

\begin{figure}
	\centering
	\includegraphics[width=1\columnwidth]{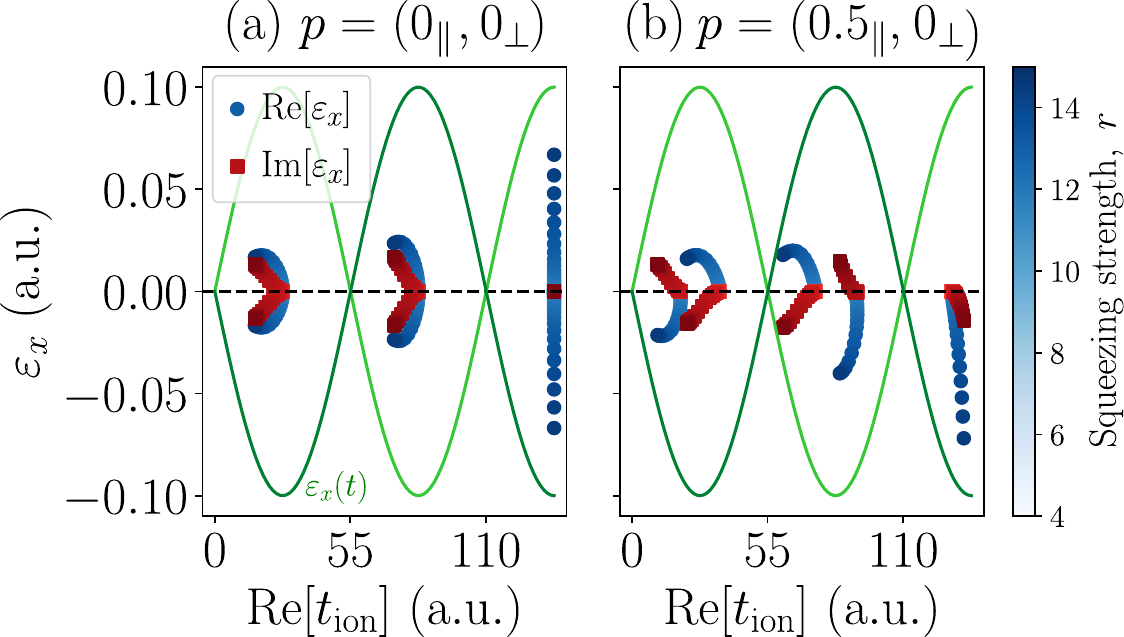}
	\caption{Solutions to Eqs.~\eqref{Eq:SP:time} and \eqref{Eq:SP:varex} for (a) $\boldsymbol{p}_f = \boldsymbol{0}$ and (b) $\boldsymbol{p}_f = (0.5_{\parallel},0_{\perp})$.~The real (blue circles) and imaginary (red squares) parts of $\varepsilon_x$ are shown as a function of the real part of the ionization time, with the opacity indicating the squeezing strength $r$.~The electric field is depicted by the green curves: the darker curve corresponds to $\alpha > 0$, while the lighter one to $\alpha < 0$, highlighting the presence of two solutions associated with different regions of phase-space.~In this analysis, we set $g(\omega_L) = 10^{-8}$ a.u., $\omega_L = 0.057$ a.u.,~and used hydrogen ($I_p \simeq 0.5$ a.u.) as the atomic system.}
	\label{Fig:saddles}
\end{figure}

Figure~\ref{Fig:saddles} shows the solutions to Eqs.~\eqref{Eq:SP:time} and \eqref{Eq:SP:varex} when setting $g(\omega_L) = 10^{-8}$ a.u., for two values of $\boldsymbol{p}_f$. Ionization times are found to occur around the extrema of the field, displayed in green color for both positive (light green) and negative (dark green) phase-space regions, with contributions arising from positive and negative field configurations. For zero final momentum [panel (a)], the symmetry of the ionization process leads to degenerate solutions for each ionization time with opposite phase, whereas this symmetry is lifted for finite longitudinal momentum [panel (b)], for which the sign of the field affects when ionization happens. Importantly, these stationary configurations correspond to field strengths $\abs{\varepsilon} \sim 10^{-2}$ a.u., which in turn correspond to coherent state amplitudes $\abs{\alpha}\sim  10^6$. This indicates that the relevant phase-space regions associated with the ATI amplitudes are located far from the origin and are distributed on both sides of phase-space.

We also observe a non-vanishing imaginary component of the field amplitude $\varepsilon_x$ for these solutions, which vanishes only when the ionization time coincides with the detection time or when squeezing is effectively absent [Fig.~\ref{Fig:saddles}~(a)]. Furthermore, this imaginary component increases with the squeezing strength, while at the same time the corresponding ionization times shift away from the extrema of the field towards earlier values. These observations indicate that the presence of squeezing continuously modifies the structure of the stationary solutions, in a manner consistent with previous studies showing that squeezing can influence the effective electron propagation dynamics~\cite{rivera-dean_microscopic_2025}.

With this in mind, we emphasize that this analysis is intended to provide analytical intuition on the macroscopic character of the generated high-photon-number superpositions.~The numerical results presented hereupon are instead obtained using the approach described at the beginning of this section.

\subsection{Non-Gaussian features of the coherent state superposition}
Optical Schrödinger cat states represent a paradigmatic class of non-Gaussian quantum states~\cite{walschaers_non-gaussian_2021} and, as such, phase-space representations provide a natural framework for characterizing their properties.~Among the various representations available in the literature~\cite{SchleichBookCh12}, the Wigner function is particularly well-suited for non-Gaussian states, as it enables both the explicit resolution of their phase-space features~\cite{SchleichBookCh3}, and the assessment of their potential advantages in quantum information applications over classical protocols~\cite{mari_positive_2012,rahimi-keshari_sufficient_2016}.~Formally, the Wigner function is defined as~\cite{royer_wigner_1977}
\begin{equation}\label{Eq:Wigner:function}
	W(\beta)
		= \frac{2}{\pi}
			\tr(\hat{\rho} \hat{D}(\beta)\hat{\Pi}\hat{D}(-\beta)),
\end{equation}
where $\hat{\Pi}$ is the parity operator, and $\text{Re}[\beta] \equiv X_1$ and $\text{Im}[\beta] \equiv X_2$ denote the optical quadratures.~While not restricted solely to classical states, for this class the Wigner function is non-negative and reduces to a Liouville probability distribution. However, in general, the Wigner function does not satisfy all the axioms of a classical probability distribution, a primary reason it is categorized as a quasiprobability distribution.~In particular, it can take negative values, an essential feature of any pure non-Gaussian state as established by Hudson's theorem~\cite{hudson_when_1974}. This negativity constitutes a definite signature of non-classicality that, in optical Schrödinger cat states, manifests as characteristic interference fringes in phase-space, making it especially well-suited for describing the states generated via our ATI protocol.

\begin{figure}
	\centering
	\includegraphics[width=1\columnwidth]{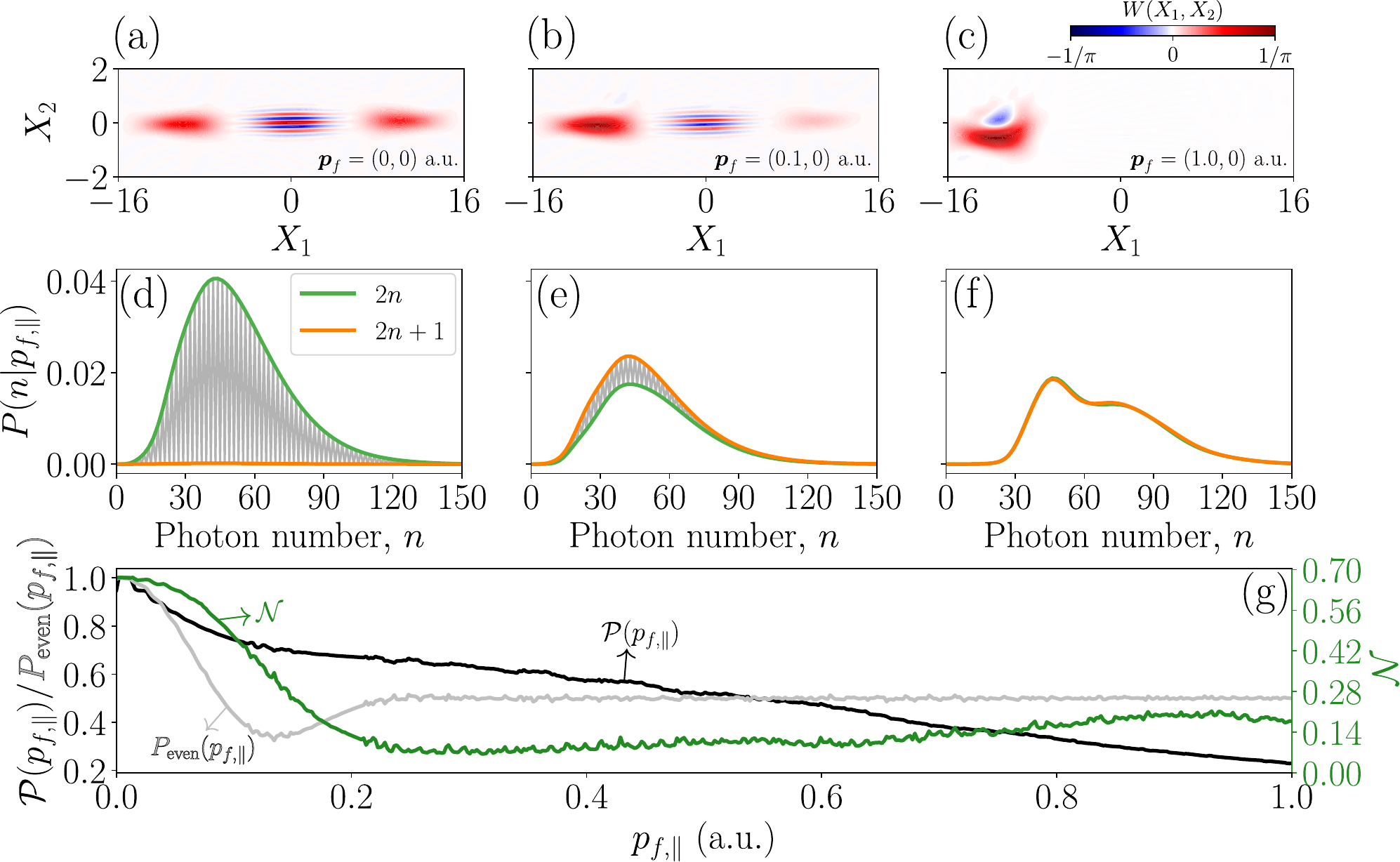}
	\caption{Non-Gaussian properties of the generated state, where each row presents a different measure.~(a)-(c) Wigner functions of the ATI-heralded state for photoelectrons with $p_{\parallel} = 0, 0.1, 1.0$ a.u., respectively, with $p_{\perp} = 0$ a.u.~held fixed in all cases.~(d)-(f) Photon number probability distributions corresponding to the Wigner functions shown in the upper row, with the green (orange) curve connecting the even (odd) photon numbers.~(g) Wigner negativity (green curve) of the resulting states as a function of the photoelectron energy, with the dark curve showing the photoelectron probability (normalized to maximum), and the gray one the probability of having even photon numbers in the state.~In this analysis, we fixed $g(\omega_L) = 5 \times 10^{-3}$ a.u., considering a light-matter interaction lasting four optical cycles ($\omega_L = 0.057$ a.u.) with hydrogen ($I_p \simeq 0.5$ a.u.) as the atomic medium.}
	\label{Fig:props:state}
\end{figure}

Figure~\ref{Fig:props:state}~(a)-(c) displays the Wigner function of the heralded quantum optical states for various values of $\boldsymbol{p}_f$. As observed, the resulting states exhibit characteristic features of large optical Schrödinger cat states, with two pronounced lobes around $X_1 \approx \pm 10$ and a central interference pattern reaching both positive and negative extremal values, as expected from the quantum superposition of macroscopically distinct lobes.~However, unlike standard even and odd optical cat states, the outer regions do not always correspond to well-defined Gaussian lobes, especially in Fig.~\ref{Fig:props:state}~(c), displaying patterns with both positive and negative regions, whose structure and relative weight can be tuned by varying $\boldsymbol{p}_f$.~This highlights that the generated states do not generally correspond to a simple superposition of two coherent states of opposite phase, $a \ket{\alpha} + b \ket{\beta}$, but rather to the continuous superposition in Eq.~\eqref{Eq:heralded:state}, which we rewrite as
\begin{equation}\label{Eq:heralded:simple:form}
	\ket{\Phi(\boldsymbol{p}_f,t)}
		= 
		\int \dd^2 \alpha 
			\big[
				a(\boldsymbol{p}_f;\alpha) \ket{\alpha}
				 + a(\boldsymbol{p}_f;-\alpha) \ket{-\alpha}
			\!
			\big]
\end{equation}
where we define
\begin{equation}
	a(\boldsymbol{p}_f;\pm\alpha)
		= \int \dd t' c(\pm \alpha) b(\boldsymbol{p}_f,t';\pm\alpha).
\end{equation}

The amplitudes and relative phases of the coefficients $a(\boldsymbol{p}_f;\pm \alpha)$ determine the specific interference pattern found in each lobe and between them, with the value of $\boldsymbol{p}_f$ strongly influencing these coefficients.~In particular, for $\boldsymbol{p}_f = \boldsymbol{0}$, the Wigner function is invariant under a $\pi$-rotation in phase-space [panel (a)], while for $p_{f,\parallel}\neq 0$ this symmetry is broken [panels (b) and (c)].~This behavior is directly related to the ATI dynamics.~For a given $\abs{\alpha}$, $a(\boldsymbol{p}_f;\alpha)$ represents the ATI probability amplitude associated with a coherent field of strength $\varepsilon(\alpha)$.~When $\boldsymbol{p}_f = \boldsymbol{0}$, ATI happens symmetrically along the time-duration of the driving field, specifically at each half-cycle of the field, so that $a(\boldsymbol{p}_f;\alpha) = a(\boldsymbol{p}_f;-\alpha)$.~Thereby, switching $\alpha \to -\alpha$, i.e., rotating the state in phase-space by an angle $\pi$, does not alter the process.~In contrast, for $p_{f,\parallel} \neq 0$ the ATI response for a given $\abs{\alpha}$ becomes time-asymmetric, as different half-cycles of the field contribute unequally due to the introduced photoelectron  momentum bias.~Consequently, $a(\boldsymbol{p}_f;\alpha) \neq a(\boldsymbol{p}_f;-\alpha)$, thus breaking the $\pi$-rotation symmetry, as reflected in panels (b) and (c).

This breaking of the symmetry is further illustrated in Fig.~\ref{Fig:props:state}~(d)-(f), which display the photon number probability $P(n\vert \boldsymbol{p}_f) = \lvert \braket{n}{\Phi(\boldsymbol{p}_f,t)}\rvert^2$ for the same states as in panels (a)-(c).~For $\boldsymbol{p}_f = \boldsymbol{0}$ [panel (d)], the distribution is strongly peaked at photon numbers with even parity (connected with the green curve), while the odd-parity channels (connected with the orange curve) are suppressed.~When $p_{f,\parallel} \neq 0$, this parity selection is broken, and peaks and troughs appear with alternating parity [panels (e) and (f)].~This behavior can be explicitly understood from the Fock expansion of Eq.~\eqref{Eq:heralded:simple:form}
\begin{equation}
	\begin{aligned}
	\ket{\Phi(\boldsymbol{p}_f,t)}
		&= \sum_{n=0}^\infty
			\int \dd^2\alpha\
				e^{-\abs{\alpha}^2/2} \dfrac{\alpha^n}{\sqrt{n!}}
				\\&\hspace{0.7cm}\times
					\big[
						a(\boldsymbol{p}_f;\alpha)
						+ (-1)^n a(\boldsymbol{p}_f;-\alpha)
					\big]\ket{n}.
	\end{aligned}
\end{equation}
When $a(\boldsymbol{p}_f;\alpha) = 	a(\boldsymbol{p}_f;-\alpha)$, as for $\boldsymbol{p}_f=\boldsymbol{0}$, the factor $[1+(-1)^n]$ vanishes for odd $n$, allowing only even photon numbers~[Fig.~\ref{Fig:props:state}~(d)].~For $p_{f,\parallel}\neq 0$ the coefficients differ, lifting this parity restriction and producing the more complex photon-number distributions observed~[Fig.~\ref{Fig:props:state}~(e),(f)].~This is further highlighted in Fig.~\ref{Fig:props:state}~(g) with the light gray curve, displaying the probability of having even photon numbers in the state, i.e.,
\begin{equation}
    P_{\text{even}}(p_{f,\parallel})
        = \sum^{\infty}_{n=0}
                \abs{\braket{2n}{\Phi(\boldsymbol{p}_{f,\parallel},t)}}^2,
\end{equation}
which becomes maximum at $\boldsymbol{p}_f = \boldsymbol{0}$, and diminishes as the symmetry of the state breaks, stabilizing at $P_{\text{even}}(p_{f,\parallel}) = 0.5$ for large values of momentum.

Finally, to emphasize the non-Gaussian nature of the state, Fig.~\ref{Fig:props:state}~(g) displays the Wigner negative volume (green curve), defined as
\begin{equation}
	\mathcal{N}
		= -1+\int \dd X_1 \int \dd X_2 
				\abs{W(X_1,X_2)},
\end{equation}
together with the probability of detecting an electron with the corresponding momentum (black curve).~As observed, the negativity remains greater than zero across all momenta, highlighting the non-Gaussian character of the generated states, while the photoelectron probability decreases and flattens compared to the case of coherent states~\cite{lyu_effect_2025,mao_benchmarking_2025,singh_interferometrically_2026}. 

\subsection{Effect of limited heralding resolution}
\begin{figure}
	\centering
	\includegraphics[width=1\columnwidth]{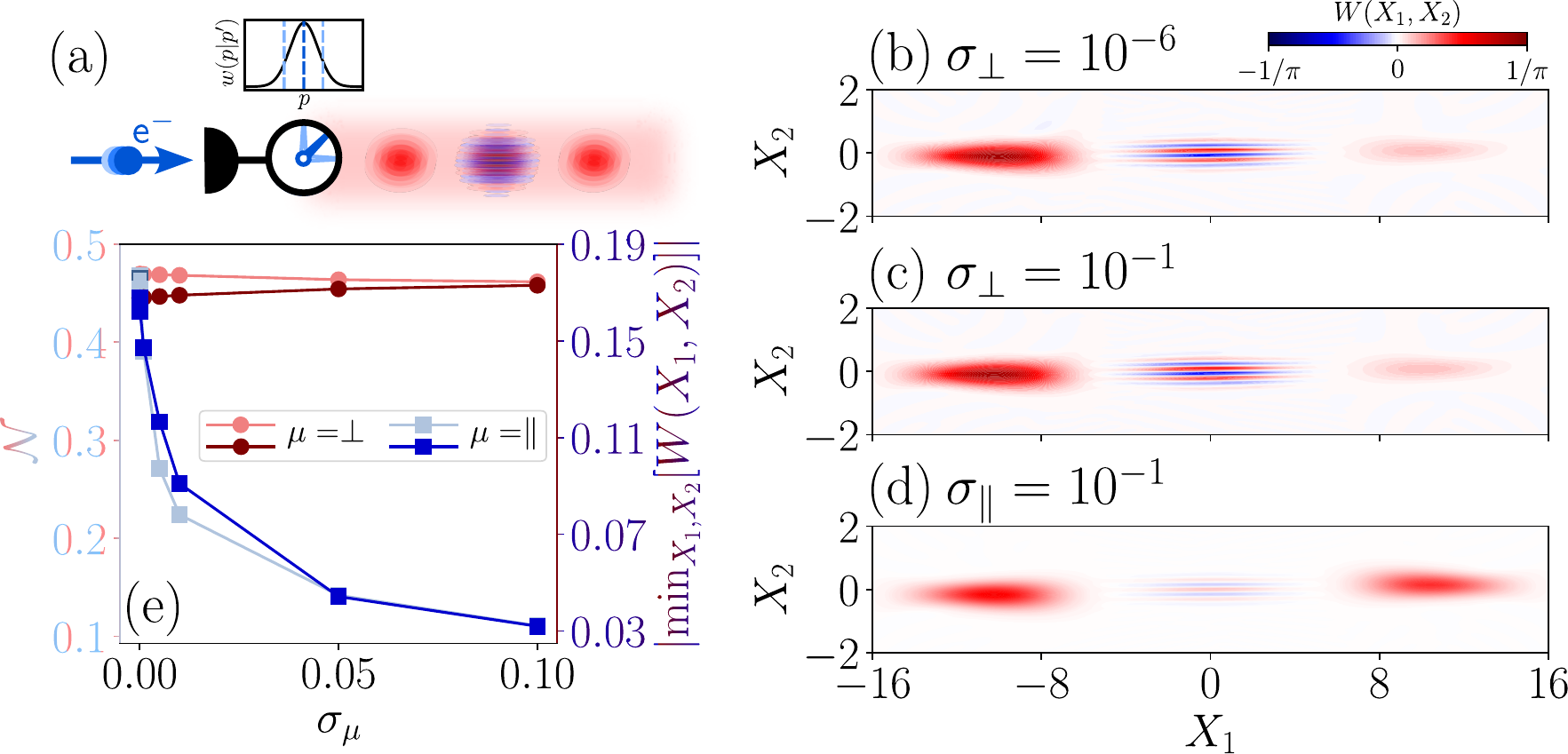}
	\caption{Effect of finite resolution in the heralding step.~(a) Schematic illustration of the effect of finite resolution, with the detector having the possibility of giving an outcome $p'$ instead of $p = 0.1$ a.u.~for the measured photoelectron momentum according to the probability $w(p\vert p')$.~(b)-(d) Wigner functions for different values of the detection resolution, with the first showing the case of close-to-ideal measurement, while the other two of imperfect detection along the $\parallel$- and $\perp$-components.~(e) Wigner negativity (light curves, left axis) and minimum value of the Wigner function (solid curves, right axis) as a function of $\sigma$.~In this analysis, we fixed $g(\omega_L) = 5 \times 10^{-3}$ a.u., considering a light-matter interaction lasting four optical cycles ($\omega_L = 0.057$ a.u.) with hydrogen ($I_p \simeq 0.5$ a.u.) as the atomic medium.}
	\label{Fig:Approx:Meas:p0=0.25}
\end{figure}

One of the main mechanisms degrading non-classical states of light is decoherence, whereby a pure state evolves into a statistical mixture with reduced coherence and diminished non-classical features. In heralding-based protocols, an important source of decoherence arises from the finite resolution of the measurement used for conditioning.~In our case, this conditioning relies on measuring the final momentum of the photoelectron which, ideally, corresponds to a projective measurement described by the operator $\hat{\Pi}_{\text{ideal}}(\boldsymbol{p}_f) = \dyad{\boldsymbol{p}_f}$.~In practice, however, detectors possess finite resolution and cannot perfectly distinguish nearby outcomes $\boldsymbol{p}_f$ and $\boldsymbol{p}'_f$.~As a result, the reported measurement outcome $\boldsymbol{p}_f$ may correspond to a range of true values $\{\boldsymbol{p}'_f\}$ each of them appearing with some probability~[Fig.~\ref{Fig:Approx:Meas:p0=0.25}~(a)].

To account for this effect, we model the measurement using the framework of approximate measurements~\cite{breuer_quantum_2007}, described by positive operator valued measure (POVM) elements~\cite{NielsenBookCh1} given by
\begin{equation}
	\hat{\Pi}(p_{f,\mu})
		= \int \dd p'_{f,\mu} w(p_{f,\mu} \vert p_{f,\mu}')
			\relaxket{p'_{f,\mu}}\!\relaxbra{p'_{f,\mu}},
\end{equation}
which satisfy the completeness relation $\int \dd p_{f,\mu} \hat{\Pi}(p_{f,\mu}) = \mathbbm{1}$, provided that $\int \dd p_{f,\mu} w(p_{f,\mu}|p_{f,\mu}') = 1$.~Here, $w(p_{f,\mu}|p_{f,\mu}')$ represents the conditional probability that the detector reports outcome $p_{f,\mu}$ given a true value $p_{f,\mu}'$. We model this response function as a Gaussian function
\begin{equation}
	w(p_{f,\mu}\vert p_{f,\mu}')
	= \dfrac{1}{\sigma_\mu\sqrt{2\pi}}
	\exp[
	- \dfrac{(p_{f,\mu}-p_{f,\mu}')^2}{2\sigma_\mu^2}],
\end{equation}
where $\sigma_\mu$ characterizes the detector resolution, with $\sigma_\mu \to 0$ corresponding to the ideal limit of perfect resolution.~For simplicity, we restrict the analysis to decoherence along a single momentum component $\mu$, allowing us to isolate the effect of measurement resolution along either the parallel or perpendicular direction.~Assuming a minimally disturbing measurement model, i.e., one that introduces classical uncertainty in the measurement outcomes without additional unitary backaction~\cite{breuer_quantum_2007}, the post-selected optical state is given, up to a normalization factor, by
\begin{equation}
	\begin{aligned}
		\hat{\rho}(\boldsymbol{p})
		&= 	\tr[\hat{\Pi}(p_{f,\mu})\dyad{\Psi(t)}]
		\\&= \int \dd p_{f,\mu}' 
		w(p_{f,\mu}\vert p_{f,\mu}')
		\dyad{\Phi(\boldsymbol{p}',t)},
	\end{aligned}
\end{equation}
where $\boldsymbol{p}' = (p_{f,\mu}',p_{f,\bar{\mu}})$.

One of the main effects of imperfect heralding on non-Gaussian states is the degradation of Wigner negativity.~This behavior is illustrated in Fig.~\ref{Fig:Approx:Meas:p0=0.25}~(b)-(d), which shows the Wigner functions obtained by post-selecting on a fixed momentum value $\boldsymbol{p}_f = (0.1_{\parallel},0_\perp)$, as a function of the detector resolution $\sigma_\mu$.~In the near-ideal case [panel~(b)], the Wigner function exhibits pronounced negative regions and, as the resolution parameter $\sigma_\mu$ increases, these negative features are progressively suppressed.~This degradation depends on the direction along which the measurement is imperfect, with finite resolution along the $\perp$-component [panel (c)] leading to an almost absent reduction of negativity, whereas imperfections along the $\parallel$-component [panel (d)] result in a much stronger suppression.~This behavior is quantified in panel (e), where we show both the Wigner negativity (light curves) and the minimum value of the Wigner function (dark curves) as a function of $\sigma_\mu$. Particularly for the $\parallel$-direction, a small increase in $\sigma_\parallel$ with respect to the ideal scenario leads to a rapid decay of non-classical features, followed by a saturation at larger values of $\sigma_\parallel$.

This saturation can be understood from the fact that, although the detector may misidentify a momentum value $p_{f,\mu}'$ as $p_{f,\mu}$, contributions from $\lvert\boldsymbol{p}_{f,\mu}'\rvert \gg \lvert\boldsymbol{p}_{f,\mu}\rvert$  are strongly suppressed due to the rapidly decreasing probability of generating electrons with large momenta.~As a result, the effective mixing by the detector becomes bounded, leading to a stabilization of the Wigner features.~This also explains the stronger sensitivity to imperfections along the $\parallel$-direction, as the linear polarization of the driving field along that direction predominantly accelerates electrons along $\mu = \parallel$.~Consequently, large momentum components are more likely along this direction than along $\mu=\perp$, making the state more susceptible to resolution-induced mixing in the $\parallel$-component.

\subsection{CHSH inequality violations}
While experimental imperfections are inevitable, non-Gaussian states remain a vital resource for both quantum information applications~\cite{mari_positive_2012,rahimi-keshari_sufficient_2016} and foundational tests of quantum mechanics~\cite{zavatta_quantum--classical_2004,paavola_finite-time_2011}.~At the nexus of these two fields lie Bell inequalities, which provide a powerful and rigorous tool to certify non-classical correlations that cannot be explained by local-hidden variable models~\cite{brunner_bell_2014} (see Supplementary Material~\ref{Sec:SM:Bell}).~By demonstrating that quantum correlations can transcend shared classical randomness while being independent of distant measurement choices, Bell tests provide a benchmark for both the utility of quantum protocols~\cite{acin_device-independent_2007,pironio_random_2010,brunner_bell_2014,supic_self-testing_2020} and the nonlocal nature of reality~\cite{einstein_can_1935,bell_einstein_1964}.

\begin{figure}
	\centering
	\includegraphics[width=1\columnwidth]{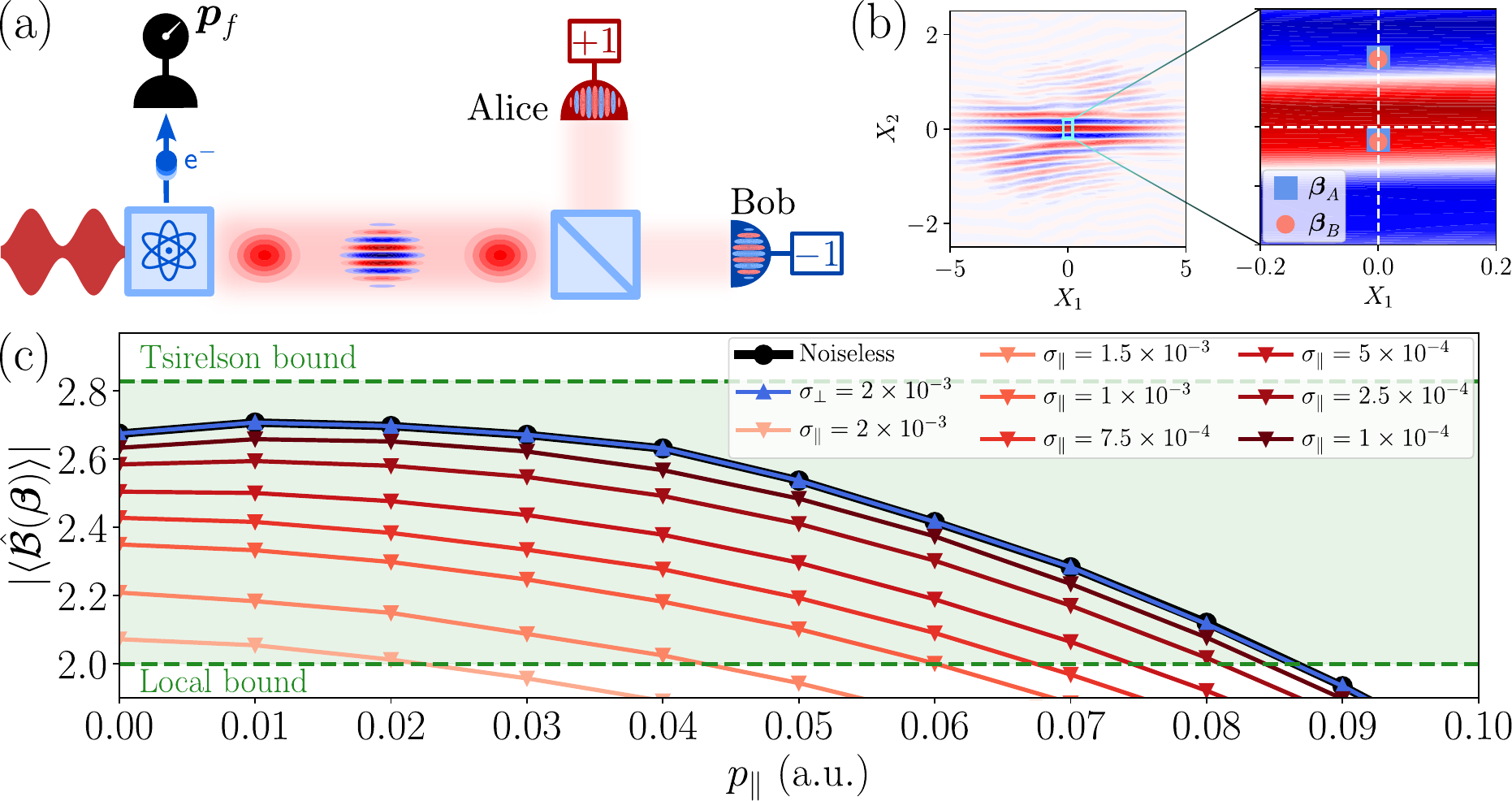}
	\caption{Analysis of the CHSH violation induced by the ATI-heralded states.~(a) Schematic illustration of the setup considered here, in which the ATI-heralded state passes through a beam splitter and is distributed to two parties, Alice and Bob, each effectively probing the Wigner function of the state. (b) Example of optimized measurements performed by Alice and Bob to violate a Bell inequality, probing the interference region arising from the superposition.~(c) CHSH inequality violation as a function of $p_{\parallel}$ for different values of the detection resolution $\sigma$, with the shaded green region indicating the Bell-nonlocal regime.~In this analysis, we fixed $g(\omega_L) = 5 \times 10^{-3}$ a.u., considering a light-matter interaction lasting four optical cycles ($\omega_L = 0.057$ a.u.) with hydrogen ($I_p \simeq 0.5$ a.u.) as the atomic medium.}
	\label{Fig:CHSH}
\end{figure}

Here, we consider a Bell test based on the standard Clauser-Horne-Shimony-Holt (CHSH) scenario~\cite{clauser_proposed_1969}, using the setup illustrated in Fig.~\ref{Fig:CHSH}~(a):~the ATI-heralded state is first split by a 50:50 beam splitter, with the two output modes sent to two parties, Alice and Bob.~Each party independently performs one of two possible measurements, chosen at random, which we denote $\beta_{A}^{(i)}$ and $\beta_{B}^{(j)}$ ($i,j \in \{0,1\})$. Following Refs.~\cite{banaszek_nonlocality_1998,banaszek_testing_1999}, we consider displaced parity measurements of the form
\begin{equation}
	\hat{\Pi}(\beta)
		= \hat{D}(\beta) (-1)^{\hat{a}^\dagger \hat{a}} \hat{D}^\dagger(\beta),
\end{equation}
which yield dichotomic outcomes $\pm1$.~Comparing with Eq.~\eqref{Eq:Wigner:function}, the expectation value of this operator is directly proportional to the Wigner function evaluated at the phase-space point $\beta$.~Thus, the underlying physical picture of the considered protocol is that Alice and Bob effectively probe the Wigner function of their respective share of the state at the phase-space points $\beta_{A}^{(i)}$ and $\beta_{B}^{(j)}$, respectively.~In terms of these measurement operators, we introduce the CHSH functional 
\begin{equation}\label{Eq:Bell:score}
	\langle \hat{\mathcal{B}}(\boldsymbol{\beta})\rangle
		= \sum_{i,j} (-1)^{i j}
			\langle \hat{\Pi}(\beta_A^{(i)}) \otimes \hat{\Pi}(\beta_B^{(j)}) \rangle, 
\end{equation}
which is bounded by $\lvert\!\langle \hat{\mathcal{B}}(\boldsymbol{\beta})\rangle\!\rvert \leq 2$ for correlations admitting a local-hidden variable description~\cite{bell_einstein_1964,brunner_bell_2014}, while quantum mechanics allows values up to the Tsirelson bound $2\sqrt{2}$~\cite{cirelson_quantum_1980}.

Importantly, for the measurements considered here, CHSH violations require two key ingredients.~First, entanglement between the modes distributed to Alice and Bob is necessary to generate nonlocal correlations.~Second, the presence of Wigner negativities enables negative contributions to the measured correlators, which are required to surpass the local bound in Eq.~\eqref{Eq:Bell:score} when Alice and Bob properly select their measurement settings [Fig.~\ref{Fig:CHSH}~(b)] (see Supplementary Material~\ref{Sec:SM:Bell}).~In the present scenario, such contributions are provided by the ATI-heralded state which, together with the beam splitter operation, yields an entangled state capable of correlating the measurement outcomes of Alice and Bob.

Figure~\ref{Fig:CHSH}~(c) summarizes the results of our analysis. In the absence of heralding imperfections, we find that it is possible to achieve clear violations of the CHSH inequality across various values of $p_{f,\parallel}$, with values approaching the Tsirelson bound after optimizing $\lvert \!\langle\hat{\mathcal{B}}\rangle\!\rvert$ over complex-valued displacements $\beta$ [Fig.~\ref{Fig:CHSH}~(b)] (see Supplementary Material~\ref{Sec:SM:Numerics:CHSH} for details).~As the final momentum $p_{f,\parallel}$ increases, the magnitude of the violation is reduced, despite the persistence of clear Wigner negativities [Fig.~\ref{Fig:props:state}~(g)]. This behavior reflects the strong sensitivity of the interference structure of the state to the electron momentum, which modifies its symmetry as well as the location and contrast of the negative regions in phase-space.~Consequently, the optimal measurement settings depend non-trivially on $p_{\parallel,f}$, suggesting that more general measurement strategies could further enhance the observed violations.

We further assess the robustness of these violations against imperfect heralding.~As shown by the red down-triangle curves, increasing the detector uncertainty $\sigma_{\parallel}$ leads to a rapid degradation of the CHSH violation when the measurement settings are kept fixed to their optimal noiseless values.~In contrast, imperfections along the $\perp$-direction have a significantly weaker impact, with the corresponding violations remaining close to the ideal case for values of $\sigma_{\perp}$ for which $\sigma_{\parallel}$ already suppresses the Bell inequality violation.~This asymmetry further highlights the dominant role of the parallel momentum component in shaping the quantum state, and the weak influence of the perpendicular component.

\section{DISCUSSION}\label{Sec:Dicussion}
Strong-field physics naturally operates in highly nonlinear regimes requiring extreme intensities---conditions that have, until recently, remained largely unexplored within the framework of standard quantum optics. With the advent of bright squeezed sources~\cite{spasibko_multiphoton_2017,manceau_indefinite-mean_2019} capable of driving strong-field processes~\cite{rasputnyi_high_2024,lemieux_photon_2025,tzur_measuring_2025,heimerl_multiphoton_2024,heimerl_driving_2025}, it is now possible to manipulate the extreme nonlinearities of strong-field physics under the control of quantum states of light.~This work explores this interface, demonstrating that ATI can be harnessed as a powerful heralding mechanism, conceptually analogous to photon subtraction protocols~[Fig.~\ref{Fig:Scheme}], to generate optical Schrödinger-like cat states with exceptionally large amplitude~[Figs.~\ref{Fig:C_matrix} to \ref{Fig:props:state}].~Beyond state preparation, we have characterized the robustness of these states against finite heralding resolution~[Fig.~\ref{Fig:Approx:Meas:p0=0.25}] and certified their non-classicality and potential foundational and practical capabilities through the violation of a Bell inequality~[Fig.~\ref{Fig:CHSH}].~Taken together, these results highlight a core strength of Strong-field Quantum Optics:~the potential ability to translate standard quantum optical approaches into the high-intensity regime to access non-classical resources of unprecedented scale.

In the following, we discuss the main limitations of the present approach and outline potential extensions from both theoretical and experimental perspectives.

\subsection{Theory perspective}
The theoretical analysis presented in this work relies on a coherent state expansion of the driving field~\cite{wang_high-order_2025}, which we implement numerically using a discrete coherent state lattice that preserves the (over)completeness of the basis~\cite{perelomov_completeness_1971,bargmann_completeness_1971}.~For the high-intensity regimes relevant to strong-field physics, this approach requires handling large coefficient matrices and coherent states amplitudes that, while allowing access to realistic parameter regimes~[Fig.~\ref{Fig:C_matrix}], they often rely on renormalization procedures~\cite{wang_high-order_2025} to extract phase-space properties~[Fig.~\ref{Fig:props:state}].~Although this strategy captures the qualitative features of the generated optical Schrödinger-like cat states, it tends to underestimate their intrinsically large coherent state nature.~An alternative approach, used in Ref.~\cite{rivera-dean_microscopic_2025}, consists in effectively squeezing phase-space such that the initial state looks as a coherent state.~While this enables calculations at realistic values of the light-matter coupling, which gets exponentially enhanced by the amount of squeezing in the initial state, and yields exact values for quantum optical measures invariant under Gaussian transformations (e.g.,~the Wigner negativity), it does not allow one to recover the true phase-space structure of the generated states neither reduce their computational overhead.

These limitations highlight the need for quantum optical methods capable of faithfully capturing the large scale phase-space structures induced by strong-field dynamics. In this work, we observe that the energy conservation in ATI leading to the so-called ATI rings~\cite{lewenstein_rings_1995}, translate into localized regions of enhanced weight in phase-space~[Fig.~\ref{Fig:C_matrix}]. This suggests that the ATI-heralded states can be effectively interpreted as superpositions of a finite set of coherent state components associated with these dominant regions.~Such a picture naturally connects to recent approaches based on finite Gaussian decompositions~\cite{braccini_superpositions_2025,hahn_classical_2025}, which may provide an efficient and exact representation for these large cat-like states~\cite{centrone_gaussian_2026}.~A complementary direction, motivated by the analysis in Sec.~\ref{Sec:SPA}, is to explore whether the integration contour defining the quantum optical state can be systematically deformed to better capture the dominant contributions.~In this context, recently developed contour-deformation techniques in semiclassical strong-field studies~\cite{weber_universal_2025} may offer a promising route toward more accurate and scalable descriptions.~

Finally, in this work we have focused on the simplest ATI configuration, namely a monochromatic, linearly polarized driving field and direct ionization pathways. However, strong-field physics offers a much richer landscape: at higher energies, additional processes such as rescattering become relevant~\cite{lewenstein_rings_1995,amini_symphony_2019}, while more structured driving fields are known to strongly shape the photoelectron spectra~\cite{milosevic_above-threshold_2006,faria_it_2020,rook_exploring_2022}.~These additional degrees of freedom provide powerful handles to engineer the structure of the emitted electrons and, consequently, the properties of the heralded state.~Exploring how such control can be leveraged to tailor non-classical features, enhance robustness, or access new classes of quantum states in light and matter~\cite{imai_heralded_2026} constitutes a promising direction for future work. 

\subsection{Experiments perspective}
While in this work we have considered single-mode squeezed driving fields, current high-intensity BSV generation techniques are typically based on spontaneous parametric down-conversion~\cite{spasibko_multiphoton_2017,manceau_indefinite-mean_2019,rasputnyi_high_2024,heimerl_multiphoton_2024,heimerl_driving_2025,lemieux_photon_2025,tzur_measuring_2025,kern_single-shot_2026}, which inherently produces broadband, multimode squeezing.~Nevertheless, these studies have also demonstrated that spatial and spectral filtering can effectively reduce the number of relevant modes, yielding states that are well approximated by a small number of Schmidt modes and, in some cases, approach an effective single-mode description~\cite{heimerl_multiphoton_2024}. Given that such sources have already been shown to drive strong-field processes in matter, these developments suggest that the generation of the large optical Schrödinger cat-like states predicted here may be within experimental reach.

A more significant challenge, however, lies in the characterization of these ATI-heralded states.~The large coherent amplitudes involved placed them deep in the high-photon number regime, where standard quantum optical tomography techniques become impractical~\cite{lvovsky_continuous-variable_2009}.~In this context, nonlinear and strong-field-based characterization methods may provide a viable alternative.~For instance, nonlinear frequency conversion processes have been shown to transfer statistical features of an optical field to a higher harmonic mode, enabling indirect characterization of the high-photon number driving field~\cite{lamprou_nonlinear_2025}. Similarly, high-harmonic generation driven by a bichromatic quantum field composed of a strong coherent field and a weaker bright squeezed field has been proposed~\cite{rivera-dean_attosecond_2026} and experimentally explored~\cite{tzur_measuring_2025} as an attosecond analogue of homodyne detection, capable of accessing phase-space information of the quantum states.~Although such approaches do not yet provide full quantum state tomography, they offer high-fidelity qualitative evidence of their phase-space properties.~Extending these techniques to the present setting, where the generated states exhibit both large and strong non-Gaussian features, therefore constitutes an important open challenge.~Achieving this would open the door to experimentally accessing advanced protocols, such as the Bell inequality violations demonstrated here, in a quantitative and controlled manner.

\section*{ACKNOWLEDGMENTS}
J.~R.-D.~gratefully acknowledges Paris Tzallas for insightful discussions and his interest in this work, Anne Weber for stimulating discussions and her kind hospitality at King's College London, and Federico Centrone and Alessio Serafini for insightful discussions.

J.~R.-D. and C.~F.~M.~F.~acknowledge funding from UK Engineering and Physical Sciences Research Council (EPSRC) Funding, Grant UKRI2300 - Attosecond Photoelectron Imaging with Quantum Light.

ICFO-QOT group acknowledges support from: European Research Council AdG NOQIA; MCIN/AEI (PGC2018-0910.13039/501100011033, CEX2019-000910-S/10.13039/501100011033, Plan National FIDEUA PID2019-106901GB-I00, Plan National STAMEENA PID2022-139099NB, I00, project funded by MCIN/AEI/10.13039/501100011033 and by the “European Union NextGenerationEU/PRTR" (PRTR-C17.I1), FPI); QUANTERA DYNAMITE PCI2022-132919, QuantERA II Programme co-funded by European Union’s Horizon 2020 program under Grant Agreement No 101017733; Ministry for Digital Transformation and of Civil Service of the Spanish Government through the QUANTUM ENIA project call - Quantum Spain project, and by the European Union through the Recovery, Transformation and Resilience Plan - NextGenerationEU within the framework of the Digital Spain 2026 Agenda; CEX2024-001490-S [MICIU/AEI/10.13039/501100011033]; Fundació Cellex; Fundació Mir-Puig; Generalitat de Catalunya (European Social Fund FEDER and CERCA program; Barcelona Supercomputing Center MareNostrum (FI-2023-3-0024); Funded by the European Union. Views and opinions expressed are however those of the author(s) only and do not necessarily reflect those of the European Union, European Commission, European Climate, Infrastructure and Environment Executive Agency (CINEA), or any other granting authority. Neither the European Union nor any granting authority can be held responsible for them (HORIZON-CL4-2022-QUANTUM-02-SGA, PASQuanS2.1, 101113690, EU Horizon 2020 FET-OPEN OPTOlogic, Grant No 899794, QU-ATTO, 101168628), EU Horizon Europe Program (This project has received funding from the European Union’s Horizon Europe research and innovation program under grant agreement No 101080086 NeQSTGrant Agreement 101080086 — NeQST); ICFO Internal “QuantumGaudi” project. 

E.P. acknowledges Royal Society funding under URF$\backslash$R1$\backslash$211390.

M.K. acknowledges Royal Society funding under URF$\backslash$R1$\backslash$231460.

\bibliography{References.bib}{}

@misc{imai_heralded_2026,
	title = {Heralded ultrafast generation of macroscopic quantum states in matter with bright squeezed vacuum light},
	copyright = {arXiv.org perpetual, non-exclusive license},
	url = {https://arxiv.org/abs/2605.30224},
	doi = {10.48550/ARXIV.2605.30224},
	urldate = {2026-05-29},
	publisher = {arXiv},
    number = {{arXiv}:2605.30224},
	author = {Imai, Shohei},
	year = {2026},
	note = {Version Number: 1},
}

@misc{lyu_attosecond_2026,
	title = {Attosecond quantum spectroscopy with entangled photon pairs},
	rights = {{arXiv}.org perpetual, non-exclusive license},
	url = {https://arxiv.org/abs/2604.06707},
	doi = {10.48550/ARXIV.2604.06707},
	number = {{arXiv}:2604.06707},
	publisher = {{arXiv}},
	author = {Lyu, Zijian and Sun, Fengxiao and Yi, Sili and Li, Jingze and Liu, Haodong and He, Qiongyi and Gong, Qihuang and Ivanov, Misha and Liu, Yunquan},
	urldate = {2026-05-18},
	year = {2026},
}

@misc{singh_interferometrically_2026,
	title = {Interferometrically {Enhanced} {Asymmetry} in {Strong}-field {Ionization} with {Bright} {Squeezed} {Vacuum}},
	copyright = {Creative Commons Attribution 4.0 International},
	url = {https://arxiv.org/abs/2604.12646},
	doi = {10.48550/ARXIV.2604.12646},
	urldate = {2026-05-19},
	publisher = {arXiv},
    note = {{arXiv}:2604.12646 [quant-ph]},
	author = {Singh, G. and Rook, T. and Rivera-Dean, J. and Faria, C. Figueira de Morisson},
	year = {2026},
}

@misc{stammer_information-theoretic_2026,
	title = {Information-theoretic perspective on energy conservation in high harmonic generation},
	url = {http://arxiv.org/abs/2410.15503},
	doi = {10.48550/arXiv.2410.15503},
	urldate = {2026-04-13},
	publisher = {arXiv},
	author = {Stammer, Philipp},
	month = mar,
	year = {2026},
	note = {arXiv:2410.15503 [quant-ph]
version: 2},
}

@article{laghaout_amplification_2013,
	title = {Amplification of realistic {Schrö}dinger-cat-state-like states by homodyne heralding},
	volume = {87},
	url = {https://link.aps.org/doi/10.1103/PhysRevA.87.043826},
	doi = {10.1103/PhysRevA.87.043826},
	number = {4},
	urldate = {2022-06-02},
	journal = {Physical Review A},
	publisher = {American Physical Society},
	author = {Laghaout, Amine and Neergaard-Nielsen, Jonas S. and Rigas, Ioannes and Kragh, Christian and Tipsmark, Anders and Andersen, Ulrik L.},
	month = apr,
	year = {2013},
	pages = {043826},
}

@article{sychev_enlargement_2017,
	title = {Enlargement of optical {Schrödinger}'s cat states},
	volume = {11},
	copyright = {2017 Springer Nature Limited},
	issn = {1749-4893},
	url = {https://www.nature.com/articles/nphoton.2017.57},
	doi = {10.1038/nphoton.2017.57},
	number = {6},
	urldate = {2026-04-10},
	journal = {Nature Photonics},
	publisher = {Nature Publishing Group},
	author = {Sychev, Demid V. and Ulanov, Alexander E. and Pushkina, Anastasia A. and Richards, Matthew W. and Fedorov, Ilya A. and Lvovsky, Alexander I.},
	month = jun,
	year = {2017},
	pages = {379--382},
}

@article{becker_plateau_2018,
	title = {The plateau in above-threshold ionization: the keystone of rescattering physics},
	volume = {51},
	issn = {0953-4075, 1361-6455},
	shorttitle = {The plateau in above-threshold ionization},
	url = {https://iopscience.iop.org/article/10.1088/1361-6455/aad150},
	doi = {10.1088/1361-6455/aad150},
	number = {16},
	urldate = {2025-07-30},
	journal = {Journal of Physics B: Atomic, Molecular and Optical Physics},
	author = {Becker, W and Goreslavski, S P and Milošević, D B and Paulus, G G},
	month = aug,
	year = {2018},
	pages = {162002},
}

@article{milosevic_above-threshold_2006,
	title = {Above-threshold ionization by few-cycle pulses},
	volume = {39},
	url = {https://iopscience.iop.org/article/10.1088/0953-4075/39/14/R01/meta},
	number = {14},
	urldate = {2023-09-30},
	journal = {Journal of Physics B: Atomic, Molecular and Optical Physics},
	author = {Milošević, D. B. and Paulus, G. G. and Bauer, D. and Becker, W.},
	year = {2006},
	pages = {R203},
}

@article{rook_exploring_2022,
	title = {Exploring symmetries in photoelectron holography with two-color linearly polarized fields},
	volume = {55},
	issn = {0953-4075},
	url = {https://doi.org/10.1088/1361-6455/ac7bbf},
	doi = {10.1088/1361-6455/ac7bbf},
	number = {16},
	urldate = {2026-04-08},
	journal = {Journal of Physics B: Atomic, Molecular and Optical Physics},
	author = {Rook, T and Figueira de Morisson Faria, C},
	month = jul,
	year = {2022},
	pages = {165601},
}

@article{cirelson_quantum_1980,
	title = {Quantum generalizations of {Bell}'s inequality},
	volume = {4},
	issn = {1573-0530},
	url = {https://doi.org/10.1007/BF00417500},
	doi = {10.1007/BF00417500},
	number = {2},
	urldate = {2023-12-02},
	journal = {Letters in Mathematical Physics},
	author = {Cirel'son, B. S.},
	month = mar,
	year = {1980},
	pages = {93--100},
}

@article{einstein_can_1935,
	title = {Can {Quantum}-{Mechanical} {Description} of {Physical} {Reality} {Be} {Considered} {Complete}?},
	volume = {47},
	url = {https://link.aps.org/doi/10.1103/PhysRev.47.777},
	doi = {10.1103/PhysRev.47.777},
	number = {10},
	urldate = {2022-03-24},
	journal = {Physical Review},
	author = {Einstein, A. and Podolsky, B. and Rosen, N.},
	month = may,
	year = {1935},
	pages = {777--780},
}

@article{supic_self-testing_2020,
	title = {Self-testing of quantum systems: a review},
	volume = {4},
	shorttitle = {Self-testing of quantum systems},
	url = {https://quantum-journal.org/papers/q-2020-09-30-337/},
	doi = {10.22331/q-2020-09-30-337},
	urldate = {2022-03-24},
	journal = {Quantum},
	publisher = {Verein zur Förderung des Open Access Publizierens in den Quantenwissenschaften},
	author = {Šupić, Ivan and Bowles, Joseph},
	month = sep,
	year = {2020},
	pages = {337},
}

@article{pironio_random_2010,
	title = {Random numbers certified by {Bell}’s theorem},
	volume = {464},
	copyright = {2010 Macmillan Publishers Limited. All rights reserved},
	issn = {1476-4687},
	url = {https://www.nature.com/articles/nature09008},
	doi = {10.1038/nature09008},
	number = {7291},
	urldate = {2022-04-22},
	journal = {Nature},
	author = {Pironio, S. and Acín, A. and Massar, S. and de la Giroday, A. Boyer and Matsukevich, D. N. and Maunz, P. and Olmschenk, S. and Hayes, D. and Luo, L. and Manning, T. A. and Monroe, C.},
	month = apr,
	year = {2010},
	pages = {1021--1024},
}

@article{clauser_proposed_1969,
	title = {Proposed {Experiment} to {Test} {Local} {Hidden}-{Variable} {Theories}},
	volume = {23},
	url = {https://link.aps.org/doi/10.1103/PhysRevLett.23.880},
	doi = {10.1103/PhysRevLett.23.880},
	number = {15},
	urldate = {2022-01-18},
	journal = {Physical Review Letters},
	author = {Clauser, John F. and Horne, Michael A. and Shimony, Abner and Holt, Richard A.},
	month = oct,
	year = {1969},
	pages = {880--884},
}

@article{acin_device-independent_2007,
	title = {Device-{Independent} {Security} of {Quantum} {Cryptography} against {Collective} {Attacks}},
	volume = {98},
	url = {https://link.aps.org/doi/10.1103/PhysRevLett.98.230501},
	doi = {10.1103/PhysRevLett.98.230501},
	number = {23},
	urldate = {2023-12-03},
	journal = {Physical Review Letters},
	publisher = {American Physical Society},
	author = {Acín, Antonio and Brunner, Nicolas and Gisin, Nicolas and Massar, Serge and Pironio, Stefano and Scarani, Valerio},
	month = jun,
	year = {2007},
	pages = {230501},
}

@inbook{SchleichBookCh3,
	author = {Schleich, Wolfgang P.},
	publisher = {John Wiley \& Sons, Berlin, Germany},
	isbn = {978-3-527-60297-1},
	title = {{Wigner Function}},
	booktitle = {{Quantum Optics in Phase Space}},
	chapter = {3},
	pages = {67--98},
	year = {2001},
}

@article{hudson_when_1974,
	title = {When is the wigner quasi-probability density non-negative?},
	volume = {6},
	issn = {0034-4877},
	url = {https://www.sciencedirect.com/science/article/pii/003448777490007X},
	doi = {10.1016/0034-4877(74)90007-X},
	number = {2},
	urldate = {2022-04-20},
	journal = {Reports on Mathematical Physics},
	author = {Hudson, R. L.},
	month = oct,
	year = {1974},
	pages = {249--252},
}

@article{royer_wigner_1977,
	title = {Wigner function as the expectation value of a parity operator},
	volume = {15},
	url = {https://link.aps.org/doi/10.1103/PhysRevA.15.449},
	doi = {10.1103/PhysRevA.15.449},
	number = {2},
	urldate = {2022-06-18},
	journal = {Physical Review A},
	publisher = {American Physical Society},
	author = {Royer, Antoine},
	month = feb,
	year = {1977},
	pages = {449--450},
}

@article{rahimi-keshari_sufficient_2016,
	title = {Sufficient {Conditions} for {Efficient} {Classical} {Simulation} of {Quantum} {Optics}},
	volume = {6},
	url = {https://link.aps.org/doi/10.1103/PhysRevX.6.021039},
	doi = {10.1103/PhysRevX.6.021039},
	number = {2},
	urldate = {2026-04-08},
	journal = {Physical Review X},
	publisher = {American Physical Society},
	author = {Rahimi-Keshari, Saleh and Ralph, Timothy C. and Caves, Carlton M.},
	month = jun,
	year = {2016},
	pages = {021039},
}

@article{mari_positive_2012,
	title = {Positive {Wigner} {Functions} {Render} {Classical} {Simulation} of {Quantum} {Computation} {Efficient}},
	volume = {109},
	url = {https://link.aps.org/doi/10.1103/PhysRevLett.109.230503},
	doi = {10.1103/PhysRevLett.109.230503},
	number = {23},
	urldate = {2026-04-08},
	journal = {Physical Review Letters},
	publisher = {American Physical Society},
	author = {Mari, A. and Eisert, J.},
	month = dec,
	year = {2012},
	pages = {230503},
}

@article{walschaers_non-gaussian_2021,
	title = {Non-{Gaussian} {Quantum} {States} and {Where} to {Find} {Them}},
	volume = {2},
	issn = {2691-3399},
	url = {http://arxiv.org/abs/2104.12596},
	doi = {10.1103/PRXQuantum.2.030204},
	number = {3},
	urldate = {2025-01-30},
	journal = {PRX Quantum},
	author = {Walschaers, Mattia},
	month = sep,
	year = {2021},
	pages = {030204},
}

@article{agostini_free-free_1979,
	title = {Free-{Free} {Transitions} {Following} {Six}-{Photon} {Ionization} of {Xenon} {Atoms}},
	volume = {42},
	url = {https://link.aps.org/doi/10.1103/PhysRevLett.42.1127},
	doi = {10.1103/PhysRevLett.42.1127},
	number = {17},
	urldate = {2022-06-16},
	journal = {Physical Review Letters},
	author = {Agostini, P. and Fabre, F. and Mainfray, G. and Petite, G. and Rahman, N. K.},
	month = apr,
	year = {1979},
	pages = {1127--1130},
}

@article{chin_observation_1983,
	title = {Observation of {Kr} and {Xe} ions created by intense nanosecond {CO2} laser pulses},
	volume = {16},
	issn = {0022-3700},
	url = {https://dx.doi.org/10.1088/0022-3700/16/8/003},
	doi = {10.1088/0022-3700/16/8/003},
	number = {8},
	urldate = {2023-09-25},
	journal = {Journal of Physics B: Atomic and Molecular Physics},
	author = {Chin, S. L. and Farkas, Gy and Yergeau, F.},
	month = apr,
	year = {1983},
	pages = {L223},
}

@article{hansch_resonant_1997,
	title = {Resonant hot-electron production in above-threshold ionization},
	volume = {55},
	url = {https://link.aps.org/doi/10.1103/PhysRevA.55.R2535},
	doi = {10.1103/PhysRevA.55.R2535},
	number = {4},
	urldate = {2023-06-26},
	journal = {Physical Review A},
	author = {Hansch, P. and Walker, M. A. and Van Woerkom, L. D.},
	month = apr,
	year = {1997},
	pages = {R2535--R2538},
}

@inbook{SchleichBookCh12,
	author = {Schleich, Wolfgang P.},
	publisher = {John Wiley \& Sons, Berlin, Germany},
	isbn = {978-3-527-60297-1},
	title = {{Phase Space Functions}},
	booktitle = {{Quantum Optics in Phase Space}},
	chapter = {12},
	pages = {321-348},
	year = {2001},
}

@misc{stammer_fluctuation-induced_2026,
	title = {Fluctuation-induced symmetry breaking in high harmonic generation for bicircular quantum light},
	url = {http://arxiv.org/abs/2603.24377},
	doi = {10.48550/arXiv.2603.24377},
	urldate = {2026-04-08},
	publisher = {arXiv},
	author = {Stammer, Philipp and Granados, Camilo and Rivera-Dean, Javier},
	month = apr,
	year = {2026},
	note = {arXiv:2603.24377 [quant-ph]},
}

@article{lange_electron-correlation-induced_2024,
	title = {Electron-correlation-induced nonclassicality of light from high-order harmonic generation},
	volume = {109},
	url = {https://link.aps.org/doi/10.1103/PhysRevA.109.033110},
	doi = {10.1103/PhysRevA.109.033110},
	number = {3},
	urldate = {2024-07-23},
	journal = {Physical Review A},
	author = {Lange, Christian Saugbjerg and Hansen, Thomas and Madsen, Lars Bojer},
	month = mar,
	year = {2024},
	pages = {033110},
}

@article{theidel_evidence_2024,
	title = {Evidence of the {Quantum} {Optical} {Nature} of {High}-{Harmonic} {Generation}},
	volume = {5},
	url = {https://link.aps.org/doi/10.1103/PRXQuantum.5.040319},
	doi = {10.1103/PRXQuantum.5.040319},
	number = {4},
	urldate = {2024-11-14},
	journal = {PRX Quantum},
	publisher = {American Physical Society},
	author = {Theidel, David and Cotte, Viviane and Sondenheimer, René and Shiriaeva, Viktoriia and Froidevaux, Marie and Severin, Vladislav and Merdji-Larue, Adam and Mosel, Philip and Fröhlich, Sven and Weber, Kim-Alessandro and Morgner, Uwe and Kovacev, Milutin and Biegert, Jens and Merdji, Hamed},
	month = nov,
	year = {2024},
	pages = {040319},
}

@article{theidel_observation_2025,
	title = {Observation of a displaced squeezed state in high-harmonic generation},
	volume = {7},
	issn = {2643-1564},
	url = {https://link.aps.org/doi/10.1103/6r6n-pxfp},
	doi = {10.1103/6r6n-pxfp},
	number = {3},
	urldate = {2025-10-29},
	journal = {Physical Review Research},
	author = {Theidel, David and Cotte, Viviane and Heinzel, Philip and Griguer, Houssna and Weis, Mateusz and Sondenheimer, René and Merdji, Hamed},
	month = sep,
	year = {2025},
	pages = {033223},
}

@misc{theidel_sub-poissonian_2026,
	title = {Sub-{Poissonian} {Statistics} and {Quantum} {Non}-{Gaussianity} from {High}-{Harmonic} {Generation}},
	url = {http://arxiv.org/abs/2602.10882},
	doi = {10.48550/arXiv.2602.10882},
	urldate = {2026-04-08},
	publisher = {arXiv},
	author = {Theidel, David and Nahra, Mackrine and Karuseichyk, Ilya and Griguer, Houssna and Weis, Mateusz and Merdji, Hamed},
	month = mar,
	year = {2026},
	note = {arXiv:2602.10882 [quant-ph]},
}

@article{stammer_high_2022,
	title = {High {Photon} {Number} {Entangled} {States} and {Coherent} {State} {Superposition} from the {Extreme} {Ultraviolet} to the {Far} {Infrared}},
	volume = {128},
	url = {https://link.aps.org/doi/10.1103/PhysRevLett.128.123603},
	doi = {10.1103/PhysRevLett.128.123603},
	number = {12},
	urldate = {2022-04-22},
	journal = {Physical Review Letters},
	author = {Stammer, Philipp and Rivera-Dean, Javier and Lamprou, Theocharis and Pisanty, Emilio and Ciappina, Marcelo F. and Tzallas, Paraskevas and Lewenstein, Maciej},
	month = mar,
	year = {2022},
	pages = {123603},
}

@article{lewenstein_generation_2021,
	title = {Generation of optical {Schrödinger} cat states in intense laser–matter interactions},
	volume = {17},
	copyright = {2021 The Author(s), under exclusive licence to Springer Nature Limited},
	issn = {1745-2481},
	url = {http://www.nature.com/articles/s41567-021-01317-w},
	doi = {10.1038/s41567-021-01317-w},
	number = {10},
	urldate = {2022-01-15},
	journal = {Nature Physics},
	author = {Lewenstein, M. and Ciappina, M. F. and Pisanty, E. and Rivera-Dean, J. and Stammer, P. and Lamprou, Th and Tzallas, P.},
	month = oct,
	year = {2021},
	pages = {1104--1108},
}

@article{gorlach_high-harmonic_2023,
	title = {High-harmonic generation driven by quantum light},
	copyright = {2023 The Author(s), under exclusive licence to Springer Nature Limited},
	issn = {1745-2481},
	url = {https://www.nature.com/articles/s41567-023-02127-y},
	doi = {10.1038/s41567-023-02127-y},
	urldate = {2023-10-09},
	journal = {Nature Physics},
	author = {Gorlach, Alexey and Tzur, Matan Even and Birk, Michael and Krüger, Michael and Rivera, Nicholas and Cohen, Oren and Kaminer, Ido},
	month = aug,
	year = {2023},
	pages = {1--8},
}

@article{stammer_metrological_2024,
	title = {Metrological robustness of high photon number optical cat states},
	volume = {9},
	issn = {2058-9565},
	url = {https://dx.doi.org/10.1088/2058-9565/ad7881},
	doi = {10.1088/2058-9565/ad7881},
	number = {4},
	urldate = {2024-11-01},
	journal = {Quantum Science and Technology},
	publisher = {IOP Publishing},
	author = {Stammer, Philipp and Martos, Tomás Fernández and Lewenstein, Maciej and Rajchel-Mieldzioć, Grzegorz},
	month = sep,
	year = {2024},
	pages = {045047},
}

@article{sennary_attosecond_2025,
	title = {Attosecond quantum uncertainty dynamics and ultrafast squeezed light for quantum communication},
	volume = {14},
	copyright = {2025 The Author(s)},
	issn = {2047-7538},
	url = {https://www.nature.com/articles/s41377-025-02055-x},
	doi = {10.1038/s41377-025-02055-x},
	number = {1},
	urldate = {2025-10-21},
	journal = {Light: Science \& Applications},
	author = {Sennary, Mohamed and Rivera-Dean, Javier and ElKabbash, Mohamed and Pervak, Vladimir and Lewenstein, Maciej and Hassan, Mohammed Th},
	month = oct,
	year = {2025},
	pages = {350},
}

@misc{sennary_attosecond_2026,
	title = {Attosecond quantum optics},
	url = {http://arxiv.org/abs/2601.08671},
	doi = {10.48550/arXiv.2601.08671},
	urldate = {2026-04-08},
	publisher = {arXiv},
	author = {Sennary, Mohamed and Rivera-Dean, Javier and Wange, Yihe and Lewenstein, Maciej and Hassan, Mohammed Th},
	month = mar,
	year = {2026},
	note = {arXiv:2601.08671 [physics]},
}

@misc{stammer_colloquium_2025,
	title = {Colloquium: {Quantum} optics of intense light--matter interaction},
	shorttitle = {Colloquium},
	url = {http://arxiv.org/abs/2510.19045},
	doi = {10.48550/arXiv.2510.19045},
	urldate = {2025-12-03},
	publisher = {arXiv},
	author = {Stammer, P. and Rivera-Dean, J. and Tzallas, P. and Ciappina, M. F. and Lewenstein, M.},
	month = oct,
	year = {2025},
	note = {arXiv:2510.19045 [quant-ph]},
}

@article{cruz-rodriguez_quantum_2024,
	title = {Quantum phenomena in attosecond science},
	volume = {6},
	copyright = {2024 Springer Nature Limited},
	issn = {2522-5820},
	url = {https://www.nature.com/articles/s42254-024-00769-2},
	doi = {10.1038/s42254-024-00769-2},
	number = {11},
	urldate = {2024-11-11},
	journal = {Nature Reviews Physics},
	author = {Cruz-Rodriguez, Lidice and Dey, Diptesh and Freibert, Antonia and Stammer, Philipp},
	month = nov,
	year = {2024},
	pages = {691--704},
}

@article{dimauro_50_2014,
	title = {50 years of optical tunneling},
	volume = {47},
	issn = {0953-4075},
	url = {https://dx.doi.org/10.1088/0953-4075/47/20/200301},
	doi = {10.1088/0953-4075/47/20/200301},
	number = {20},
	urldate = {2023-09-25},
	journal = {Journal of Physics B: Atomic, Molecular and Optical Physics},
	publisher = {IOP Publishing},
	author = {DiMauro, Louis and Frolov, Mikhail and Ishikawa, Kenichi L. and Ivanov, Misha},
	month = oct,
	year = {2014},
	pages = {200301},
}

@article{keldysh_ionization_1965,
	title = {{Ionization of atoms in an alternating electric field}},
	author = {Keldysh, L. V.},
	year = {1965},
	journal = {Sov. Phys. JETP},
	volume = {20},
	number = {5},
	pages = {1307--1314},
	url = {http://www.jetp.ac.ru/cgi-bin/e/index/e/20/5/p1307?a=list}
}

@article{antoine_attosecond_1996,
	title = {Attosecond {Pulse} {Trains} {Using} {High}--{Order} {Harmonics}},
	volume = {77},
	url = {https://link.aps.org/doi/10.1103/PhysRevLett.77.1234},
	doi = {10.1103/PhysRevLett.77.1234},
	number = {7},
	urldate = {2024-03-08},
	journal = {Physical Review Letters},
	author = {Antoine, Philippe and L'Huillier, Anne and Lewenstein, Maciej},
	month = aug,
	year = {1996},
	pages = {1234--1237},
}

@article{drescher_x-ray_2001,
	title = {X-ray {Pulses} {Approaching} the {Attosecond} {Frontier}},
	volume = {291},
	url = {https://www.science.org/doi/10.1126/science.1058561},
	doi = {10.1126/science.1058561},
	number = {5510},
	urldate = {2023-08-24},
	journal = {Science},
	author = {Drescher, Markus and Hentschel, Michael and Kienberger, Reinhard and Tempea, Gabriel and Spielmann, Christian and Reider, Georg A. and Corkum, Paul B. and Krausz, Ferenc},
	month = mar,
	year = {2001},
	pages = {1923--1927},
}

@article{paul_observation_2001,
	title = {Observation of a {Train} of {Attosecond} {Pulses} from {High} {Harmonic} {Generation}},
	volume = {292},
	url = {https://www.science.org/doi/10.1126/science.1059413},
	doi = {10.1126/science.1059413},
	number = {5522},
	urldate = {2023-09-03},
	journal = {Science},
	author = {Paul, P. M. and Toma, E. S. and Breger, P. and Mullot, G. and Augé, F. and Balcou, Ph. and Muller, H. G. and Agostini, P.},
	month = jun,
	year = {2001},
	pages = {1689--1692},
}

@article{corkum_attosecond_2007,
	title = {Attosecond science},
	volume = {3},
	copyright = {2007 Springer Nature Limited},
	issn = {1745-2481},
	url = {https://www.nature.com/articles/nphys620},
	doi = {10.1038/nphys620},
	number = {6},
	urldate = {2023-06-15},
	journal = {Nature Physics},
	author = {Corkum, P. B. and Krausz, Ferenc},
	month = jun,
	year = {2007},
	pages = {381--387},
}

@article{krausz_attosecond_2009,
	title = {Attosecond physics},
	volume = {81},
	url = {https://link.aps.org/doi/10.1103/RevModPhys.81.163},
	doi = {10.1103/RevModPhys.81.163},
	number = {1},
	urldate = {2022-10-18},
	journal = {Reviews of Modern Physics},
	author = {Krausz, Ferenc and Ivanov, Misha},
	month = feb,
	year = {2009},
	pages = {163--234},
}

@article{van_enk_entangled_2001,
	title = {Entangled coherent states: {Teleportation} and decoherence},
	volume = {64},
	shorttitle = {Entangled coherent states},
	url = {https://link.aps.org/doi/10.1103/PhysRevA.64.022313},
	doi = {10.1103/PhysRevA.64.022313},
	number = {2},
	urldate = {2023-06-20},
	journal = {Physical Review A},
	author = {van Enk, S. J. and Hirota, O.},
	month = jul,
	year = {2001},
	pages = {022313},
}

@article{lee_teleportation_2011,
	title = {Teleportation of {Nonclassical} {Wave} {Packets} of {Light}},
	volume = {332},
	url = {https://www.science.org/doi/full/10.1126/science.1201034},
	doi = {10.1126/science.1201034},
	number = {6027},
	urldate = {2023-09-17},
	journal = {Science},
	author = {Lee, Noriyuki and Benichi, Hugo and Takeno, Yuishi and Takeda, Shuntaro and Webb, James and Huntington, Elanor and Furusawa, Akira},
	month = apr,
	year = {2011},
	pages = {330--333},
}

@article{lund_fault-tolerant_2008,
	title = {Fault-{Tolerant} {Linear} {Optical} {Quantum} {Computing} with {Small}-{Amplitude} {Coherent} {States}},
	volume = {100},
	url = {https://link.aps.org/doi/10.1103/PhysRevLett.100.030503},
	doi = {10.1103/PhysRevLett.100.030503},
	number = {3},
	urldate = {2023-09-17},
	journal = {Physical Review Letters},
	author = {Lund, A. P. and Ralph, T. C. and Haselgrove, H. L.},
	month = jan,
	year = {2008},
	pages = {030503},
}

@article{neergaard-nielsen_generation_2006,
	title = {Generation of a {Superposition} of {Odd} {Photon} {Number} {States} for {Quantum} {Information} {Networks}},
	volume = {97},
	url = {https://link.aps.org/doi/10.1103/PhysRevLett.97.083604},
	doi = {10.1103/PhysRevLett.97.083604},
	number = {8},
	urldate = {2023-09-17},
	journal = {Physical Review Letters},
	author = {Neergaard-Nielsen, J. S. and Nielsen, B. Melholt and Hettich, C. and Mølmer, K. and Polzik, E. S.},
	month = aug,
	year = {2006},
	pages = {083604},
}

@article{zavatta_quantum--classical_2004,
	title = {Quantum-to-{Classical} {Transition} with {Single}-{Photon}-{Added} {Coherent} {States} of {Light}},
	volume = {306},
	url = {https://www.science.org/doi/10.1126/science.1103190},
	doi = {10.1126/science.1103190},
	number = {5696},
	urldate = {2022-04-20},
	journal = {Science},
	author = {Zavatta, Alessandro and Viciani, Silvia and Bellini, Marco},
	month = oct,
	year = {2004},
	pages = {660--662},
}

@article{paavola_finite-time_2011,
	title = {Finite-time quantum-to-classical transition for a {Schr}ödinger-cat state},
	volume = {84},
	url = {https://link.aps.org/doi/10.1103/PhysRevA.84.012121},
	doi = {10.1103/PhysRevA.84.012121},
	number = {1},
	urldate = {2026-04-08},
	journal = {Physical Review A},
	publisher = {American Physical Society},
	author = {Paavola, Janika and Hall, Michael J. W. and Paris, Matteo G. A. and Maniscalco, Sabrina},
	month = jul,
	year = {2011},
	pages = {012121},
}

@article{gilchrist_schrodinger_2004,
	title = {Schrödinger cats and their power for quantum information processing},
	volume = {6},
	issn = {1464-4266},
	url = {https://dx.doi.org/10.1088/1464-4266/6/8/032},
	doi = {10.1088/1464-4266/6/8/032},
	number = {8},
	urldate = {2022-12-19},
	journal = {Journal of Optics B: Quantum and Semiclassical Optics},
	author = {Gilchrist, A. and Nemoto, Kae and Munro, W. J. and Ralph, T. C. and Glancy, S. and Braunstein, Samuel L. and Milburn, G. J.},
	month = jul,
	year = {2004},
	pages = {S828},
}

@article{ourjoumtsev_generation_2007,
	title = {Generation of optical ‘{Schrödinger} cats’ from photon number states},
	volume = {448},
	copyright = {2007 Springer Nature Limited},
	issn = {1476-4687},
	url = {https://www.nature.com/articles/nature06054},
	doi = {10.1038/nature06054},
	number = {7155},
	urldate = {2023-06-20},
	journal = {Nature},
	author = {Ourjoumtsev, Alexei and Jeong, Hyunseok and Tualle-Brouri, Rosa and Grangier, Philippe},
	month = aug,
	year = {2007},
	note = {Number: 7155},
	pages = {784--786},
}

@article{hacker_deterministic_2019,
	title = {Deterministic creation of entangled atom–light {Schrödinger}-cat states},
	volume = {13},
	copyright = {2019 The Author(s), under exclusive licence to Springer Nature Limited},
	issn = {1749-4893},
	url = {https://www.nature.com/articles/s41566-018-0339-5},
	doi = {10.1038/s41566-018-0339-5},
	number = {2},
	urldate = {2023-09-17},
	journal = {Nature Photonics},
	author = {Hacker, Bastian and Welte, Stephan and Daiss, Severin and Shaukat, Armin and Ritter, Stephan and Li, Lin and Rempe, Gerhard},
	month = feb,
	year = {2019},
	pages = {110--115},
}

@article{burnham_observation_1970,
	title = {Observation of {Simultaneity} in {Parametric} {Production} of {Optical} {Photon} {Pairs}},
	volume = {25},
	url = {https://link.aps.org/doi/10.1103/PhysRevLett.25.84},
	doi = {10.1103/PhysRevLett.25.84},
	number = {2},
	urldate = {2026-04-08},
	journal = {Physical Review Letters},
	author = {Burnham, David C. and Weinberg, Donald L.},
	month = jul,
	year = {1970},
	pages = {84--87},
}

@article{hong_measurement_1987,
	title = {Measurement of subpicosecond time intervals between two photons by interference},
	volume = {59},
	url = {https://link.aps.org/doi/10.1103/PhysRevLett.59.2044},
	doi = {10.1103/PhysRevLett.59.2044},
	number = {18},
	urldate = {2026-04-08},
	journal = {Physical Review Letters},
	author = {Hong, C. K. and Ou, Z. Y. and Mandel, L.},
	month = nov,
	year = {1987},
	pages = {2044--2046},
}

@article{andersen_hybrid_2015,
	title = {Hybrid discrete- and continuous-variable quantum information},
	volume = {11},
	copyright = {2015 Springer Nature Limited},
	issn = {1745-2481},
	url = {https://www.nature.com/articles/nphys3410},
	doi = {10.1038/nphys3410},
	number = {9},
	urldate = {2026-04-08},
	journal = {Nature Physics},
	publisher = {Nature Publishing Group},
	author = {Andersen, Ulrik L. and Neergaard-Nielsen, Jonas S. and van Loock, Peter and Furusawa, Akira},
	month = sep,
	year = {2015},
	pages = {713--719},
}

@article{van_loock_optical_2011,
	title = {Optical hybrid approaches to quantum information},
	volume = {5},
	issn = {1863-8899},
	url = {https://onlinelibrary.wiley.com/doi/abs/10.1002/lpor.201000005},
	doi = {10.1002/lpor.201000005},
	number = {2},
	urldate = {2022-04-22},
	journal = {Laser \& Photonics Reviews},
	author = {van Loock, P.},
	year = {2011},
	pages = {167--200},
}

@article{wu_squeezed_1987,
	title = {Squeezed states of light from an optical parametric oscillator},
	volume = {4},
	copyright = {© 1987 Optical Society of America},
	issn = {1520-8540},
	url = {https://opg.optica.org/josab/abstract.cfm?uri=josab-4-10-1465},
	doi = {10.1364/JOSAB.4.001465},
	number = {10},
	urldate = {2024-07-23},
	journal = {JOSA B},
	author = {Wu, Ling-An and Xiao, Min and Kimble, H. J.},
	month = oct,
	year = {1987},
	pages = {1465--1475},
}

@article{lam_optimization_1999,
	title = {Optimization and transfer of vacuum squeezing from an optical parametric oscillator},
	volume = {1},
	issn = {1464-4266},
	url = {https://dx.doi.org/10.1088/1464-4266/1/4/319},
	doi = {10.1088/1464-4266/1/4/319},
	number = {4},
	urldate = {2024-07-23},
	journal = {Journal of Optics B: Quantum and Semiclassical Optics},
	author = {Lam, P. K. and Ralph, T. C. and Buchler, B. C. and McClelland, D. E. and Bachor, H.-A. and Gao, J.},
	month = aug,
	year = {1999},
	pages = {469},
}

@article{schneider_generation_1998,
	title = {Generation of strongly squeezed continuous-wave light at 1064 nm},
	volume = {2},
	copyright = {© 1998 Optical Society of America},
	issn = {1094-4087},
	url = {https://opg.optica.org/oe/abstract.cfm?uri=oe-2-3-59},
	doi = {10.1364/OE.2.000059},
	number = {3},
	urldate = {2024-07-23},
	journal = {Optics Express},
	author = {Schneider, K. and Lang, M. and Mlynek, J. and Schiller, S.},
	month = feb,
	year = {1998},
	pages = {59--64},
}

@article{abadie_gravitational_2011,
	title = {A gravitational wave observatory operating beyond the quantum shot-noise limit},
	volume = {7},
	copyright = {2011 Springer Nature Limited},
	issn = {1745-2481},
	url = {https://www.nature.com/articles/nphys2083},
	doi = {10.1038/nphys2083},
	number = {12},
	urldate = {2023-08-21},
	journal = {Nature Physics},
	author = {{The LIGO Scientific Collaboration}},
	month = dec,
	year = {2011},
	pages = {962--965},
}

@article{xiao_precision_1987,
	title = {Precision measurement beyond the shot-noise limit},
	volume = {59},
	url = {https://link.aps.org/doi/10.1103/PhysRevLett.59.278},
	doi = {10.1103/PhysRevLett.59.278},
	number = {3},
	urldate = {2023-09-15},
	journal = {Physical Review Letters},
	author = {Xiao, Min and Wu, Ling-An and Kimble, H. J.},
	month = jul,
	year = {1987},
	pages = {278--281},
}

@article{slusher_observation_1985,
	title = {Observation of {Squeezed} {States} {Generated} by {Four}-{Wave} {Mixing} in an {Optical} {Cavity}},
	volume = {55},
	url = {https://link.aps.org/doi/10.1103/PhysRevLett.55.2409},
	doi = {10.1103/PhysRevLett.55.2409},
	number = {22},
	urldate = {2023-09-15},
	journal = {Physical Review Letters},
	author = {Slusher, R. E. and Hollberg, L. W. and Yurke, B. and Mertz, J. C. and Valley, J. F.},
	month = nov,
	year = {1985},
	pages = {2409--2412},
}

@article{walls_squeezed_1983,
	title = {Squeezed states of light},
	volume = {306},
	copyright = {1983 Springer Nature Limited},
	issn = {1476-4687},
	url = {https://www.nature.com/articles/306141a0},
	doi = {10.1038/306141a0},
	number = {5939},
	urldate = {2023-09-15},
	journal = {Nature},
	author = {Walls, D. F.},
	month = nov,
	year = {1983},
	pages = {141--146},
}

@article{portmann_security_2022,
	title = {Security in quantum cryptography},
	volume = {94},
	url = {https://link.aps.org/doi/10.1103/RevModPhys.94.025008},
	doi = {10.1103/RevModPhys.94.025008},
	number = {2},
	urldate = {2026-04-08},
	journal = {Reviews of Modern Physics},
	publisher = {American Physical Society},
	author = {Portmann, Christopher and Renner, Renato},
	month = jun,
	year = {2022},
	pages = {025008},
}

@article{horodecki_quantum_2009,
	title = {Quantum entanglement},
	volume = {81},
	url = {https://link.aps.org/doi/10.1103/RevModPhys.81.865},
	doi = {10.1103/RevModPhys.81.865},
	number = {2},
	urldate = {2022-11-03},
	journal = {Reviews of Modern Physics},
	publisher = {American Physical Society},
	author = {Horodecki, Ryszard and Horodecki, Paweł and Horodecki, Michał and Horodecki, Karol},
	month = jun,
	year = {2009},
	pages = {865--942},
}

@article{degen_quantum_2017,
	title = {Quantum sensing},
	volume = {89},
	url = {https://link.aps.org/doi/10.1103/RevModPhys.89.035002},
	doi = {10.1103/RevModPhys.89.035002},
	number = {3},
	urldate = {2022-02-16},
	journal = {Reviews of Modern Physics},
	publisher = {American Physical Society},
	author = {Degen, C. L. and Reinhard, F. and Cappellaro, P.},
	month = jul,
	year = {2017},
	pages = {035002},
}

@article{georgescu_quantum_2014,
	title = {Quantum simulation},
	volume = {86},
	url = {https://link.aps.org/doi/10.1103/RevModPhys.86.153},
	doi = {10.1103/RevModPhys.86.153},
	number = {1},
	urldate = {2022-02-16},
	journal = {Reviews of Modern Physics},
	publisher = {American Physical Society},
	author = {Georgescu, I. M. and Ashhab, S. and Nori, Franco},
	month = mar,
	year = {2014},
	pages = {153--185},
}

@article{gisin_quantum_2007,
	title = {Quantum communication},
	volume = {1},
	copyright = {2007 Nature Publishing Group},
	issn = {1749-4893},
	url = {https://www.nature.com/articles/nphoton.2007.22},
	doi = {10.1038/nphoton.2007.22},
	number = {3},
	urldate = {2022-12-19},
	journal = {Nature Photonics},
	author = {Gisin, Nicolas and Thew, Rob},
	month = mar,
	year = {2007},
	pages = {165--171}
}

@article{kok_linear_2007,
	title = {Linear optical quantum computing with photonic qubits},
	volume = {79},
	url = {https://link.aps.org/doi/10.1103/RevModPhys.79.135},
	doi = {10.1103/RevModPhys.79.135},
	number = {1},
	urldate = {2023-08-21},
	journal = {Reviews of Modern Physics},
	author = {Kok, Pieter and Munro, W. J. and Nemoto, Kae and Ralph, T. C. and Dowling, Jonathan P. and Milburn, G. J.},
	month = jan,
	year = {2007},
	pages = {135--174},
}

@article{aspuru-guzik_photonic_2012,
	title = {Photonic quantum simulators},
	volume = {8},
	copyright = {2012 Springer Nature Limited},
	issn = {1745-2481},
	url = {https://www.nature.com/articles/nphys2253},
	doi = {10.1038/nphys2253},
	number = {4},
	urldate = {2023-08-21},
	journal = {Nature Physics},
	author = {Aspuru-Guzik, Alán and Walther, Philip},
	month = apr,
	year = {2012},
	pages = {285--291},
}

@article{giovannetti_advances_2011,
	title = {Advances in quantum metrology},
	volume = {5},
	issn = {1749-4893},
	url = {http://www.nature.com/articles/nphoton.2011.35},
	doi = {10.1038/nphoton.2011.35},
	number = {4},
	urldate = {2022-02-16},
	journal = {Nature Photonics},
	author = {Giovannetti, Vittorio and Lloyd, Seth and Maccone, Lorenzo},
	month = apr,
	year = {2011},
	pages = {222--229},
}

@article{pirandola_advances_2018,
	title = {Advances in photonic quantum sensing},
	volume = {12},
	copyright = {2018 Springer Nature Limited},
	issn = {1749-4893},
	url = {https://www.nature.com/articles/s41566-018-0301-6},
	doi = {10.1038/s41566-018-0301-6},
	number = {12},
	urldate = {2023-09-07},
	journal = {Nature Photonics},
	author = {Pirandola, S. and Bardhan, B. R. and Gehring, T. and Weedbrook, C. and Lloyd, S.},
	month = dec,
	year = {2018},
	pages = {724--733},
}

@article{usenko_continuous-variable_2026,
	title = {Continuous-variable quantum communication},
	volume = {98},
	url = {https://link.aps.org/doi/10.1103/mgj7-t6d3},
	doi = {10.1103/mgj7-t6d3},
	number = {1},
	urldate = {2026-04-08},
	journal = {Reviews of Modern Physics},
	publisher = {American Physical Society},
	author = {Usenko, Vladyslav C. and Acín, Antonio and Alléaume, Romain and Andersen, Ulrik L. and Diamanti, Eleni and Gehring, Tobias and Hajomer, Adnan A. E. and Kanitschar, Florian and Pacher, Christoph and Pirandola, Stefano and Pruneri, Valerio},
	month = mar,
	year = {2026},
	pages = {015003},
}

@article{acin_quantum_2018,
	title = {The quantum technologies roadmap: a {European} community view},
	volume = {20},
	issn = {1367-2630},
	shorttitle = {The quantum technologies roadmap},
	url = {https://dx.doi.org/10.1088/1367-2630/aad1ea},
	doi = {10.1088/1367-2630/aad1ea},
	number = {8},
	urldate = {2023-05-29},
	journal = {New Journal of Physics},
	publisher = {IOP Publishing},
	author = {Acín, Antonio and Bloch, Immanuel and Buhrman, Harry and Calarco, Tommaso and Eichler, Christopher and Eisert, Jens and Esteve, Daniel and Gisin, Nicolas and Glaser, Steffen J. and Jelezko, Fedor and Kuhr, Stefan and Lewenstein, Maciej and Riedel, Max F. and Schmidt, Piet O. and Thew, Rob and Wallraff, Andreas and Walmsley, Ian and Wilhelm, Frank K.},
	month = aug,
	year = {2018},
	pages = {080201},
}

@article{lvovsky_continuous-variable_2009,
	title = {Continuous-variable optical quantum-state tomography},
	volume = {81},
	url = {https://link.aps.org/doi/10.1103/RevModPhys.81.299},
	doi = {10.1103/RevModPhys.81.299},
	number = {1},
	urldate = {2023-09-11},
	journal = {Reviews of Modern Physics},
	publisher = {American Physical Society},
	author = {Lvovsky, A. I. and Raymer, M. G.},
	month = mar,
	year = {2009},
	pages = {299--332},
}

@article{rivera-dean_attosecond_2026,
	title = {Attosecond quantum optical interferometry},
	volume = {89},
	issn = {0034-4885},
	url = {https://doi.org/10.1088/1361-6633/ae5847},
	doi = {10.1088/1361-6633/ae5847},
	number = {4},
	urldate = {2026-05-25},
	journal = {Reports on Progress in Physics},
	publisher = {IOP Publishing},
	author = {Rivera-Dean, Javier and Petrovic, Lidija and Lewenstein, Maciej and Stammer, Philipp},
	month = apr,
	year = {2026},
	pages = {047901},
}

@article{lamprou_nonlinear_2025,
	title = {Nonlinear {Optics} {Using} {Intense} {Optical} {Coherent} {State} {Superpositions}},
	volume = {134},
	url = {https://link.aps.org/doi/10.1103/PhysRevLett.134.013601},
	doi = {10.1103/PhysRevLett.134.013601},
	number = {1},
	urldate = {2025-02-04},
	journal = {Physical Review Letters},
	publisher = {American Physical Society},
	author = {Lamprou, Th. and Rivera-Dean, J. and Stammer, P. and Lewenstein, M. and Tzallas, P.},
	month = jan,
	year = {2025},
	pages = {013601},
}

@article{kern_single-shot_2026,
	title = {Single-shot pulse retrieval of femtosecond bright squeezed vacuum},
	volume = {13},
	copyright = {© 2026 Optica Publishing Group},
	issn = {2334-2536},
	url = {https://opg.optica.org/optica/abstract.cfm?uri=optica-13-3-395},
	doi = {10.1364/OPTICA.580767},
	number = {3},
	urldate = {2026-04-06},
	journal = {Optica},
	publisher = {Optica Publishing Group},
	author = {Kern, Yuval and Nisim, Ido and Birk, Michael and Rasputnyi, Andrei and Behar, Doron and Chen, Zhaopin and Kaminer, Ido and Sidorenko, Pavel and Cohen, Oren and Krüger, Michael},
	month = mar,
	year = {2026},
	pages = {395--399},
}

@misc{weber_universal_2025,
	title = {A universal approach to saddle-point methods in attosecond science},
	url = {http://arxiv.org/abs/2510.12545},
	doi = {10.48550/arXiv.2510.12545},
	urldate = {2026-04-05},
	publisher = {arXiv},
	author = {Weber, Anne and Feldbrugge, Job and Pisanty, Emilio},
	month = oct,
	year = {2025},
	note = {arXiv:2510.12545 [quant-ph]},
}

@misc{braccini_superpositions_2025,
	title = {Superpositions of {Quantum} {Gaussian} {Processes}},
	url = {http://arxiv.org/abs/2510.01156},
	doi = {10.48550/arXiv.2510.01156},
	urldate = {2026-04-05},
	publisher = {arXiv},
	author = {Braccini, Lorenzo and Bose, Sougato and Serafini, Alessio},
	month = oct,
	year = {2025},
	note = {arXiv:2510.01156 [quant-ph]},
}

@article{hahn_classical_2025,
	title = {Classical simulation and quantum resource theory of non-{Gaussian} optics},
	volume = {9},
	url = {https://quantum-journal.org/papers/q-2025-10-13-1881/},
	doi = {10.22331/q-2025-10-13-1881},
	urldate = {2026-04-05},
	journal = {Quantum},
	publisher = {Verein zur Förderung des Open Access Publizierens in den Quantenwissenschaften},
	author = {Hahn, Oliver and Takagi, Ryuji and Ferrini, Giulia and Yamasaki, Hayata},
	month = oct,
	year = {2025},
	pages = {1881},
}

@misc{centrone_gaussian_2026,
	title = {Gaussian superpositions for bosonic encodings},
	url = {http://arxiv.org/abs/2603.15258},
	doi = {10.48550/arXiv.2603.15258},
	urldate = {2026-04-05},
	publisher = {arXiv},
	author = {Centrone, Federico and Paz, Juan Pablo and Roncaglia, Augusto},
	month = mar,
	year = {2026},
	note = {arXiv:2603.15258 [quant-ph]},
}

@inbook{olga_simpleman,
	author = {Smirnova, Olga and Ivanov, Misha},
	publisher = {John Wiley \& Sons, Ltd},
	isbn = {9783527677689},
	title = {Multielectron High Harmonic Generation: Simple Man on a Complex Plane},
	booktitle = {Attosecond and XUV Physics},
	chapter = {7},
	pages = {201-256},
	doi = {https://doi.org/10.1002/9783527677689.ch7},
	year = {2014}
}

@article{stammer_quantum_2023,
	title = {Quantum {Electrodynamics} of {Intense} {Laser}-{Matter} {Interactions}: {A} {Tool} for {Quantum} {State} {Engineering}},
	volume = {4},
	shorttitle = {Quantum {Electrodynamics} of {Intense} {Laser}-{Matter} {Interactions}},
	url = {https://link.aps.org/doi/10.1103/PRXQuantum.4.010201},
	doi = {10.1103/PRXQuantum.4.010201},
	number = {1},
	urldate = {2023-01-25},
	journal = {PRX Quantum},
	publisher = {American Physical Society},
	author = {Stammer, Philipp and Rivera-Dean, Javier and Maxwell, Andrew and Lamprou, Theocharis and Ordóñez, Andrés and Ciappina, Marcelo F. and Tzallas, Paraskevas and Lewenstein, Maciej},
	month = jan,
	year = {2023},
	pages = {010201},
}

@article{lemieux_photon_2025,
	title = {Photon bunching in high-harmonic emission controlled by quantum light},
	volume = {19},
	copyright = {2025 Crown},
	issn = {1749-4893},
	url = {https://www.nature.com/articles/s41566-025-01673-6},
	doi = {10.1038/s41566-025-01673-6},
	number = {7},
	urldate = {2025-12-01},
	journal = {Nature Photonics},
	author = {Lemieux, Samuel and Jalil, Sohail A. and Purschke, David N. and Boroumand, Neda and Hammond, T. J. and Villeneuve, David and Naumov, Andrei and Brabec, Thomas and Vampa, Giulio},
	month = jul,
	year = {2025},
	pages = {767--771},
}

@misc{tzur_measuring_2025,
	title = {Measuring and controlling the birth of quantum attosecond pulses},
	url = {http://arxiv.org/abs/2502.09427},
	doi = {10.48550/arXiv.2502.09427},
	urldate = {2025-02-24},
	publisher = {arXiv},
	author = {Tzur, Matan Even and Mor, Chen and Yaffe, Noa and Birk, Michael and Rasputnyi, Andrei and Kneller, Omer and Nisim, Ido and Kaminer, Ido and Krüger, Michael and Dudovich, Nirit and Cohen, Oren},
	month = feb,
	year = {2025},
	note = {arXiv:2502.09427 [physics]},
}

@article{heimerl_multiphoton_2024,
	title = {Multiphoton electron emission with non-classical light},
	volume = {20},
	copyright = {2024 The Author(s), under exclusive licence to Springer Nature Limited},
	issn = {1745-2481},
	url = {https://www.nature.com/articles/s41567-024-02472-6},
	doi = {10.1038/s41567-024-02472-6},
	number = {6},
	urldate = {2025-08-12},
	journal = {Nature Physics},
	author = {Heimerl, Jonas and Mikhaylov, Alexander and Meier, Stefan and Höllerer, Henrick and Kaminer, Ido and Chekhova, Maria and Hommelhoff, Peter},
	month = jun,
	year = {2024},
	pages = {945--950},
}

@article{heimerl_driving_2025,
	title = {Quantum light drives electrons strongly at metal needle tips},
	volume = {21},
	copyright = {2025 The Author(s)},
	issn = {1745-2481},
	url = {https://www.nature.com/articles/s41567-025-03087-1},
	doi = {10.1038/s41567-025-03087-1},
	number = {12},
	urldate = {2026-04-10},
	journal = {Nature Physics},
	author = {Heimerl, Jonas and Rasputnyi, Andrei and Pölloth, Jonathan and Meier, Stefan and Chekhova, Maria and Hommelhoff, Peter},
	month = dec,
	year = {2025},
	pages = {1899--1904},
}

@article{spasibko_multiphoton_2017,
	title = {Multiphoton {Effects} {Enhanced} due to {Ultrafast} {Photon}-{Number} {Fluctuations}},
	volume = {119},
	url = {https://link.aps.org/doi/10.1103/PhysRevLett.119.223603},
	doi = {10.1103/PhysRevLett.119.223603},
	number = {22},
	urldate = {2025-02-06},
	journal = {Physical Review Letters},
	publisher = {American Physical Society},
	author = {Spasibko, Kirill Yu. and Kopylov, Denis A. and Krutyanskiy, Victor L. and Murzina, Tatiana V. and Leuchs, Gerd and Chekhova, Maria V.},
	month = nov,
	year = {2017},
	pages = {223603},
}

@article{manceau_indefinite-mean_2019,
	title = {Indefinite-{Mean} {Pareto} {Photon} {Distribution} from {Amplified} {Quantum} {Noise}},
	volume = {123},
	url = {https://link.aps.org/doi/10.1103/PhysRevLett.123.123606},
	doi = {10.1103/PhysRevLett.123.123606},
	number = {12},
	urldate = {2025-02-06},
	journal = {Physical Review Letters},
	publisher = {American Physical Society},
	author = {Manceau, Mathieu and Spasibko, Kirill Yu. and Leuchs, Gerd and Filip, Radim and Chekhova, Maria V.},
	month = sep,
	year = {2019},
	pages = {123606},
}

@article{rasputnyi_high_2024,
	title = {High-harmonic generation by a bright squeezed vacuum},
	copyright = {2024 The Author(s)},
	issn = {1745-2481},
	url = {https://www.nature.com/articles/s41567-024-02659-x},
	doi = {10.1038/s41567-024-02659-x},
	urldate = {2024-11-11},
	journal = {Nature Physics},
	author = {Rasputnyi, Andrei and Chen, Zhaopin and Birk, Michael and Cohen, Oren and Kaminer, Ido and Krüger, Michael and Seletskiy, Denis and Chekhova, Maria and Tani, Francesco},
	month = oct,
	year = {2024},
}

@article{dakna_generating_1997,
	title = {Generating {Schödinger-cat-like} states by means of conditional measurements on a beam splitter},
	volume = {55},
	url = {https://link.aps.org/doi/10.1103/PhysRevA.55.3184},
	doi = {10.1103/PhysRevA.55.3184},
	number = {4},
	urldate = {2023-09-17},
	journal = {Physical Review A},
	publisher = {American Physical Society},
	author = {Dakna, M. and Anhut, T. and Opatrný, T. and Knöll, L. and Welsch, D.-G.},
	month = apr,
	year = {1997},
	pages = {3184--3194},
}

@article{banaszek_nonlocality_1998,
	title = {Nonlocality of the {Einstein}-{Podolsky}-{Rosen} state in the {Wigner} representation},
	volume = {58},
	url = {https://link.aps.org/doi/10.1103/PhysRevA.58.4345},
	doi = {10.1103/PhysRevA.58.4345},
	number = {6},
	urldate = {2026-04-08},
	journal = {Physical Review A},
	author = {Banaszek, Konrad and Wódkiewicz, Krzysztof},
	month = dec,
	year = {1998},
	pages = {4345--4347},
}

@article{banaszek_testing_1999,
	title = {Testing {Quantum} {Nonlocality} in {Phase} {Space}},
	volume = {82},
	copyright = {http://link.aps.org/licenses/aps-default-license},
	issn = {0031-9007, 1079-7114},
	url = {https://link.aps.org/doi/10.1103/PhysRevLett.82.2009},
	doi = {10.1103/PhysRevLett.82.2009},
	number = {10},
	urldate = {2026-03-09},
	journal = {Physical Review Letters},
	author = {Banaszek, Konrad and Wódkiewicz, Krzysztof},
	month = mar,
	year = {1999},
	pages = {2009--2013},
}

@article{bell_einstein_1964,
	title = {On the {Einstein} {Podolsky} {Rosen} paradox},
	volume = {1},
	url = {https://link.aps.org/doi/10.1103/PhysicsPhysiqueFizika.1.195},
	doi = {10.1103/PhysicsPhysiqueFizika.1.195},
	number = {3},
	urldate = {2023-08-21},
	journal = {Physics Physique Fizika},
	author = {Bell, J. S.},
	month = nov,
	year = {1964},
	pages = {195--200},
}

@article{born_zur_1926,
	title = {Zur {Quantenmechanik} der {Stoßvorgänge}},
	volume = {37},
	issn = {0044-3328},
	url = {https://doi.org/10.1007/BF01397477},
	doi = {10.1007/BF01397477},
	number = {12},
	urldate = {2023-12-02},
	journal = {Zeitschrift für Physik},
	author = {Born, Max},
	month = dec,
	year = {1926},
	pages = {863--867},
}

@article{pitowsky_range_1986,
	title = {The range of quantum probability},
	volume = {27},
	issn = {0022-2488},
	url = {https://doi.org/10.1063/1.527066},
	doi = {10.1063/1.527066},
	number = {6},
	urldate = {2023-09-20},
	journal = {Journal of Mathematical Physics},
	author = {Pitowsky, Itamar},
	month = jun,
	year = {1986},
	pages = {1556--1565},
}

@inbook{NielsenBookCh1,
	booktitle = {Quantum {Computation} and {Quantum} {Information}: 10th {Anniversary} {Edition}},
	isbn = {978-1-139-49548-6},
	title = {Introduction to quantum mechanics},
	publisher = {Cambridge University Press, Cambridge, UK},
	author = {Nielsen, Michael A. and Chuang, Isaac L.},
	month = dec,
	year = {2010},
	chapter = {2},
	pages = {60-119},
}

@article{brunner_bell_2014,
	title = {Bell nonlocality},
	volume = {86},
	url = {https://link.aps.org/doi/10.1103/RevModPhys.86.419},
	doi = {10.1103/RevModPhys.86.419},
	number = {2},
	urldate = {2022-01-18},
	journal = {Reviews of Modern Physics},
	publisher = {American Physical Society},
	author = {Brunner, Nicolas and Cavalcanti, Daniel and Pironio, Stefano and Scarani, Valerio and Wehner, Stephanie},
	month = apr,
	year = {2014},
	pages = {419--478},
}

@article{liu_atomic_2025,
	title = {Atomic {Double} {Ionization} with {Quantum} {Light}},
	volume = {134},
	url = {https://link.aps.org/doi/10.1103/PhysRevLett.134.123202},
	doi = {10.1103/PhysRevLett.134.123202},
	number = {12},
	urldate = {2025-06-11},
	journal = {Physical Review Letters},
	author = {Liu, Haoyu and Zhang, Hanxu and Wang, Xu and Yuan, Jianmin},
	month = mar,
	year = {2025},
	pages = {123202},
}

@article{wang_high-order_2023,
	title = {High-order above-threshold ionization of an atom in intense quantum light},
	volume = {108},
	url = {https://link.aps.org/doi/10.1103/PhysRevA.108.063101},
	doi = {10.1103/PhysRevA.108.063101},
	number = {6},
	urldate = {2025-06-11},
	journal = {Physical Review A},
	author = {Wang, ShiJun and Lai, XuanYang},
	month = dec,
	year = {2023},
	pages = {063101},
}

@article{lyu_effect_2025,
	title = {Effect of photon quantum statistics on electrons in above-threshold ionization},
	volume = {7},
	url = {https://link.aps.org/doi/10.1103/PhysRevResearch.7.L012072},
	doi = {10.1103/PhysRevResearch.7.L012072},
	number = {1},
	urldate = {2025-06-11},
	journal = {Physical Review Research},
	author = {Lyu, Zijian and Sun, Fengxiao and Fang, Yiqi and He, Qiongyi and Liu, Yunquan},
	month = mar,
	year = {2025},
	pages = {L012072},
}

@article{drummond_generalised_1980,
	title = {Generalised {P}-representations in quantum optics},
	volume = {13},
	issn = {0305-4470},
	url = {https://dx.doi.org/10.1088/0305-4470/13/7/018},
	doi = {10.1088/0305-4470/13/7/018},
	number = {7},
	urldate = {2024-10-31},
	journal = {Journal of Physics A: Mathematical and General},
	author = {Drummond, P. D. and Gardiner, C. W.},
	month = jul,
	year = {1980},
	pages = {2353},
}

@article{stammer_squeezing_2023,
	title = {Entanglement and {Squeezing} of the {Optical} {Field} {Modes} in {High} {Harmonic} {Generation}},
	volume = {132},
	url = {https://link.aps.org/doi/10.1103/PhysRevLett.132.143603},
	doi = {10.1103/PhysRevLett.132.143603},
	number = {14},
	urldate = {2024-05-14},
	journal = {Physical Review Letters},
	author = {Stammer, Philipp and Rivera-Dean, Javier and Maxwell, Andrew S. and Lamprou, Theocharis and Argüello-Luengo, Javier and Tzallas, Paraskevas and Ciappina, Marcelo F. and Lewenstein, Maciej},
	month = apr,
	year = {2024},
	pages = {143603},
}

@article{magnus_exponential_1954,
	title = {On the exponential solution of differential equations for a linear operator},
	volume = {7},
	issn = {0010-3640, 1097-0312},
	url = {https://onlinelibrary.wiley.com/doi/10.1002/cpa.3160070404},
	doi = {10.1002/cpa.3160070404},
	number = {4},
	urldate = {2026-03-09},
	journal = {Communications on Pure and Applied Mathematics},
	author = {Magnus, Wilhelm},
	month = nov,
	year = {1954},
	pages = {649--673},
}

@article{milosevic_phase_2013,
	title = {Phase space path-integral formulation of the above-threshold ionization},
	volume = {54},
	issn = {0022-2488},
	url = {https://doi.org/10.1063/1.4797476},
	doi = {10.1063/1.4797476},
	number = {4},
	urldate = {2025-06-13},
	journal = {Journal of Mathematical Physics},
	author = {Milošević, D. B.},
	month = apr,
	year = {2013},
	pages = {042101},
}

@article{lewenstein_rings_1995,
	title = {Rings in above-threshold ionization: {A} quasiclassical analysis},
	volume = {51},
	shorttitle = {Rings in above-threshold ionization},
	url = {https://link.aps.org/doi/10.1103/PhysRevA.51.1495},
	doi = {10.1103/PhysRevA.51.1495},
	number = {2},
	urldate = {2022-06-16},
	journal = {Physical Review A},
	author = {Lewenstein, M. and Kulander, K. C. and Schafer, K. J. and Bucksbaum, P. H.},
	month = feb,
	year = {1995},
	pages = {1495--1507},
}

@article{lohr_above-threshold_1997,
	title = {Above-threshold ionization in the tunneling regime},
	volume = {55},
	url = {https://link.aps.org/doi/10.1103/PhysRevA.55.R4003},
	doi = {10.1103/PhysRevA.55.R4003},
	number = {6},
	urldate = {2025-04-08},
	journal = {Physical Review A},
	author = {Lohr, A. and Kleber, M. and Kopold, R. and Becker, W.},
	month = jun,
	year = {1997},
	pages = {R4003--R4006},
}

@mastersthesis{rivera_dean_quantum-optical_2019,
	type = {Master thesis},
	title = {Quantum-optical analysis of high-order harmonic generation},
	copyright = {S'autoritza la difusió de l'obra mitjançant la llicència Creative Commons o similar 'Reconeixement-NoComercial- SenseObraDerivada'},
	url = {https://upcommons.upc.edu/handle/2117/168580},
	urldate = {2023-10-09},
	school = {Universitat Politècnica de Catalunya},
	author = {Rivera-Dean, Javier},
	month = sep,
	year = {2019}
}

@article{smirnova_anatomy_2007,
	title = {Anatomy of strong field ionization {II}: to dress or not to dress?},
	volume = {54},
	issn = {0950-0340, 1362-3044},
	shorttitle = {Anatomy of strong field ionization {II}},
	url = {http://www.tandfonline.com/doi/abs/10.1080/09500340701234656},
	doi = {10.1080/09500340701234656},
	number = {7},
	urldate = {2025-04-05},
	journal = {Journal of Modern Optics},
	author = {Smirnova, Olga and Spanner, Michael and Ivanov, Misha},
	month = may,
	year = {2007},
	pages = {1019--1038},
}

@misc{mao_benchmarking_2025,
	title = {Benchmarking {Atomic} {Ionization} {Driven} by {Strong} {Quantum} {Light}},
	url = {http://arxiv.org/abs/2512.15458},
	doi = {10.48550/arXiv.2512.15458},
	urldate = {2025-12-26},
	publisher = {arXiv},
	author = {Mao, Yi-Jia and Zhou, En-Rui and Li, Yang and He, Pei-Lun and He, Feng},
	month = dec,
	year = {2025},
	note = {arXiv:2512.15458 [quant-ph]},
}

@article{wang_high-order_2025,
	title = {High-order harmonic generation in quantum light by a generalized von {Neumann} lattice method},
	volume = {111},
	url = {https://link.aps.org/doi/10.1103/PhysRevA.111.043111},
	doi = {10.1103/PhysRevA.111.043111},
	number = {4},
	urldate = {2025-04-20},
	journal = {Physical Review A},
	publisher = {American Physical Society},
	author = {Wang, Yi-Ben and Bian, Xue-Bin},
	month = apr,
	year = {2025},
	pages = {043111},
}

@incollection{breuer_quantum_2007,
	title = {Quantum {Probability}},
	isbn = {978-0-19-921390-0},
	url = {https://doi.org/10.1093/acprof:oso/9780199213900.003.02},
	doi = {10.1093/acprof:oso/9780199213900.003.02},
	urldate = {2026-02-25},
	booktitle = {The {Theory} of {Open} {Quantum} {Systems}},
	publisher = {Oxford University Press},
	author = {Breuer, Heinz-Peter and Petruccione, Francesco},
	editor = {Breuer, Heinz-Peter and Petruccione, Francesco},
	month = jan,
	year = {2007},
	pages = {59--105},
}

@article{rivera-dean_microscopic_2025,
	title = {Microscopic analysis of above-threshold ionization driven by squeezed light},
	volume = {112},
	url = {https://link.aps.org/doi/10.1103/hb3n-h2zy},
	doi = {10.1103/hb3n-h2zy},
	number = {6},
	urldate = {2026-01-05},
	journal = {Physical Review A},
	author = {Rivera-Dean, J. and Stammer, P. and Faria, C. Figueira de Morisson and Lewenstein, M.},
	month = dec,
	year = {2025},
	pages = {063101},
}

@article{yi_generation_2025,
	title = {Generation of {Massively} {Entangled} {Bright} {States} of {Light} during {Harmonic} {Generation} in {Resonant} {Media}},
	volume = {15},
	url = {https://link.aps.org/doi/10.1103/PhysRevX.15.011023},
	doi = {10.1103/PhysRevX.15.011023},
	number = {1},
	urldate = {2025-02-24},
	journal = {Physical Review X},
	author = {Yi, Sili and Klimkin, Nikolai D. and Brown, Graham Gardiner and Smirnova, Olga and Patchkovskii, Serguei and Babushkin, Ihar and Ivanov, Misha},
	month = feb,
	year = {2025},
	pages = {011023},
}

@misc{rivera-dean_condition_202X,
	title =  {Non-classiality criteria using coherent state expansions},
	author = {Rivera-Dean, J. and Stammer, P.},
	month = dec,
	year = {202X},
}

@article{lewenstein_theory_1994,
	title = {Theory of high-harmonic generation by low-frequency laser fields},
	volume = {49},
	url = {https://link.aps.org/doi/10.1103/PhysRevA.49.2117},
	doi = {10.1103/PhysRevA.49.2117},
	number = {3},
	urldate = {2022-06-16},
	journal = {Physical Review A},
	author = {Lewenstein, M. and Balcou, Ph. and Ivanov, M. Yu. and L’Huillier, Anne and Corkum, P. B.},
	month = mar,
	year = {1994},
	pages = {2117--2132},
}

@article{amini_symphony_2019,
	title = {Symphony on strong field approximation},
	volume = {82},
	issn = {0034-4885},
	url = {https://doi.org/10.1088/1361-6633/ab2bb1},
	doi = {10.1088/1361-6633/ab2bb1},
	number = {11},
	urldate = {2021-12-01},
	journal = {Reports on Progress in Physics},
	author = {Amini, Kasra and Biegert, Jens and Calegari, Francesca and Chacón, Alexis and Ciappina, Marcelo F. and Dauphin, Alexandre and Efimov, Dmitry K. and Faria, Carla Figueira de Morisson and Giergiel, Krzysztof and Gniewek, Piotr and Landsman, Alexandra S. and Lesiuk, Micha{\textbackslash}l and Mandrysz, Micha{\textbackslash}l and Maxwell, Andrew S. and Moszyński, Robert and Ortmann, Lisa and Pérez-Hernández, Jose Antonio and Picón, Antonio and Pisanty, Emilio and Prauzner-Bechcicki, Jakub and Sacha, Krzysztof and Suárez, Noslen and Zaïr, Amelle and Zakrzewski, Jakub and Lewenstein, Maciej},
	month = oct,
	year = {2019},
	pages = {116001},
}

@article{ourjoumtsev_generating_2006,
	title = {Generating {Optical} {Schrödinger} {Kittens} for {Quantum} {Information} {Processing}},
	volume = {312},
	url = {https://www.science.org/doi/10.1126/science.1122858},
	doi = {10.1126/science.1122858},
	number = {5770},
	urldate = {2023-06-20},
	journal = {Science},
	author = {Ourjoumtsev, Alexei and Tualle-Brouri, Rosa and Laurat, Julien and Grangier, Philippe},
	month = apr,
	year = {2006},
	pages = {83--86},
}

@article{rivera-dean_structured_2025,
	title = {Structured {Squeezed} {Light} {Allows} for {High}-{Harmonic} {Generation} in {Classical} {Forbidden} {Geometries}},
	volume = {135},
	url = {https://link.aps.org/doi/10.1103/4hdl-bdwj},
	doi = {10.1103/4hdl-bdwj},
	number = {1},
	urldate = {2025-07-02},
	journal = {Physical Review Letters},
	author = {Rivera-Dean, J. and Stammer, P. and Ciappina, M. F. and Lewenstein, M.},
	month = jul,
	year = {2025},
	pages = {013801},
}

@article{even_tzur_photon-statistics_2023,
	title = {Photon-statistics force in ultrafast electron dynamics},
	volume = {17},
	copyright = {2023 The Author(s), under exclusive licence to Springer Nature Limited},
	issn = {1749-4893},
	url = {https://www.nature.com/articles/s41566-023-01209-w},
	doi = {10.1038/s41566-023-01209-w},
	number = {6},
	urldate = {2023-06-15},
	journal = {Nature Photonics},
	author = {Even Tzur, Matan and Birk, Michael and Gorlach, Alexey and Krüger, Michael and Kaminer, Ido and Cohen, Oren},
	month = jun,
	year = {2023},
	pages = {501--509},
}

@article{perelomov_completeness_1971,
	title = {On the completeness of a system of coherent states},
	volume = {6},
	issn = {1573-9333},
	url = {https://doi.org/10.1007/BF01036577},
	doi = {10.1007/BF01036577},
	number = {2},
	urldate = {2025-05-13},
	journal = {Theoretical and Mathematical Physics},
	author = {Perelomov, A. M.},
	month = feb,
	year = {1971},
	pages = {156--164},
}

@article{bargmann_completeness_1971,
	title = {On the completeness of the coherent states},
	volume = {2},
	issn = {0034-4877},
	url = {https://www.sciencedirect.com/science/article/pii/0034487771900061},
	doi = {10.1016/0034-4877(71)90006-1},
	number = {4},
	urldate = {2025-04-20},
	journal = {Reports on Mathematical Physics},
	author = {Bargmann, V. and Butera, P. and Girardello, L. and Klauder, John R.},
	month = dec,
	year = {1971},
	pages = {221--228},
}

@article{rivera-dean_light-matter_2022,
	title = {Light-matter entanglement after above-threshold ionization processes in atoms},
	volume = {106},
	url = {https://link.aps.org/doi/10.1103/PhysRevA.106.063705},
	doi = {10.1103/PhysRevA.106.063705},
	number = {6},
	urldate = {2023-06-05},
	journal = {Physical Review A},
	author = {Rivera-Dean, J. and Stammer, P. and Maxwell, A. S. and Lamprou, Th. and Tzallas, P. and Lewenstein, M. and Ciappina, M. F.},
	month = dec,
	year = {2022},
	pages = {063705},
}

@article{rivera-dean_role_2024,
	title = {Role of short and long trajectories on the quantum-optical state after high-order harmonic generation},
	volume = {110},
	url = {https://link.aps.org/doi/10.1103/PhysRevA.110.063704},
	doi = {10.1103/PhysRevA.110.063704},
	number = {6},
	urldate = {2025-01-06},
	journal = {Physical Review A},
	author = {Rivera-Dean, Javier},
	month = dec,
	year = {2024},
	pages = {063704},
}

@article{faria_it_2020,
	title = {It is all about phases: ultrafast holographic photoelectron imaging},
	volume = {83},
	issn = {0034-4885},
	shorttitle = {It is all about phases},
	url = {https://dx.doi.org/10.1088/1361-6633/ab5c91},
	doi = {10.1088/1361-6633/ab5c91},
	number = {3},
	urldate = {2023-09-25},
	journal = {Reports on Progress in Physics},
	author = {{Figueira de Morisson Faria}, C. and Maxwell, A. S.},
	month = jan,
	year = {2020},
	pages = {034401},
}

@phdthesis{maxwell_strong-field_2019,
	type = {PhD thesis},
	title = {Strong-{Field} {Interference} of {Quantum} {Trajectories} with {Coulomb} {Distortion} and {Electron} {Correlation}},
	copyright = {open},
	url = {https://discovery.ucl.ac.uk/id/eprint/10064744/},
	urldate = {2025-09-19},
	school = {UCL (University College London)},
	author = {Maxwell, Andrew S.},
	month = jan,
	year = {2019},
}

@article{lai_influence_2015,
	title = {Influence of the {Coulomb} potential on above-threshold ionization: {A} quantum-orbit analysis beyond the strong-field approximation},
	volume = {92},
	shorttitle = {Influence of the {Coulomb} potential on above-threshold ionization},
	url = {https://link.aps.org/doi/10.1103/PhysRevA.92.043407},
	doi = {10.1103/PhysRevA.92.043407},
	number = {4},
	urldate = {2023-09-26},
	journal = {Physical Review A},
	author = {Lai, X.-Y. and Poli, C. and Schomerus, H. and {Figueira de Morisson Faria}, C.},
	month = oct,
	year = {2015},
	pages = {043407},
}

@book{kleinert_path_2009,
	edition = {5},
	title = {Path {Integrals} in {Quantum} {Mechanics}, {Statistics}, {Polymer} {Physics}, and {Financial} {Markets}},
	isbn = {978-981-4273-55-8 978-981-4273-57-2},
	url = {https://www.worldscientific.com/worldscibooks/10.1142/7305},
	urldate = {2025-09-19},
	publisher = {{World Scientific}},
	author = {Kleinert, Hagen},
	month = may,
	year = {2009},
	doi = {10.1142/7305},
}

@article{johansson_qutip_2013,
	title = {{QuTiP} 2: {A} {Python} framework for the dynamics of open quantum systems},
	volume = {184},
	issn = {0010-4655},
	shorttitle = {{QuTiP} 2},
	url = {https://www.sciencedirect.com/science/article/pii/S0010465512003955},
	doi = {10.1016/j.cpc.2012.11.019},
	number = {4},
	urldate = {2023-03-06},
	journal = {Computer Physics Communications},
	author = {Johansson, J. R. and Nation, P. D. and Nori, Franco},
	month = apr,
	year = {2013},
	pages = {1234--1240},
}

@article{johansson_qutip_2012,
	title = {{QuTiP}: {An} open-source {Python} framework for the dynamics of open quantum systems},
	volume = {183},
	issn = {0010-4655},
	shorttitle = {{QuTiP}},
	url = {https://www.sciencedirect.com/science/article/pii/S0010465512000835},
	doi = {10.1016/j.cpc.2012.02.021},
	number = {8},
	urldate = {2023-03-06},
	journal = {Computer Physics Communications},
	author = {Johansson, J. R. and Nation, P. D. and Nori, Franco},
	month = aug,
	year = {2012},
	pages = {1760--1772},
}

\newpage
\clearpage
\appendix
\onecolumngrid
\begin{center}
	\large \textbf{\textsc{Supplementary Material}}
\end{center}

\section{THEORETICAL ANALYSIS}
\subsection{Preliminaries}\label{Sec:SM:Preliminaries}
In this work, we describe the light-matter interaction dynamics between an atomic system and a strong driving field within the single-active-electron and dipole approximations.~In the following, we work with SI units, and retake atomic units in Sec.~\ref{Sec:App:Semiclassical:SPA}.~The dynamics are governed by the time-dependent Schrödinger equation (TDSE)
\begin{equation}\label{Eq:SM:TDSE}
	i\hbar \pdv{\relaxket{\Tilde{\Psi}(t)}}{t}
		= \big[
				\hat{H}_{\text{at}} 
				+ \mathsf{e}\hat{\boldsymbol{r}}\cdot \hat{\boldsymbol{E}}
				+ \hat{H}_{\text{field}}
			\big] \relaxket{\Tilde{\Psi}(t)},
\end{equation}
where $\hat{H}_{\text{at}}$ denotes the atomic Hamiltonian and $\mathsf{e}\hat{\boldsymbol{r}}\cdot \hat{\boldsymbol{E}}$ describes the light-matter interaction in the length gauge.~The electric field operator is given by
\begin{equation}
	\hat{\boldsymbol{E}}
		= -i\sum_{\mu,q=1} \boldsymbol{\epsilon}_{\mu}g(\omega_q)
				\big[
					\hat{a}_{q,\mu}
					- \hat{a}_{q,\mu}^\dagger
				\big]
		\equiv \sum_{\mu,q=1} \hat{\boldsymbol{E}}_{q,\mu}
\end{equation}
where $\boldsymbol{\epsilon}_\mu$ denotes the polarization direction ($\mu = \perp, \parallel$), and $g(\omega_q) = \sqrt{\hbar \omega_q/(2\epsilon_0 V)}$ is a mode-dependent coupling constant arising from the quantization of the electric field operator, with $V$ denoting the quantization volume.~Finally, $\hat{H}_{\text{field}} = \sum_{q,\mu} \hbar \omega_q \hat{a}^\dagger_{q,\mu} \hat{a}_{q,\mu}$ is the free-field Hamiltonian.~For convenience, we move to the interaction picture with respect to $\hat{H}_{\text{field}}$, i.e., $\relaxket{\Tilde{\Psi}(t)} = e^{-i\hat{H}_{\text{field}}t/\hbar}\ket{\Psi(t)}$, such that the annihilation operators transform as $\hat{a}_{q,\mu} \to \hat{a}_{q,\mu} e^{-i\omega_q t}$. As a consequence, the electric field operator becomes explicitly time dependent, $\hat{\boldsymbol{E}} \to \hat{\boldsymbol{E}}(t)$, and Eq.~\eqref{Eq:SM:TDSE} then takes the form
\begin{equation}\label{Eq:SM:TDSE:II}
	i\hbar \pdv{\ket{\Psi(t)}}{t}
		= \big[
				\hat{H}_{\text{at}} 
					+ \mathsf{e}\hat{\boldsymbol{r}}\cdot \hat{\boldsymbol{E}}
			\big] \ket{\Psi(t)}.
\end{equation}

As the initial condition for Eq.~\eqref{Eq:SM:TDSE:II}, we consider an atomic system initially prepared in its ground state $\ket{\text{g}}$, while the field is initialized in a linearly polarized bright-squeezed vacuum (BSV) state $\hat{S}_L(r)\relaxket{0_L}$ for the fundamental mode $(\{q=1,\parallel\}\equiv L$), with $\hat{S}_{q,\mu}(r) = \exp[r(\hat{a}^2_{q,\mu}-\hat{a}^{\dagger 2}_{q,\mu})]$, and in the vacuum state for all remaining modes $\bigotimes_{q>2} \ket{0_q}$. In the following, we restrict to real squeezing parameters $r\in \mathbbm{R}$ which, in the absence of a phase reference such as a coherent displacement or an additional non-vacuum field, does not affect the generality of our results.~The initial joint light-matter state thus reads
\begin{equation}
	\ket{\Psi(t_0)}
		= \ket{\text{g}} \otimes \hat{S}_L(r) \ket{\boldsymbol{0}},
\end{equation}
where $\ket{\boldsymbol{0}} = \bigotimes_{q=1} \ket{0_q}$.~Using the resolution of the identity in the coherent state basis, $\mathbbm{1} = \pi^{-1} \int \dd^2 \alpha \dyad{\alpha}$, with $\alpha \in \mathbbm{C}$, we may express the initial state equivalently in the coherent state representation by inserting the identity for the fundamental mode only,
\begin{equation}\label{Eq:SM:init:state}
	\ket{\Psi(t_0)}
		= \pi^{-1}
				\int \dd^2\alpha
					 \mel{\alpha}{\hat{S}_L(r)}{0_L} \ket{\text{g}}\otimes \ket{\alpha}\bigotimes_{q>2} \ket{0_q}.
\end{equation}
Employing the expansion of the BSV state in the Fock basis, we obtain
\begin{equation}
	\begin{aligned}
	\mel{\alpha}{\hat{S}_L(r)}{0_L}
		&= \dfrac{1}{\sqrt{\cosh(r)}}
				\sum_{n=0}^\infty
					\big[\!- \tanh(r)\big]^n
						\dfrac{\sqrt{(2n)!}}{2^n n!}
						\braket{\alpha}{2n}
		=  \dfrac{e^{-\abs{\alpha}^2/2}}{\sqrt{\cosh(r)}}
		\sum_{n=0}^\infty
			\dfrac{1}{n!}
				\bigg[
					- \dfrac{\alpha^{*2}\tanh(r)}{2}
				\bigg]^n
		\\&=\dfrac{1}{\sqrt{\cosh(r)}}
				\exp{-\dfrac12
						 \Big[
						 	\abs{\alpha}^2 + \alpha^{*2} \tanh(r)
						 \Big]}
			\equiv \pi c(\alpha),
	\end{aligned}
\end{equation}
which allows us to rewrite Eq.~\eqref{Eq:SM:init:state} as
\begin{equation}\label{Eq:SM:init:state:II}
	\ket{\Psi(t_0)}
		= \int \dd^2\alpha\ c(\alpha) \ket{\text{g}}\otimes \hat{D}_L(\alpha)\ket{\boldsymbol{0}},
\end{equation}
where $\hat{D}_{q,\mu}(\alpha) = \exp[\alpha \hat{a}_{q,\mu}^\dagger - \alpha \hat{a}_{q,\mu}]$ is the displacement operator acting on the optical mode $\{q,\mu\}$.~

At any later time $t \geq t_0$, the state of the joint system can be written as $\ket{\Psi(t)} = \hat{U}(t,t_0)\ket{\Psi(t_0)}$, where $\hat{U}(t,t_0)$ is the time-evolution operator, also referred to as propagator, connecting the initial and final times.~From Eq.~\eqref{Eq:SM:TDSE:II}, it follows that this operator satisfies
\begin{equation}\label{Eq:SM:Prop:TDSE}
	i \hbar \pdv{\hat{U}(t)}{t}
		= \big[
				\hat{H}_{\text{at}}
				+ \mathsf{e} \hat{\boldsymbol{r}}\cdot \hat{\boldsymbol{E}}(t)
			\big]
			\hat{U}(t).
\end{equation}
Acting on the initial state in Eq.~\eqref{Eq:SM:init:state:II}, this evolution can be equivalently expressed as
\begin{equation}
	\ket{\Psi(t)}
		= \int \dd^2\alpha\ c(\alpha) \hat{U}(t,t_0) \hat{D}_L(\alpha)
					\ket{\text{g}}\otimes \ket{\boldsymbol{0}}
		= \int \dd^2 \alpha \ c(\alpha)
			\hat{D}_L(\alpha) \hat{U}_{\alpha}(t,t_0)
					\ket{\text{g}}\otimes \ket{\boldsymbol{0}},
\end{equation}
where we have defined $\hat{U}_{\alpha}(t,t_0) \equiv \hat{D}_L^\dagger(\alpha) \hat{U}(t,t_0)\hat{D}_L(\alpha)$ as a displaced version of the propagator.~Using the identity $\hat{D}_L^\dagger(\alpha)\hat{\boldsymbol{E}}_L(t)\hat{D}_L(\alpha) = \hat{\boldsymbol{E}}_L(t) + \boldsymbol{E}_{\text{cl}}(t)$, where $\boldsymbol{E}_{\text{cl}}(t) = \mel{\alpha}{\hat{\boldsymbol{E}}_L(t)}{\alpha}$, it follows from Eq.~\eqref{Eq:SM:Prop:TDSE} that 
\begin{equation}\label{Eq:SM:Prop:TDSE:II}
	i\hbar \pdv{\hat{U}_{\alpha}(t)}{t}
		= \Big[
				\hat{H}_{\text{at}} 
				+ \mathsf{e}\hat{\boldsymbol{r}}
						\cdot 
						\big(
							\hat{\boldsymbol{E}}(t)
							+ \boldsymbol{E}_{\text{cl}}(t)
						\big)
			\Big]\hat{U}_{\alpha}(t).
\end{equation}
In the following subsections, we solve this differential equation explicitly, thereby justifying the expressions employed in the main text.~We further compare our results with those obtained in related scenarios previously discussed in the literature~\cite{rivera-dean_light-matter_2022,rivera-dean_role_2024,rivera-dean_microscopic_2025,mao_benchmarking_2025}.

\subsection{Dyson expansion of the time-evolution operator}\label{Sec:SM:Dyson}
To solve Eq.~\eqref{Eq:SM:Prop:TDSE:II}, we adopt a strategy similar to that of Refs.~\cite{rivera_dean_quantum-optical_2019,rivera-dean_microscopic_2025}, where a Dyson expansion~\cite{smirnova_anatomy_2007} of the propagator $\hat{U}_{\alpha}(t,t_0)$ is employed to solve the corresponding differential equation.~In general, given a Hamiltonian decomposed as the sum of two contributions $\hat{H}_0(t)$ and $\hat{V}(t)$, the time-evolution operator can be written in the integral Dyson form
\begin{equation}\label{Eq:SM:Dyson:integral}
	\hat{U}(t,t_0)
		= U_0(t,t_0)
			- \dfrac{i}{\hbar}
				\int^{t}_{t_0} \dd t'
					\hat{U}(t,t')
						\hat{V}(t')
							\hat{U}_0(t',t_0),
\end{equation}
where $\hat{U}_0(t,t_0)$ denotes the propagator generated by $\hat{H}_0(t)$, satisfying $i\hbar \partial\hat{U}_0(t)/\partial t = \hat{H}_0(t) \hat{U}(t,t_0)$.~Equation~\eqref{Eq:SM:Dyson:integral} provides the basis for a recursive expansion, since the propagator $\hat{U}(t,t')$ appearing inside the integral governs the full dynamics generated by $\hat{H}_0(t) + \hat{V}(t)$.~Following Refs.~\cite{rivera_dean_quantum-optical_2019,rivera-dean_microscopic_2025,lohr_above-threshold_1997}, we exploit this structure by adopting different partitions of the Hamiltonian at successive stages of the recursive procedure.

 For the first iteration, leading explicitly to Eq.~\eqref{Eq:SM:Dyson:integral}, we choose $\hat{H}_0(t) = \hat{H}_{\text{at}}$ and $\hat{V}(t) = \mathsf{e}\hat{r}\cdot \hat{\boldsymbol{E}}(t;\alpha)$ where for notational convenience, we define $\hat{\boldsymbol{E}}(t;\alpha) \equiv \hat{\boldsymbol{E}}(t) + \boldsymbol{E}_{\text{cl}}(t)$.~With this choice, Eq.~\eqref{Eq:SM:Dyson:integral} yields
\begin{equation}\label{Eq:SM:Dyson:partition:1}
	\hat{U}_{\alpha}(t,t_0)
		= \hat{U}_{\text{at}}(t,t_0)
			- \dfrac{i\mathsf{e}}{\hbar}
					\int^t_{t_0} \dd t_1
						\hat{U}_\alpha(t,t_1)
							\hat{\boldsymbol{r}}\cdot \hat{\boldsymbol{E}}(t_1;\alpha)
								\hat{U}_{\text{at}}(t_1,t_0).
\end{equation}
However, since the total Hamiltonian reads
\begin{equation}
	\hat{H}(t)
		= \dfrac{\hat{\boldsymbol{p}}^2}{2m_{\mathsf{e}}}
			+ V_{\text{at}}(\hat{\boldsymbol{r}})
			+ \mathsf{e} \hat{\boldsymbol{r}}
					\cdot
						\hat{\boldsymbol{E}}(t;\alpha),
\end{equation}
an equally valid choice consists in taking $\hat{H}_0(t) = \hat{\boldsymbol{p}}^2/(2m_{\mathsf{e}}) + \mathsf{e}\hat{\boldsymbol{r}}\cdot \hat{\boldsymbol{E}}(t;\alpha) \equiv \hat{H}_L(t)$ and $\hat{V}(t) = \hat{V}_{\text{at}}$, which allows the propagator to be expressed as
\begin{equation}\label{Eq:SM:Dyson:partition:2}
	\hat{U}_\alpha(t,t_0)
		= \hat{U}_L(t,t_0)
			- \dfrac{i}{\hbar}
					\int^t_{t_0}
						\dd t_1 \hat{U}_{\alpha}(t,t_1)
							\hat{V}_{\text{at}}
								\hat{U}_L(t_1,t_0).
\end{equation}
For the second iteration, we substitute Eq.~\eqref{Eq:SM:Dyson:partition:2} into \eqref{Eq:SM:Dyson:partition:1}, leading to the combined expression
\begin{equation}
	\begin{aligned}
	\hat{U}_{\alpha}(t,t_0)
		&= \hat{U}_{\text{at}}(t,t_0)
				- \dfrac{i\mathsf{e}}{\hbar}
					\int^t_{t_0}
						\dd t_1
							\hat{U}_{L}(t,t_1)
								\hat{\boldsymbol{r}}\cdot \hat{\boldsymbol{E}}(t_1;\alpha)
									\hat{U}_{\text{at}}(t_1,t_0)
		\\& \quad 
				- \dfrac{\mathsf{e}}{\hbar^2}
					\int^{t}_{t_0} \dd t_1
						\int^{t}_{t_1} \dd t_2
							\hat{U}_{\alpha}(t,t_2)
								\hat{V}_{\text{at}}
									\hat{U}_L(t_2,t_1)
										\hat{\boldsymbol{r}}\cdot\hat{\boldsymbol{E}}(t_1;\alpha)
											\hat{U}_{\text{at}}(t_1,t_0).
	\end{aligned}
\end{equation}

In this expression, the first term describes unperturbed propagation governed solely by the atomic Hamiltonian. The second term accounts for ionization at time $t_1$, followed by Coulomb-free propagation in the continuum driven by the laser field, while the third term incorporates rescattering processes~\cite{lewenstein_rings_1995,lohr_above-threshold_1997}.~When applied to the initial state, this yields
\begin{equation}
	\begin{aligned}
	\ket{\Psi(t)}
		&= e^{iI_p(t-t_0)}
			\int \dd^2\alpha\ c(\alpha)
				 \ket{\text{g}}\otimes \hat{D}_L(\alpha)\ket{\boldsymbol{0}}
			\\&\quad 
			- \dfrac{i\mathsf{e}}{\hbar}
				\int \dd^2 \alpha
					\int^{t}_{t_0}
						\dd t_1 \ c(\alpha)\hat{D}_L(\alpha)
							 \hat{U}_{L}(t,t_1)
							 	\hat{\boldsymbol{r}}\cdot \hat{\boldsymbol{E}}(t_1;\alpha)
							 		e^{iI_p(t_1-t_0)/\hbar}\ket{\text{g}}\otimes \ket{\alpha}
			\\&\quad 
			- \dfrac{\mathsf{e}}{\hbar^2}
				\int \dd \alpha \int^t_{t_0} \dd t_1 \int^t_{t_1} \dd t_2
				\ c(\alpha)\hat{D}_L(\alpha)
					\hat{U}_{\alpha}(t,t_2)
						\hat{V}_{\text{at}}
							\hat{U}_L(t_2,t_1)
								\hat{\boldsymbol{r}}
									\cdot \hat{\boldsymbol{E}}(t_1;\alpha)
									e^{iI_p(t_1-t_0)/\hbar}
									\ket{\text{g}}\otimes \ket{\boldsymbol{0}},
	\end{aligned}
\end{equation}
where $I_p$ denotes the atomic ionization potential.~Projecting onto a final photoelectron momentum $\ket{\boldsymbol{p}_f}$, we can approximately write
\begin{equation}
	\ket{\Phi(\boldsymbol{p}_f,t)}
		\equiv \braket{\boldsymbol{p}_f}{\Psi(t)}
		\simeq
			-\dfrac{i\mathsf{e}}{\hbar}
				\int \dd^2 \alpha
					\int^{t}_{t_0}\dd t_1\
							c(\alpha)
							\hat{D}_L(\alpha)
							\bra{\boldsymbol{p}_f}
								\hat{U}_L(t,t_1)
								\hat{\boldsymbol{r}}\cdot \hat{\boldsymbol{E}}(t_1;\alpha)
								e^{iI_p(t_1-t_0)/\hbar}\ket{\text{g}}\otimes \ket{\alpha},
\end{equation}
where we have assumed that, for the considered values of $\boldsymbol{p}_f$, rescattering contributions are negligible compared to those obtained through direct ATI processes.~Introducing the identity in the electronic momentum representation, $\mathbbm{1} = \int \dd \boldsymbol{p}_0 \dyad{\boldsymbol{p}_0}$, to the right of $\hat{U}_L(t,t_1)$, we obtain 
\begin{equation}\label{Eq:SM:ATI:Direct}
	\ket{\Phi(\boldsymbol{p}_f,t)}
		= - \dfrac{i\mathsf{e}}{\hbar}
				\int \dd^2 \alpha
					\int^{t}_{t_0}\dd t_1
						\int \dd \boldsymbol{p}_0\
						c(\alpha)
						\hat{D}_L(\alpha)
							\mel{\boldsymbol{p}_f}{\hat{U}_L(t,t_1)}{\boldsymbol{p}_0}
							\!
							\mel{\boldsymbol{p}_0}{\hat{\boldsymbol{r}}}{\text{g}}
								\cdot \hat{\boldsymbol{E}}(t_1;\alpha)
								e^{iI_p(t_1-t_0)/\hbar}
									 \ket{\alpha}.
\end{equation}
In what follows, we focus on describing the structure of the propagator matrix element $	\mel{\boldsymbol{p}_f}{\hat{U}_L(t,t_1)}{\boldsymbol{p}_0}$.

\subsection{Path-integral derivation of the Coulomb-free propagator}\label{Sec:SM:Path:Int}
Unlike in classical analysis, it is important to emphasize that the matrix element $\mel{\boldsymbol{p}_f}{\hat{U}_L(t,t_1)}{\boldsymbol{p}_0}$ remains a quantum optical operator acting on the Hilbert space of the radiation modes, with only the electronic degrees of freedom projected onto momentum eigenstates.~To evaluate this quantity, we make use of Feynman path-integral formalisms~\cite{kleinert_path_2009}, a common practice in semiclassical strong-field approaches~\cite{milosevic_phase_2013,lai_influence_2015,maxwell_strong-field_2019,faria_it_2020}.~We note, however, that this formalism has also been recently formulated within fully quantum optical treatments~\cite{mao_benchmarking_2025}.

Path-integral formalisms rely on the fact that any unitary time-propagator operator can be expressed as a product of infinitesimal time slices,
\begin{equation}\label{Eq:SM:timeslices}
	\hat{U}_L(t,t_0)
		= \prod_{i=1}^{N+1} \hat{U}_L(t_i,t_{i-1}),
\end{equation}
where $\hat{U}(t_i,t_{i-1}) = \text{exp}[- i\epsilon \hat{H}_L(t_i)/\hbar]$, and $\epsilon$ denotes an infinitesimal time step.~For each of these single-time-step propagators, we apply the Zassenhaus formula~\cite{magnus_exponential_1954}
\begin{equation}\label{Eq:SM:Zassenhaus}
	e^{\delta(\hat{X} + \hat{Y})}
		= e^{\delta\hat{X}}e^{\delta\hat{Y}}
			e^{-\delta^2[\hat{X},\hat{Y}]/2}
			e^{-\delta^3(2[\hat{Y},[\hat{X},\hat{Y}]] + [\hat{X},[\hat{X},\hat{Y}]])/6}
			\cdots,
\end{equation}
where the ellipsis denotes exponentials involving higher-order nested commutators.~Identifying in our case $\delta = -i \epsilon/\hbar$, $\hat{X} = \hat{\boldsymbol{p}}^2/(2m_{\mathsf{e}})$ and $\hat{Y} = \mathsf{e}\hat{\boldsymbol{r}}\cdot \hat{\boldsymbol{E}}(t;\alpha)$, we obtain
\begin{align}
	&[\hat{X},\hat{Y}]
		= \dfrac{\mathsf{e}}{2m_{\mathsf{e}}} [\hat{\boldsymbol{p}}^2,\hat{\boldsymbol{r}}]\cdot \hat{\boldsymbol{E}}(t;\alpha)
		= -\dfrac{i\hbar\mathsf{e}}{m_{\mathsf{e}}} \hat{\boldsymbol{p}}\cdot \hat{\boldsymbol{E}}(t;\alpha)
	\\&\big[\hat{X},[\hat{X},\hat{Y}]\big]
		= -\dfrac{i\hbar\mathsf{e}}{2m_{\mathsf{e}}^2} [\hat{\boldsymbol{p}}^2, \hat{\boldsymbol{p}}\cdot \hat{\boldsymbol{E}}(t;\alpha)]
		= 0
	\\&\big[\hat{Y},[\hat{X},\hat{Y}]\big]
		= -\dfrac{i\hbar\mathsf{e}^2 }{m_{\mathsf{e}}}[\hat{\boldsymbol{r}}\cdot \hat{\boldsymbol{E}}(t;\alpha)
						,\hat{\boldsymbol{p}}\cdot \hat{\boldsymbol{E}}(t;\alpha)]
		= \dfrac{\hbar^2\mathsf{e}^2 }{m_{\mathsf{e}}}  \hat{\boldsymbol{E}}^2(t;\alpha),
\end{align}
with all higher-order nested commutators vanishing identically due to the above commutation relations. Substituting these results into Eq.~\eqref{Eq:SM:Zassenhaus}, the single time-step propagator can be written as
\begin{equation}\label{Eq:SM:explicit:timeslice}
	\hat{U}(t_i,t_{i-1})
		= \exp[-\dfrac{i\epsilon}{2m_{\mathsf{e}}\hbar}\hat{\boldsymbol{p}}^2]
			\exp[-\dfrac{i\epsilon\mathsf{e}}{\hbar}\hat{\boldsymbol{r}}\cdot\hat{\boldsymbol{E}}(t_i;\alpha)]
			\exp[-\dfrac{i\epsilon^2\mathsf{e}}{m_{\mathsf{e}}\hbar} \hat{\boldsymbol{p}}\cdot \hat{\boldsymbol{E}}(t_i;\alpha)]
			\exp[-i \dfrac{\epsilon^3\mathsf{e}^2}{3m_{\mathsf{e}}\hbar} 
					\hat{\boldsymbol{E}}^2(t_i;\alpha)].
\end{equation}

To proceed further, between each time-slice of Eq.~\eqref{Eq:SM:timeslices} we insert the identity in the momentum representation, yielding
\begin{equation}
	\bra{\boldsymbol{p}_f}\hat{U}_L(t,t_0)\ket{\boldsymbol{p}_0}
		= \bigg[ 
				\prod_{i=1}^N
					\int \dd \boldsymbol{p}_i
			\bigg]
			\prod_{i=1}^{N+1}			
				\bra{\boldsymbol{p}_i}
					\hat{U}_L(t_i,t_{i-1})
				\ket{\boldsymbol{p}_{i-1}},
\end{equation}
where we identify $\boldsymbol{p}_{f} \equiv \boldsymbol{p_{N}}$ and $t = t_{N+1}$.~This expression can be further expanded by introducing the identity in the position representation between the first pair of exponentials in Eq.~\eqref{Eq:SM:explicit:timeslice}, that is,
\begin{equation}
	\begin{aligned}
	\bra{\boldsymbol{p}_f}\hat{U}_L(t,t_0)\ket{\boldsymbol{p}_0}
		&= 
		\bigg[ 
			\prod_{i=1}^N
				\int \dd \boldsymbol{p}_i
		\bigg]
		\bigg[ 
			\prod_{i=1}^{N+1}
				\int \dd \boldsymbol{r}_i
		\bigg]
			\prod_{i=1}^{N+1}
				\bigg\{
					\bra{\boldsymbol{p}_i}\!
						\exp[-\dfrac{i\epsilon}{2m_{\mathsf{e}}\hbar}\hat{\boldsymbol{p}}^2]
						\exp[-\dfrac{i\epsilon\mathsf{e}}{\hbar}\hat{\boldsymbol{r}}\cdot\hat{\boldsymbol{E}}(t_i;\alpha)]\!
					\ket{\boldsymbol{r}_i}
						\\&\hspace{5cm}\times
					\bra{\boldsymbol{r}_i}\!
						\exp[-\dfrac{i\epsilon^2\mathsf{e}}{m_{\mathsf{e}}\hbar} \hat{\boldsymbol{p}}\cdot \hat{\boldsymbol{E}}(t_i;\alpha)]
						\exp[-i \dfrac{\epsilon^3\mathsf{e}^2}{3m_{\mathsf{e}}\hbar} 
					\hat{\boldsymbol{E}}^2(t_i;\alpha)]\!\!
					\ket{\boldsymbol{p}_{i-1}}
				\bigg\}.
	\end{aligned}
\end{equation}
Using $\braket{\boldsymbol{p}_i}{\boldsymbol{r}_i} = (2\pi\hbar)^{-3/2} \text{exp}[-i\boldsymbol{p}_i\cdot \boldsymbol{r}_i/\hbar]$, this expression can be written as the following quantum optical operator
\begin{align}
	\bra{\boldsymbol{p}_f}\hat{U}_L(t,t_0)\ket{\boldsymbol{p}_0}
		&= 
			\bigg[ 
				\prod_{i=1}^N
				\int \dd \boldsymbol{p}_i
			\bigg]
			\bigg[ 
				\prod_{i=1}^{N+1}
				\int \dfrac{\dd \boldsymbol{r}_i}{{(2\pi\hbar)^3}}
			\bigg]
			\nonumber
			\\&\quad \times
			\prod_{i=1}^{N+1}
			\bigg\{
			\exp[-\dfrac{i\epsilon}{2m_{\mathsf{e}}\hbar}\boldsymbol{p}_i^2
			-\dfrac{i\epsilon}{\hbar}\mathsf{e}\boldsymbol{r}_i\cdot\hat{\boldsymbol{E}}(t_i;\alpha)
			-\dfrac{i\epsilon^2}{m_{\mathsf{e}}\hbar} \mathsf{e} \boldsymbol{p}_{i-1}\cdot \hat{\boldsymbol{E}}(t_i;\alpha)
			-\dfrac{i}{\hbar}\boldsymbol{r}_i\cdot(\boldsymbol{p}_i - \boldsymbol{p}_{i-1})]
			\\&\hspace{1.7cm}\times
			\exp[-i \dfrac{\epsilon^3}{3m_{\mathsf{e}}\hbar} 
			\mathsf{e}^2\hat{\boldsymbol{E}}^2(t_i;\alpha)]
	\bigg\}.\nonumber
		\\&=	
		\bigg[ 
			\prod_{i=1}^N
				\int \dd \boldsymbol{p}_i
		\bigg]
		\bigg[ 
			\prod_{i=1}^{N+1}
				\int \dfrac{\dd \boldsymbol{r}_i}{{(2\pi\hbar)^3}}
		\bigg]
			\nonumber
		\\&\quad \times
		\prod_{i=1}^{N+1}
		\bigg\{
			\exp[-\dfrac{i\epsilon}{2m_{\mathsf{e}}\hbar}\boldsymbol{p}_i^2
			-\dfrac{i\epsilon}{\hbar}\mathsf{e}\boldsymbol{r}_i\cdot\boldsymbol{E}_{\text{cl}}(t_i;\alpha)
			-\dfrac{i\epsilon^2}{m_{\mathsf{e}}\hbar} \mathsf{e} \boldsymbol{p}_{i-1}\cdot \boldsymbol{E}_{\text{cl}}(t_i;\alpha)
			-\dfrac{i}{\hbar}\boldsymbol{r}_i\cdot(\boldsymbol{p}_i - \boldsymbol{p}_{i-1})]
			\\&\hspace{1.7cm}\times
			\hat{\boldsymbol{D}}\big(\boldsymbol{f}(t_i;\alpha)\big)
			\exp[-i \dfrac{\epsilon^3}{3m_{\mathsf{e}}\hbar} 
			\mathsf{e}^2\hat{\boldsymbol{E}}^2(t_i;\alpha)]
		\bigg\}\nonumber,
\end{align}
where we have explicitly separated between the classical and quantum optical components of the electric field operator. Furthermore, we have collected all linear terms in the field operators, yielding a multimode displacement operator $\hat{\boldsymbol{D}}(\boldsymbol{\alpha}) \equiv \bigotimes_{q=1,\mu} \hat{D}_{q,\mu}(\alpha_{q,\mu})$, with
\begin{equation}
	f_{q,\mu}(t_i;\alpha)
		= g(\omega_q)
			\boldsymbol{\epsilon}_\mu
			\cdot
			\bigg[
				\dfrac{\epsilon}{\hbar} \boldsymbol{r}_i 
				+ \dfrac{\epsilon^2}{m_{\mathsf{e}}\hbar}\boldsymbol{p}_{i-1}
			\bigg] e^{i\omega_q t}.
\end{equation}

From this expression it becomes clear that, during each propagation step, the electron acquires a phase due to its motion and, as consequence of the light-matter interaction, the electromagnetic field modes are in principle displaced and squeezed.~The multimode squeezing contribution originates from the quadratic field term and typically becomes relevant in the moderate-depletion regime, where the applied laser field significantly depletes the atomic ground state~\cite{stammer_squeezing_2023}.~In the following, we therefore neglect this contribution and approximate
\begin{equation}
	\begin{aligned}
	\bra{\boldsymbol{p}_f}\hat{U}_L(t,t_0)\ket{\boldsymbol{p}_0}
		&\approx
			\bigg[ 
				\prod_{i=1}^N
					\int \dd \boldsymbol{p}_i
			\bigg]
			\bigg[ 
				\prod_{i=1}^{N+1}
				\int \dfrac{\dd \boldsymbol{r}_i}{(2\pi\hbar)^3}
			\bigg]
			\\&\quad \times
			\prod_{i=1}^{N+1}\!
			\bigg\{\!
				\exp[-\dfrac{i\epsilon}{2m_{\mathsf{e}}\hbar}\boldsymbol{p}_i^2
				-\dfrac{i\epsilon}{\hbar}\mathsf{e}\boldsymbol{r}_i\cdot\boldsymbol{E}_{\text{cl}}(t_i;\alpha)
				-\dfrac{i\epsilon^2}{m_{\mathsf{e}}\hbar} \mathsf{e} \boldsymbol{p}_{i-1}\cdot \boldsymbol{E}_{\text{cl}}(t_i;\alpha)
				-i\boldsymbol{r}_i\cdot(\boldsymbol{p}_i - \boldsymbol{p}_{i-1})]
				\hat{\boldsymbol{D}}\big(\boldsymbol{f}(t_i;\alpha)\big)
			\!\bigg\}.
	\end{aligned}
\end{equation}
Thus, when inserting this expression inside Eq.~\eqref{Eq:SM:ATI:Direct}, we get
\begin{equation}
	\begin{aligned}
	\ket{\Phi(\boldsymbol{p}_f,t)}
		\approx -
		\dfrac{i\mathsf{e}}{\hbar}
			\int \dd^2\alpha\ 
				c(\alpha) \hat{D}_L(\alpha)
				\int^t_{t_0} \dd t'
				&	\int \dd \boldsymbol{p}_0
					\bigg[ 
						\prod_{i=1}^N
							\int \dd \boldsymbol{p}_i
					\bigg]
					\bigg[ 
						\prod_{i=1}^{N+1}
							\int \dfrac{\dd \boldsymbol{r}_i}{(2\pi\hbar)^3}
					\bigg]
						\\&\quad \times
							e^{i \mathcal{A}_N + iI_p(t'-t_0)}
							\bigg[
								\prod_{i=1}^{N+1}
								\hat{\boldsymbol{D}}\big(\boldsymbol{f}(t_i;\alpha)\big)
							\bigg]
								\mel{\boldsymbol{p}_0}{\hat{\boldsymbol{r}}}{\text{g}}
									\cdot \hat{\boldsymbol{E}}(t';\alpha)
								\ket{\boldsymbol{0}},
	\end{aligned}
\end{equation}
where we have defined
\begin{equation}
	\mathcal{A}_N
		\equiv - \dfrac{\epsilon}{\hbar} \sum_{i=1}^{N+1}
				\bigg[
					\dfrac{\boldsymbol{p}_i^2}{2m_{\mathsf{e}}}
					+ \mathsf{e} \boldsymbol{r}_i\cdot \boldsymbol{E}_{\text{cl}}(t_i,\alpha)
					+ \epsilon \dfrac{\mathsf{e}}{m_{\mathsf{e}}} \boldsymbol{p}_{i-1}\cdot \boldsymbol{E}_{\text{cl}}(t_i;\alpha)
					+ \boldsymbol{r}_i\cdot
						(
						\boldsymbol{p}_i-\boldsymbol{p}_{i-1}
						)
				\bigg].
\end{equation}

Before proceeding further, it turns convenient to recall the following property of the displacement operator
\begin{equation}
	\hat{D}(\alpha_2)\hat{D}(\alpha_1)
		= \exp[\dfrac12
					\big(
						\alpha_2\alpha_1^*
						- \alpha_2^*\alpha_1
					\big)]
			\hat{D}(\alpha_2+\alpha_1),
\end{equation}
which allows us to express
\begin{equation}
	\prod_{i=1}^{N+1}
		\hat{D}_{q,\mu}
			\big(
				f_{q,\mu}(t_i;\alpha)
			\big)
	= \exp[
			\dfrac12
				\bigg(
					\sum_{i=1}^N\sum_{j=1}^N
						f_{q,\mu}(t_{j+1};\alpha)f_{q,\mu}^*(t_i;\alpha)- \text{c.c.}
				\bigg)
			]
		\hat{D}_{q,\mu}
			\bigg(
				\sum_{i=1}^{N+1}f_{q,\mu}(t_i;\alpha)
			\bigg).
\end{equation}
Then, when taking the continuous limit $\epsilon \to 0$, the resulting quantum optical state reads
\begin{equation}
	\begin{aligned}
	\ket{\Phi(\boldsymbol{p}_f,t)}
		= -\dfrac{i\mathsf{e}}{\hbar}
			\int \dd^2 \alpha
					\ c(\alpha)
						\hat{D}_L(\alpha)
						\int^t_{t_0} \dd t'
						&\int \dd \boldsymbol{p}_0
						\int^{\boldsymbol{p}_f}_{\boldsymbol{p}_0} \mathcal{D}' \boldsymbol{p}
						\int \dfrac{\mathcal{D}\boldsymbol{r}}{(2\pi\hbar)^3}
							\exp[i \mathcal{A}(\boldsymbol{p}.\boldsymbol{r})
									+I_p(t'-t_0)/\hbar
									+ i\varphi(t')]
						\\&\hspace{1cm}\times
							\hat{\boldsymbol{D}}\big(\boldsymbol{F}(t';\alpha)\big)
							\mel{\boldsymbol{p}_0}{\hat{\boldsymbol{r}}}{\text{g}}
							\cdot \hat{\boldsymbol{E}}(t';\alpha)
							\ket{\boldsymbol{0}},
	\end{aligned}
\end{equation}
where we have defined
\begin{equation}
	\begin{aligned}
		& \mathcal{A}(\boldsymbol{r},\boldsymbol{p})
			= - \dfrac{1}{\hbar}
				\int_{t'}^t \dd \tau
					\Big[
						\dfrac{\boldsymbol{p}(\tau)^2}{2m_{\mathsf{e}}}
						+ \mathsf{e}\boldsymbol{r}(\tau)\cdot \boldsymbol{E}_{\text{cl}}(\tau;\alpha)
						+ \boldsymbol{r}(\tau)\cdot \dot{\boldsymbol{p}}(\tau)
					\Big],
		\\&
		F_{q,\mu}(t';\alpha)
			= g(\omega_q)
				\boldsymbol{\epsilon}_\mu
					\cdot
					\bigg[
						\int^t_{t'}\dd \tau\ 
							\boldsymbol{r}(\tau)e^{i\omega_q \tau}
					\bigg],
		\\&
		\varphi(t')
			= \sum_{q,\mu}
				2g(\omega_q)^2
					\int_{t'}^t\dd \tau_1
						\int^t_{\tau_1} \dd \tau_2
							\big(
								\boldsymbol{r}(\tau_2)
									\cdot \boldsymbol{\epsilon}_\mu
							\big)
							\big(
								\boldsymbol{r}(\tau_1)
									\cdot \boldsymbol{\epsilon}_\mu
							\big)
							\sin(\omega_q(\tau_2-\tau_1)).
	\end{aligned}
\end{equation}

It is worth noting that the obtained results bears similarities with those recently reported in the literature using Feynmann path-integral formalisms as well~\cite{mao_benchmarking_2025}. To the best of our knowledge and understanding, the main difference lies in the fact that, in our case, we keep the quantum optical degrees of freedom in fully operational form, whereas in Ref.~\cite{mao_benchmarking_2025} the authors introduce identity operators in the coherent state basis, in a manner analogous to how we have introduced here position- and momentum-resolved identity operators for the electronic degrees of freedom.

When restricting our analysis to the case of coherent state driving fields and set all quantum optical terms to zero, we find that the obtained expressions reduce to those commonly encountered in semiclassical analyses within the Feynmann-path integral formalism~\cite{milosevic_phase_2013,lai_influence_2015,maxwell_strong-field_2019,faria_it_2020}.~The main difference from these approaches is that, on top of the semiclassical phase acquired by the electron, it induces a displacement on the optical degrees of freedom.~In the following, we take into account that, within this coherent state expansion picture, the magnitude of this displacement for typical strong-field parameters ($I= 10^{14}$ W/cm$^2$ and $\lambda = 800$ nm) is very small, namely $\lvert F(t';\alpha)\rvert \propto 10^{-4}$~\cite{rivera-dean_light-matter_2022}.~We therefore approximate the resulting quantum optical state by approximating this displacement by the identity operator.~Under this approximation, we obtain
\begin{equation}
	\begin{aligned}
		\ket{\Phi(\boldsymbol{p}_f,t)}
		\approx -\dfrac{i\mathsf{e}}{\hbar}
			\int \dd^2 \alpha
				\ c(\alpha)
					\hat{D}_L(\alpha)
					\int^t_{t_0} \dd t'
					&\int \dd \boldsymbol{p}_0
						\int^{\boldsymbol{p}_f}_{\boldsymbol{p}_0} \mathcal{D}' \boldsymbol{p}
						\int \dfrac{\mathcal{D}\boldsymbol{r}}{(2\pi\hbar)^3}
		\exp[i \mathcal{A}(\boldsymbol{p}.\boldsymbol{r})
		+I_p(t'-t_0)/\hbar]
			\mel{\boldsymbol{p}_0}{\hat{\boldsymbol{r}}}{\text{g}}
				\cdot \hat{\boldsymbol{E}}(t';\alpha)
					\ket{\boldsymbol{0}}.
	\end{aligned}
\end{equation}

Recently, it has been shown that driving ATI with bright squeezed light can magnify the laser-matter interaction, yielding non-Gaussian states of light upon post-selection on the final electron momentum~\cite{rivera-dean_microscopic_2025}.~While the underlying mechanism explored in that work is analogous to the one considered here, the theoretical framework differs substantially. In particular, Ref.~\cite{rivera-dean_microscopic_2025} considered displaced squeezed states and introduced a transformation to a displaced and squeezed frame of reference, whereby the field quadratures are rescaled by factor $e^{\pm r}$, allowing the initial squeezed state to be mapped onto an unsqueezed one.~In contrast, here we work entirely within the unsqueezed frame of reference and instead employ a decomposition of the initial states in terms of coherent states.~This approach is more convenient in the present context, as our driving field is a bright squeezed vacuum.~In such scenario, the approach considered in Ref.~\cite{rivera-dean_microscopic_2025} to introduce strong-field approximations is not applicable here, as they relied on the presence of a large coherent displacement capable of inducing strong-field phenomena even in the absence of squeezing.

Within our framework, the strong-field approximation corresponds to the regime in which the coherent field contribution $\boldsymbol{E}_{\text{cl}}(t;\alpha) $ dominates over the contribution arising from vacuum fluctuations, whose typical amplitude satisfies $\lVert \hat{\boldsymbol{E}}(t') \rVert \sim 10^{-8}$. While this condition is not satisfied for all values of $\alpha$---in particular $\alpha = 0$ where $\boldsymbol{E}_{\text{cl}}(t;\alpha) =0$---the probability amplitude $c(\alpha)$ is broad due to the bright squeezed nature of the field.~As a consequence, the tails of the distribution contribute more strongly to strong-field ionization than its center, and therefore dominate the strong-field response.~Under these conditions, the SFA-version of the state becomes
\begin{equation}\label{Eq:SM:SFA:QO:state}
	\begin{aligned}
		\ket{\Phi(\boldsymbol{p}_f,t)}
			\approx -\dfrac{i\mathsf{e}}{\hbar}
				\int \dd^2 \alpha
					\ c(\alpha)
						\int^t_{t_0} \dd t'
						&\int \dd \boldsymbol{p}_0
							\int^{\boldsymbol{p}_f}_{\boldsymbol{p}_0} \mathcal{D}' \boldsymbol{p}
							\int \dfrac{\mathcal{D}\boldsymbol{r}}{(2\pi\hbar)^3}
								\exp[i \mathcal{A}(\boldsymbol{p},\boldsymbol{r})
										+I_p(t'-t_0)/\hbar]
							\mel{\boldsymbol{p}_0}{\hat{\boldsymbol{r}}}{\text{g}}
								\cdot \boldsymbol{E}_{\text{cl}}(t';\alpha)
							\ket{\alpha},
	\end{aligned}
\end{equation}
where, in what follows, we omit the contribution of all harmonic-field modes, as they remain in the vacuum state throughout the interaction.

As written in Eq.~\eqref{Eq:SM:SFA:QO:state}, the path-integrals with respect to the electronic position and momentum, which we collectively denote as $I_{\text{path}}$, can be evaluated analytically.~These integrals act only the exponential containing the action $e^{i\mathcal{A}(\boldsymbol{r},\boldsymbol{p})}$, and in the following we proceed to solve them explicitly. More specifically, we obtain
\begin{equation}
	\begin{aligned}
	I_{\text{path}} &= 
			\int^{\boldsymbol{p}_f}_{\boldsymbol{p}_0} 
				\mathcal{D}'\boldsymbol{p}
			\int \dfrac{\mathcal{D}\boldsymbol{r}}{(2\pi\hbar)^3}
				\exp{-\dfrac{i}{\hbar}
					\int_{t'}^t \dd \tau
						\Big[
							\dfrac{\boldsymbol{p}(\tau)^2}{2m_{\mathsf{e}}}
							+ \mathsf{e}\boldsymbol{r}(\tau)\cdot \boldsymbol{E}_{\text{cl}}(\tau;\alpha)
						+ \boldsymbol{r}(\tau)\cdot \dot{\boldsymbol{p}}(\tau)
						\Big]}
		\\&
		= \int^{\boldsymbol{p}_f}_{\boldsymbol{p}_0} 
				\mathcal{D}'\boldsymbol{p}
				\exp{-\dfrac{i}{\hbar}
					\int_{t'}^t \dd \tau
						\dfrac{\boldsymbol{p}(\tau)^2}{2m_{\mathsf{e}}}
					}
				\delta
					\Big(
						\dot{\boldsymbol{p}}(\tau) + \mathsf{e}\boldsymbol{E}_{\text{cl}}(\tau;\alpha)
					\Big).
	\end{aligned}
\end{equation}
The delta functional constrains the electronic momentum to satisfy the classical equation of motion
\begin{equation}
	\dot{\boldsymbol{p}}(\tau) 
		= -\mathsf{e}\boldsymbol{E}_{\text{cl}}(\tau;\alpha)
	\Rightarrow
		\boldsymbol{p}(\tau) - \mathsf{e}\boldsymbol{A}_{\text{cl}}(\tau;\alpha)
			= 	\boldsymbol{p}_0 - \mathsf{e}\boldsymbol{A}_{\text{cl}}(t';\alpha),
\end{equation}
which, evaluated at the final time $t$ yields $\boldsymbol{p}_0 = \boldsymbol{p}_f-\mathsf{e} \boldsymbol{A}_{\text{cl}}(t;\alpha) + \mathsf{e}\boldsymbol{A}_{\text{cl}}(t';\alpha)$.~Defining the canonical final momentum $\Tilde{\boldsymbol{p}}_f(\alpha) = \boldsymbol{p}_f-\mathsf{e}A_{\text{cl}}(t;\alpha)$, we can express the quantum state in Eq.~\eqref{Eq:SM:SFA:QO:state} after performing the path integrals as
\begin{equation}\label{Eq:SM:final:state}
	\begin{aligned}
	\ket{\Phi(\boldsymbol{p}_f,t)}
		&= -\dfrac{i\mathsf{e}}{\hbar}
			\int \dd^2 \alpha \ c(\alpha)
				\int^{t}_{t_0} \dd t'
					e^{-\tfrac{i}{2m_{\mathsf{e}}\hbar}
						\int^t_{t'}\dd \tau
							[
								\Tilde{\boldsymbol{p}}_f(\alpha) + \mathsf{e}\boldsymbol{A}_{\text{cl}}(\tau;\alpha)
							]^2 + i \frac{I_p}{\hbar}(t'-t_0)}
					\langle \Tilde{\boldsymbol{p}}_f(\alpha)+\mathsf{e}\boldsymbol{A}_{\text{cl}}(t';\alpha)\rvert
					\hat{\boldsymbol{r}}\lvert \text{g}\rangle\cdot \boldsymbol{E}_{\text{cl}}(t';\alpha)\ket{\alpha}
		\\&
		\equiv
			-\dfrac{i\mathsf{e}}{\hbar}
				\int \dd^2 \alpha \ c(\alpha)
					\int^{t}_{t_0} \dd t'
						b(\Tilde{\boldsymbol{p}}_f,t';\alpha) \ket{\alpha}.
	\end{aligned}
\end{equation}
For convenience, we express all quantities in terms of the canonical momentum $\Tilde{\boldsymbol{p}}_f(\alpha)$.~This is the state we explicitly evaluate in our analysis.~It is important to highlight that, here, the canonical momentum explicitly depends on $\alpha$ and, unlike the classical scenario, ceases to be a constant of the motion.~It remains, however, \emph{locally} constant within each realization $\alpha$ of the driving field in the coherent state superposition, while not \emph{globally} constant between different realizations.~When considering pulsed fields, however, $\boldsymbol{A}(t\to\infty;\alpha) = 0$ such that kinetic and canonical momentum asymptotically coincide, therefore making $\Tilde{\boldsymbol{p}}_f(\alpha) = \boldsymbol{p}_f$.

\subsection{Relation to Husimi-averaged methods}\label{Sec:SM:smcl:SPA}
Based on a generalized positive-$P$ representation~\cite{drummond_generalised_1980} of the driving field, it has been shown that photoelectron observables driven by BSV, as well as other quantum light sources, can be written as~\cite{wang_high-order_2023,lyu_effect_2025,liu_atomic_2025}
\begin{equation}
	P(\boldsymbol{p}_f)
		= \int \dd^2 \alpha \ Q_0(\alpha) P(\boldsymbol{p}_f;\alpha),
\end{equation}
where $P(\boldsymbol{p}_f;\alpha) = |\!\int \dd t' b(\tilde{p}_f,t';\alpha)|^2$, and $Q_0(\alpha) = \pi^{-1}\abs{\braket{\alpha}{\Phi(t_0)}}^2$ is the Husimi function of the initial field state.

We now show that this expression can be recovered directly from Eq.~\eqref{Eq:SM:final:state}. From that equation, the photoelectron probability can be written (up to normalization) as
\begin{equation}
	P(\boldsymbol{p}_f)
		= \int \dd^2 \alpha \int \dd^2 \beta\
			 c(\alpha)c^*(\beta) \braket{\beta}{\alpha}
			\bigg[	
				\int^{t}_{t_0} \dd t'
					b(\Tilde{\boldsymbol{p}}_f,t';\alpha)
			\bigg]
			\bigg[	
				\int^{t}_{t_0} \dd t'
					b(\Tilde{\boldsymbol{p}}_f,t';\beta)
			\bigg]^*,
\end{equation}
where the magnitude of the coherent state overlap decays as $\lvert\braket{\beta}{\alpha}\rvert \propto e^{-\abs{\beta - \alpha}^2/2}$. This Gaussian suppression ensures that the double integral is effectively dominated by regions where $\beta \approx \alpha$.~For instance, already for $\abs{\beta-\alpha} = 5$, the overlap is suppressed by more than six orders of magnitude.~In the regime relevant to strong-field physics, variations of this magnitude in coherent state amplitudes correspond to field strength changes of $\varepsilon_\alpha \sim 10^{-7}$ a.u., that are several orders of magnitude smaller than those required to significantly modify the electron dynamics.~Consequently, within the regime where $\braket{\beta}{\alpha}$ is appreciable, the ionization probability amplitudes barely change, and one may approximate $b(\Tilde{\boldsymbol{p}}_f,t';\beta)\simeq b(\Tilde{\boldsymbol{p}}_f,t';\alpha)$.~Under this controlled diagonal approximation, the probability becomes
\begin{equation}\label{Eq:SM:PMD:Husimi}
	\begin{aligned}
	P(\boldsymbol{p}_f)
	&= \int \dd^2 \alpha
		\bigg|	
			\int^{t}_{t_0} \dd t'
			b(\Tilde{\boldsymbol{p}}_f,t';\alpha)
			\bigg|^2
		\int \dd^2\beta\
			\pi^{-2}\braket{\alpha}{\Phi(t_0)}\braket{\Phi(t_0)}{\beta} 
				\braket{\beta}{\alpha}
	\\&	= \int \dd^2 \alpha\
			\underbrace{\pi^{-1} \abs{\braket{\alpha}{\Phi(t_0)}}^2}_{Q_0(\alpha)}
			\bigg|	
				\int^{t}_{t_0} \dd t'
				b(\Tilde{\boldsymbol{p}}_f,t';\alpha)
			\bigg|^2,
	\end{aligned}
\end{equation}
where in transitioning from the first to the second equality we have used the resolution of the identity in the coherent state basis.~It is worth noting that an expression formally identical to Eq.~\eqref{Eq:SM:PMD:Husimi} can also be obtained if the driving field is initially prepared in the incoherent mixture $\hat{\rho}_{\text{mix}} = \int \dd \alpha \ Q(\alpha) \dyad{\alpha}$. In that case, the photoelectron spectrum would be described by the same statistical average over coherent state contributions.~However, the physical origin of this averaging is fundamentally different.~When starting from Eq.~\eqref{Eq:SM:final:state}, the effective mixture arises from the entanglement generated between the light and matter degrees of freedom during the interaction, and the loss of information induced by tracing out one of the subsystems.~By contrast, for the state $\hat{\rho}_{\text{mix}}$ the averaging simply reflects the intrinsic mixedness of the initial driving field.

As a consequence, the two scenarios can be distinguished by observables or operations that are sensitive to quantum coherences in the optical subsystem. In particular, after considering the heralding protocol used in this work, phase-space observables such as the Wigner function retain interference features when the state originates from Eq.~\eqref{Eq:SM:final:state}, giving rise to the negative regions reported in the main text.~In contrast, the absence of coherences in $\hat{\rho}_{\text{mix}}$ prevents the appearance of such non-classical features.

\subsection{Semiclassical saddle-point analysis}\label{Sec:App:Semiclassical:SPA}
There are several ways in which one can evaluate the probability amplitudes weighting each of the coherent states in Eq.~\eqref{Eq:SM:final:state}.~Perhaps the more straightforward approach would involve a direct time integration evaluated at each value of $\alpha$.~Here, instead, we employ the standard semiclassical saddle-point approximation and express the time-integral as a sum over stationary points of the action
\begin{equation}\label{Eq:SM:action}
	S(\tilde{\boldsymbol{p}}_f,t,t';\alpha)
		= -I_p (t'-t_0)
			+ \dfrac{1}{2m_{\mathsf{e}}}
				\int^{t}_{t'} \dd \tau
					\big[ 
						\Tilde{\boldsymbol{p}}_f
						+ \mathsf{e} \boldsymbol{A}_{\text{cl}}(\tau;\alpha)
					\big]^2,
\end{equation}
that yields
\begin{equation}\label{Eq:SM:final:state:SPA}
	\ket{\Phi(\boldsymbol{p}_f,t)}
		\approx - \dfrac{i\mathsf{e}}{\hbar}
			\int \dd^2 \alpha
				\Bigg[
					\sum_{t_{\text{ion}}}
						\sqrt{\dfrac{2\pi i}{\partial^2_{t'}S(\Tilde{\boldsymbol{p}}_f,t,t';\alpha)\vert_{t_{\text{ion}}}}}
							c(\alpha) b(\boldsymbol{p}_f,t_{\text{ion}};\alpha)
								\ket{\alpha}
				\Bigg]
		\equiv
			- \dfrac{i\mathsf{e}}{\hbar}
				\int \dd^2 \alpha \	\Tilde{c}(\alpha)\ket{\alpha}.
\end{equation}
From Eq.~\eqref{Eq:SM:action}, the saddle-point equation reads
\begin{equation}
	\dfrac{1}{2m_{\mathsf{e}}}
		\bigg[
			\Tilde{p}_{f,\parallel}
			+ \mathsf{e}\dfrac{\varepsilon_\alpha}{\omega_L}\cos(\omega_L t_{\text{ion}}-\theta)
		\bigg]^2
	+ \dfrac{\Tilde{p}_{f,\perp}^2}{2m_{\mathsf{e}}}
	+ I_p = 0,
\end{equation}
where we have explicitly expanded the vector potential, and introduced the field strength $\varepsilon_\alpha = 2 g(\omega_L) \abs{\alpha}$, such that $\alpha = \abs{\alpha}e^{i\theta}$. Using atomic units ($\hbar = \abs{\mathsf{e}}= m_{\mathsf{e}}=1$), it follows that the saddle-point admits the solutions
\begin{equation}
	t_{\text{ion},\pm}
		= \dfrac{2\pi k}{\omega}
			+ \dfrac{\theta}{\omega}
			\pm \dfrac{1}{\omega}
				\arccos
					\bigg[
						\dfrac{-\Tilde{p}_{f,\parallel}\mp i \sqrt{\Tilde{p}^2_{f,\perp}+2I_p}}{2\sqrt{U_p}}
					\bigg],
\end{equation}
where $U_p = \varepsilon_{\alpha}^2/(4\omega_L^2)$ the ponderomotive energy and $k\in \mathbbm{Z}$.

\begin{figure}[ht]
	\centering
	\includegraphics[width=0.78\textwidth]{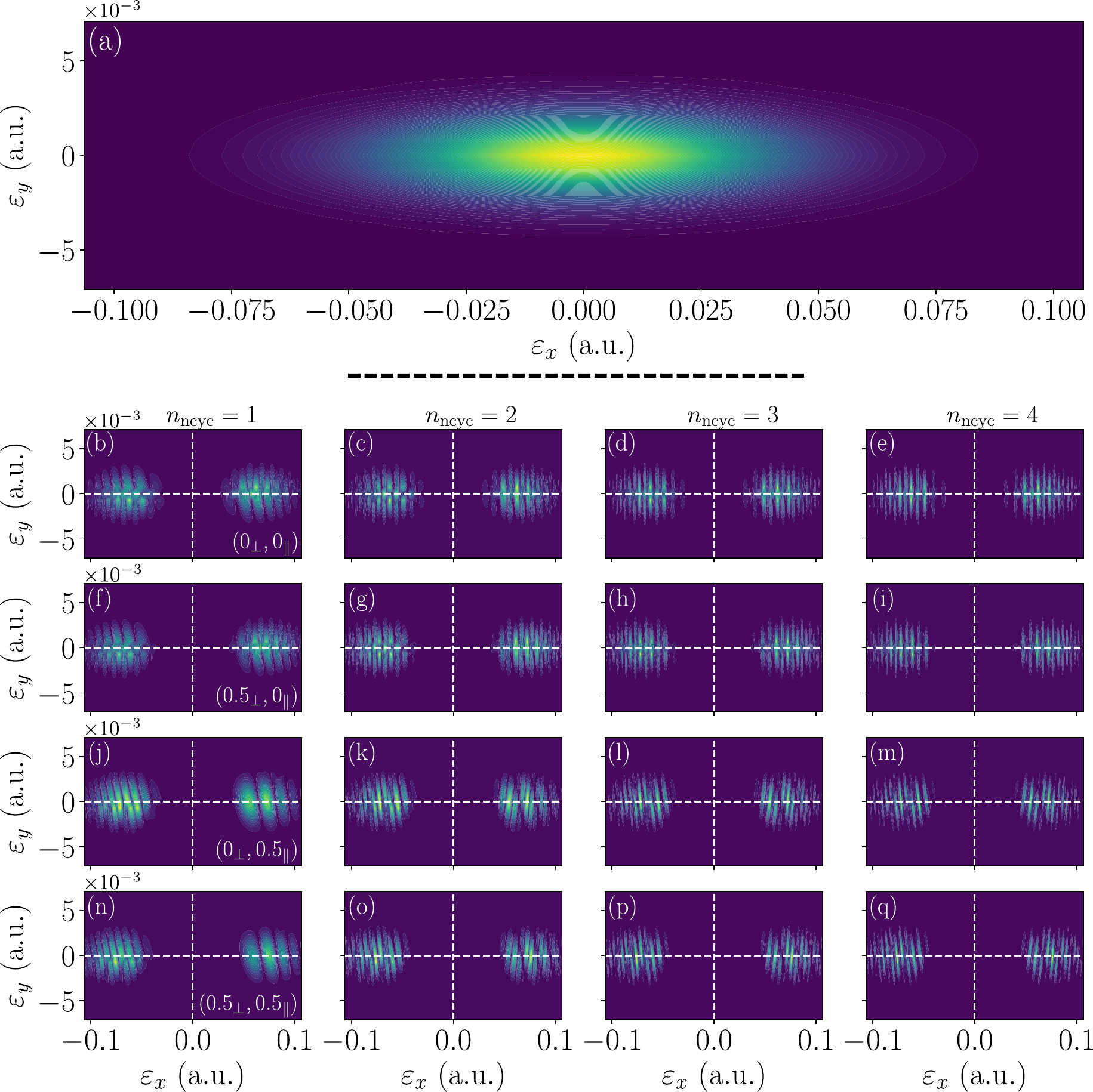}
	\caption{(a) Absolute value of the coefficient function $c(\alpha)$ before the ATI process.~(b)-(q) Absolute value of the effective coefficient function $\Tilde{c}(\alpha)$ after post-selection on various electron momenta $\boldsymbol{p}_f$ (shown as inset). Here, we have fixed the measurement time to $t= 2\pi n_{\text{cyc}}/\omega + \pi/(2\omega)$, with $\omega_L = 0.057$ a.u. ($\lambda_L = 800$ nm).~Here, we have set $g(\omega_L) = 10^{-3}$ and $r=3$, setting the lattice constant for the discrete coherent state expansion to $k= 0.2\sqrt{\pi}$.}
	\label{Fig:SM:C:before:after}
\end{figure}

Strictly speaking, the saddle-point approximation is valid when the phase factor $e^{-iS/\hbar}$ oscillates rapidly, which depends sensitively on the field amplitude $\varepsilon_\alpha$.~In the weak-field limit $\varepsilon_{\alpha}\to 0$, the action varies more slowly and the approximation formally deteriorates.~However, in this regime the imaginary part of the action evaluated at the saddle-point increases substantially, leading to $e^{-i\text{Im}[S]/\hbar} \to 0$, that is, an exponential suppression of the ionization probability. This has been further benchmarked using exact numerical integration of $b(\Tilde{\boldsymbol{p}}_f,t';\alpha)$.~For the photoelectron momenta considered here, the ionization yield becomes significant for field amplitudes of order $\varepsilon \sim 10^{-2}$ a.u., which lie within the regime where both the saddle-point and strong-field approximations remains reliable.

Figure~\ref{Fig:SM:C:before:after} compares $\abs{c(\alpha)}$ (prior to the interaction) with $\abs{\tilde{c}(\alpha)}$ for various values of the final momentum $\boldsymbol{p}_f$ and different optical-cycle durations of the driving field.~Comparing these effective coefficients, we observe that their values are not fully rotationally invariant in phase-space.~This lack of fully rotationally invariance is a necessary and sufficient condition for non-classicality in single-mode pure states~\cite{rivera-dean_condition_202X}.~Therefore, both before and after the light-matter interaction---and subsequent electronic heralding---the resulting quantum optical states are intrinsically non-classical.~Despite this shared non-classical character, important differences emerge between the pre- and post-interaction states.~Before the interaction, the coefficient function is squeezed in space, as expected for a bright squeezed input.~After the heralding, however, two pronounced lobes appear around the origin, each displaying clear interference patterns.~The structure in the interference patterns within each lobe depends on both the number of cycles considered and the post-selected final electron momentum.~In particular, the momentum component $p_{f,\parallel}$ plays a more prominent role than $p_{f,\perp}$ in shaping the intralobe structure and the symmetry between the two lobes.~For a fixed value of $\boldsymbol{p}_f$, increasing the number of optical cycles leads to sharper interference fringes~\cite{lewenstein_rings_1995}.~This behavior indicates that the emergence of well-defined peaks is associated with an increasing number of interfering electron trajectories as the interaction duration grows.

\begin{figure}
	\centering
	\includegraphics[width=1\textwidth]{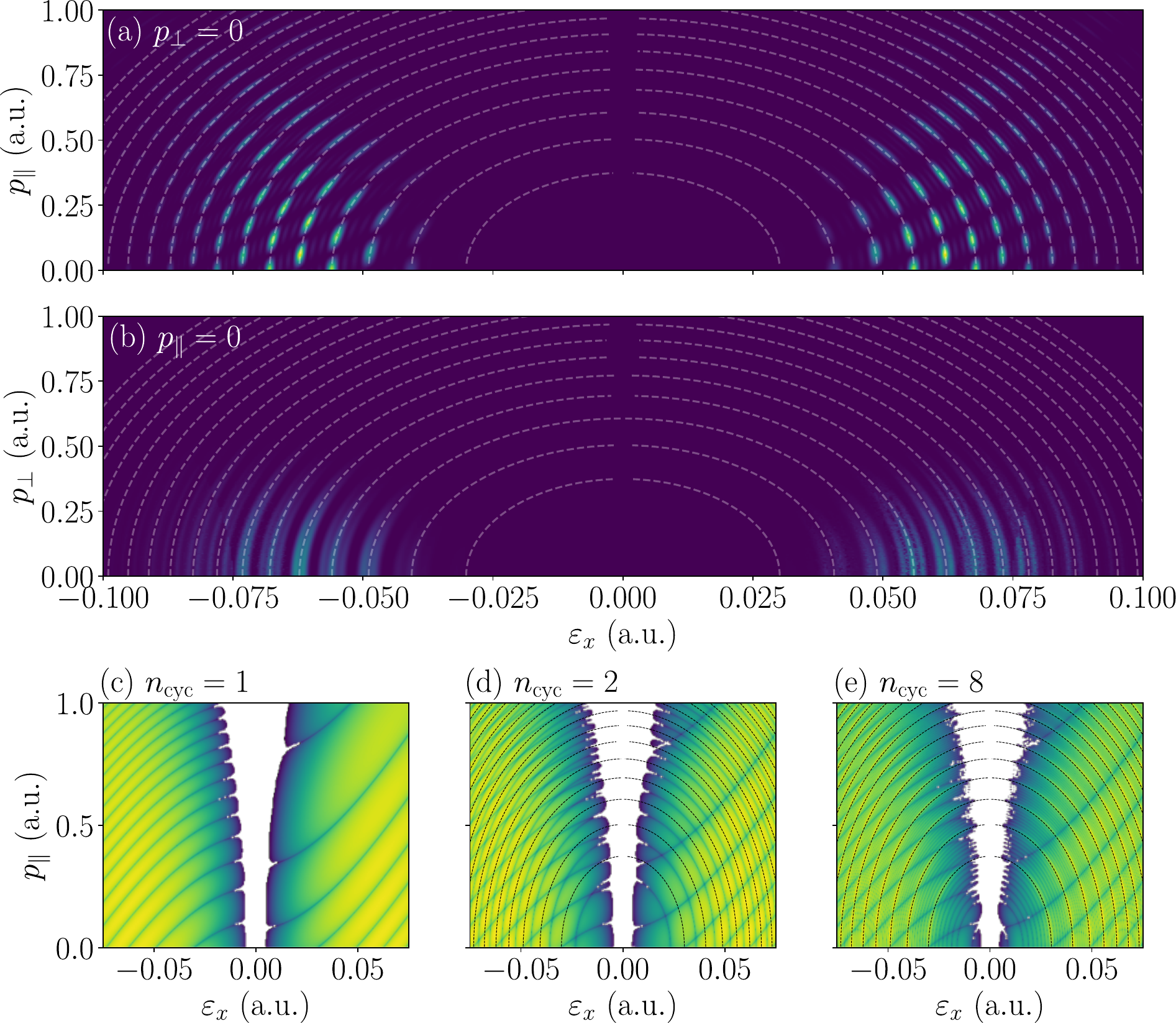}
	\caption{Effective distribution $\abs{\Tilde{c}(\alpha)}$ obtained after heralding on the electron momentum (a) $\boldsymbol{p}_f = (p_\parallel,0)$ and (b) $\boldsymbol{p}_f = (0,p_\perp)$ for $n_{\text{cyc}} = 4$. The dashed white lines indicate the ATI ring positions defined by Eq.~\eqref{Eq:SM:energy:conservation}. (c)-(e) Logarithmic-scale view of $\abs{\Tilde{c}(\alpha)}$ for $\boldsymbol{p}_f = (p_\parallel,0)$ and increasing $n_{\text{cyc}}$. In these panels, the ATI rings are indicated with the black dashed lines. The white region near $\varepsilon_x \simeq 0$ arises from the logarithmic representation combined with the exponential suppression of the ionization probability at low field amplitudes. Values of $\abs{\Tilde{c}(\alpha)}$ are numerically rounded to $10^{-13}$ for visualization purposes. The same parameters as those used in Fig.~\ref{Fig:SM:C:before:after} are used.}
	\label{Fig:SM:C:Rings}
\end{figure}
	
One can nevertheless identify a form of periodicity in the peaks observed within the intralobe structures of $\bar{c}(\alpha)$. The presence of these peaks can be related to the ATI energy conservation, provided that ATI peaks occur at energies given by~\cite{lewenstein_rings_1995}
\begin{equation}\label{Eq:SM:energy:conservation}
	 \dfrac{\Tilde{\boldsymbol{p}}^2_f}{2}
	= n\omega - I_p - \dfrac{\varepsilon_\alpha^2}{4\omega^2}.
\end{equation}
where the last term corresponds to the ponderomotive energy $U_p$.

Figure~\ref{Fig:SM:C:Rings} displays $\abs{\Tilde{c}(\alpha)}$ when considering cuts along the $\varepsilon_y=0$ in Fig.~\ref{Fig:SM:C:before:after}, after post-selection on (a) $\boldsymbol{p}_f = (p_\parallel,0)$ and (b) $\boldsymbol{p}_f = (0,p_\perp)$. The dashed white lines, correspond to the energy conservation dictated by Eq.~\eqref{Eq:SM:energy:conservation}, which defines the location of the so-called ATI rings along which $\abs{\Tilde{c}(\alpha)}$ is distributed.~The distribution along these rings differs for $\boldsymbol{p}_f = (p_\parallel,0)$ and $\boldsymbol{p}_f = (0,p_\perp)$:~in the former case, pronounced peaks appear along each ATI ring, whereas in the latter the structure is more continuous.~This difference can be attributed to stronger intracycle interference along the electric field polarization axis; indeed, within each optical cycle two complex saddle-point solutions exist, corresponding to ionization events near consecutive extrema of the driving field.~The interference between these two trajectories produces intracycle modulation, which becomes particularly visible along the polarization direction.

To make this more explicit, Fig.~\ref{Fig:SM:C:Rings}~(c)-(e) displays $\abs{\Tilde{c}(\alpha)}$ for $\boldsymbol{p}_f = (p_\parallel,0)$ and increasing number of optical cycles, shown on a logarithmic scale.~For $n_{\text{cyc}} = 1$, intercycle interference is absent and therefore ATI rings do not form. Instead, the observed beatings are only due to intracycle interference from the two saddle-points within a single optical cycle. As $n_{\text{cyc}}$ increases, the intracycle modulation persists but becomes combined with intercycle interference, giving rise to ATI rings that sharpen progressively with increasing values of $n_{\text{cyc}}$.~Along the $p_{\perp}$ direction, the final momentum does not acquire a drift from the field's vector potential and, consequently, the modulation along the ATI rings becomes absent, resulting in cleaner ring structures.

\subsection{Dominant phase-space contributions}\label{Sec:SM:SPA:V2}
In the numerical analysis presented in this work, the parameter $g(\omega_L)$ (and therefore $\alpha$) is restricted to values that allow for a tractable Hilbert-space representation of the generated states, involving on the order of hundreds of photons (see Sec.~\ref{Sec:SM:Numerics} for a more detailed discussion).~While this regime suffices to characterize their phase-space structure and nonclassical properties, direct numerical access to the experimentally relevant high-photon number limit is computationally prohibitive.

To assess the macroscopic nature of the generated states, we therefore turn to semiclassical methods.~Specifically, we analyze the stationary configurations of the phase of the integrand in a saddle-point approximation manner~\cite{lewenstein_theory_1994} by looking at regions where the phase of the probability amplitudes $\Tilde{c}(\alpha)$ remain stationary~\cite{singh_interferometrically_2026}.~This would allow us to identify the parameter-space regions where such contributions are dominant.~Thus, finding saddle-points at large values of $\abs{\alpha}$ and symmetrically distributed around the origin, would provide analytical insight into the regions of parameter space where the phase of the probability amplitude becomes stationary, and which are expected to play a relevant role in shaping the coherent state superposition.

To this end, starting from Eq.~\eqref{Eq:SM:final:state}, we note that when strong squeezing is applied along the $\alpha_y$ quadrature, the $c(\alpha)$ coefficient becomes extremely localized in that direction, allowing us to effectively reduce the integral to a single quadrature
\begin{equation}
	\ket{\Phi(\boldsymbol{p}_f,t)}
		\simeq - i
			\int \dd \alpha_x\ c(\alpha_x)
				 \int^{t}_{t_0} \dd t' b(\tilde{\boldsymbol{p}}_f,t';\alpha)\ket{\alpha}
		\equiv - i
				\int_{\mathcal{C}} \dd \mathsf{x}\ 
					\tilde{c}(\mathsf{x})
						\ket{\mathsf{x}_0}.
\end{equation}
where we have introduced the joint variable $\mathsf{x} = (\alpha,t')$, and $\mathcal{C} \subset \mathbbm{R}^2$ denoting the original integration domain.~To analyze the structure of this integral, we expand in the coherent states in the Fock basis,
\begin{equation}
		\ket{\Phi(\boldsymbol{p}_f,t)}
			= -i
				\sum_{n=0}^\infty
				\int_{\mathcal{C}} \dd \mathsf{x}\
					\Tilde{c}(\mathsf{x}) e^{-\mathsf{x}_0^2/2} 
						\dfrac{\mathsf{x}_0^n}{\sqrt{n!}}
							\ket{n},
\end{equation}
such that each coefficient of this expansion is expressed as a scalar integral whose integrand admits analytic continuation in the variables $(\alpha_x,t')$, provided that $\tilde{c}(\mathsf{x})$ is analytic. Under this condition, the integration domain $\mathcal{C}_0$ may be continuously deformed within $\mathbbm{C}^2$ without crossing singularities, leaving the value of the integral invariant.

Motivated by semiclassical strong-field methods, we introduce the decomposition of the $\tilde{c}(\mathsf{x})$ contribution to the integrand into a slowly varying prefactor and a rapidly oscillating phase $\tilde{c}(\mathsf{x}) = g(\mathsf{x}) e^{i S_{\text{QO}}(\tilde{\boldsymbol{p}}_f,\mathsf{x})}$ where
\begin{equation}
	g(\mathsf{x})
		\propto \langle \Tilde{\boldsymbol{p}}_f(\alpha)+\mathsf{e}\boldsymbol{A}_{\text{cl}}(t';\alpha)\rvert
		\hat{\boldsymbol{r}}\lvert \text{g}\rangle\cdot \boldsymbol{E}_{\text{cl}}(t';\alpha),
\end{equation}
while for the highly oscillatory exponent is given by
\begin{equation}
	\begin{aligned}
		S_{\text{QO}}(\tilde{\boldsymbol{p}}_f,\mathsf{x})
		&= -\dfrac12
			\int_{t'}^{t} \dd \tau
				\Bigg\{
				\Big[
					\tilde{p}_{f,\parallel}
					+\dfrac{\varepsilon_x}{\omega}\cos(\omega_L \tau)
				\Big]^2
			+ \tilde{p}_{f,\perp}^2
				\Bigg\}
			+ I_p(t'-t_0)
		+
		\dfrac{i\varepsilon_x^2}{4g(\omega_L)^2e^{r}\cosh(r)},
	\end{aligned}
\end{equation}
that is, an effective action incorporating the both the semiclassical propagation phase and corrections arising from the field statistics of the driver.~In this representation, we analyze the regions of parameter space where the phase of the probability amplitudes becomes stationary, i.e., $\nabla_{\mathsf{x}} S_{\text{QO}}(\tilde{\boldsymbol{p}}_f,\mathsf{x}) = \boldsymbol{0}$.~Such stationary configurations provide an indication, in analogy to stationary-phase arguments, of regions that may play a relevant role in shaping the coherent state-superposition.~Importantly, we do not explicitly evaluate the resulting saddle-point approximation; rather, we use the structure of these stationary points to identify regions of parameter space that contribute most significantly to the state.~The corresponding stationary points read
\begin{align}
	&\pdv{S_{\text{QO}}}{t'} = 0
	\Rightarrow
	\dfrac12
	\bigg[ 
	\tilde{p}_{f,\parallel}
	+ \frac{\varepsilon_x}{\omega}\cos(\omega_L t')
	\bigg]^2
	+ \dfrac{\tilde{p}_{f,\perp}^2}{2}
	+ I_p = 0,\label{Eq:SM:SP:time}
	\\&
	\begin{aligned}\label{Eq:SM:SP:varex}
		\pdv{S_{\text{QO}}}{\varepsilon_x} = 0
		&\Rightarrow
		-\dfrac{1}{\omega}
		\int^{t}_{t'} \dd \tau
		\Big[
		\tilde{p}_{f,\parallel}
		+ \dfrac{\varepsilon_x}{\omega_L}
		\cos(\omega_L \tau)
		\Big]\cos(\omega_L \tau)
		+ 
		\dfrac{i\varepsilon_x}{2g(\omega_L)^2e^{r}\cosh(r)} = 0,
	\end{aligned}
\end{align}
where $\alpha_x = \varepsilon_x/(2g(\omega_L))$, and which determine the relevant stationary configurations in the joint parameter space $(\alpha_x,t')$ for $\tilde{c}(\mathsf{x})$. 

\begin{figure}
	\centering
	\includegraphics[width=1\textwidth]{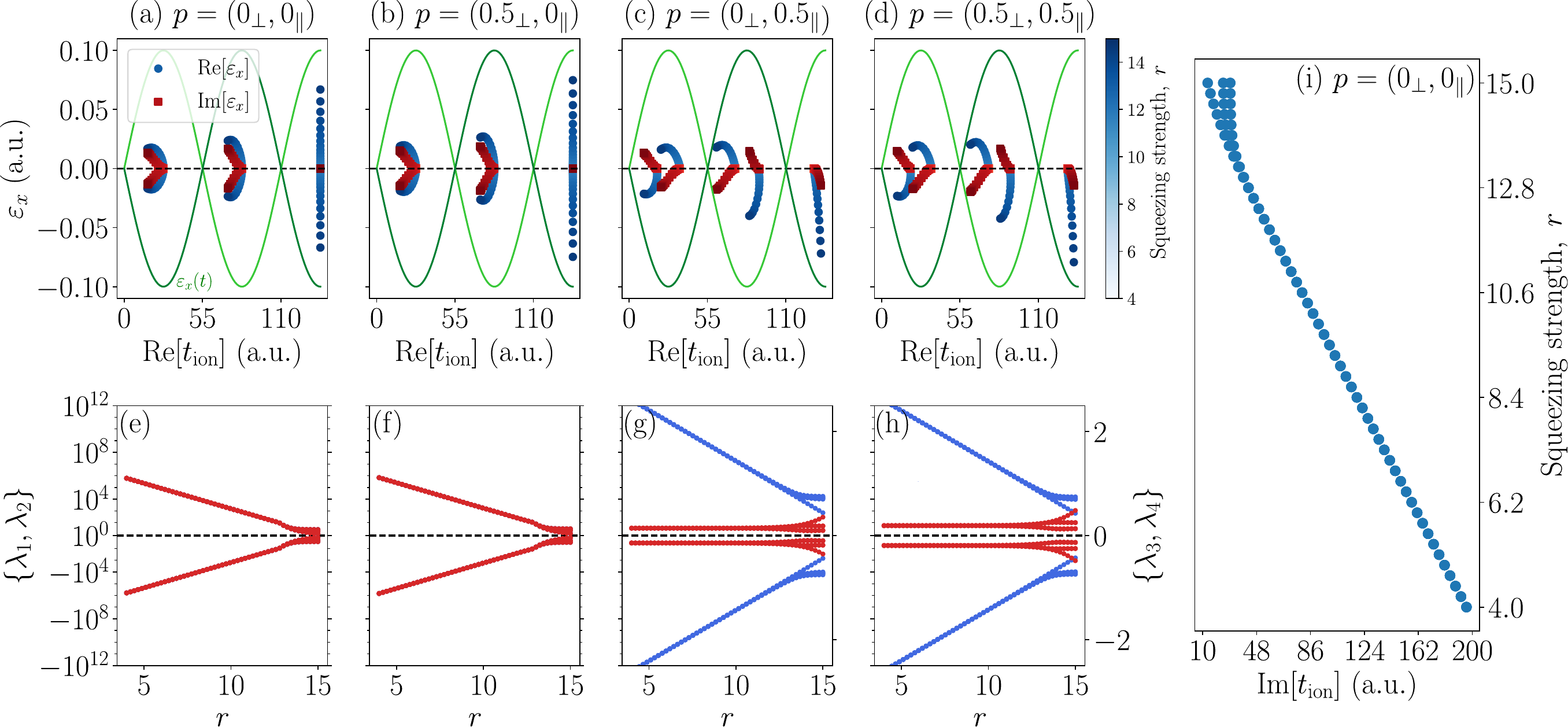}
	\caption{Saddle-point solutions for $\varepsilon_{y} = 0$. (a)-(d) real and imaginary part of $\varepsilon_x$ as a function of the real part of the ionization time for various values of $\boldsymbol{p}$. (a)-(h) eigenvalues of the Hessian matrix---note the difference in orders of magnitude between the two sets of eigenvalues. (i) Imaginary part of the ionization time as a function of the squeezing strength $r$ for the case $\boldsymbol{p} = \boldsymbol{0}$. Here, we have considered hydrogen for the atomic system ($I_p \simeq 0.5$ a.u.), $\omega = 0.057$ a.u.~and set $g(\omega_L) = 10^{-8}$ a.u.~\cite{rivera-dean_light-matter_2022}. We note that, under the present case, $\boldsymbol{p}_f = \tilde{\boldsymbol{p}}_{f}$.}
	\label{Fig:SP:sols}
\end{figure}

Figure~\ref{Fig:SP:sols} shows the solutions to the saddle-point equations \eqref{Eq:SM:SP:time} and \eqref{Eq:SM:SP:varex}. In this case, and for the evaluated times, we have that $\boldsymbol{p}_f = \Tilde{\boldsymbol{p}}_f$.~Under this condition, the fields parametrized by $\text{Re}[\varepsilon_x]$ reach their maximum amplitude at $t\simeq 27.5, 82.7 $ and $137.8$ a.u., and we thus find for $\boldsymbol{p} = \boldsymbol{0}$ and $\boldsymbol{p} = (0.5_{\perp},0_{\parallel})$ [Fig.~\ref{Fig:SP:sols}~(a)-(b)] that ionization takes place at the two later times when looking for saddle points within the interval $\text{Re}[t_{\text{ion}}] \in [0, 2\pi/\omega_L+\pi/(2\omega_L)]$. For each of these times, there exist two solutions of equal amplitude but opposite sign, corresponding to maxima and minimum of the field amplitude, i.e., the intensity maxima.~As $p_{\parallel}$ increases, this symmetry begins to break, and ionization no longer occurs exactly at the field extrema [Fig.~\ref{Fig:SP:sols}~(c)-(d)].~Instead, the ionization times shift to the left or right of the extremum depending on the sign of $\text{Re}[\varepsilon_x]$. These two solutions intrinsically reflect the generation of large coherent state superpositions in our state, with positive and negative values of $\alpha = \varepsilon_\alpha/[2g(\omega_L)]$. Since $\varepsilon_\alpha \propto \pm 10^{-2}$, we thus get $\alpha \propto \pm 5 \times 10^{5}$.

We also observe a non-vanishing imaginary component of the field amplitude $\varepsilon_x$ for these solutions, which vanishes only when the ionization time coincides with the detection time or when squeezing is effectively absent [Fig.~\ref{Fig:saddles}~(a)]. Furthermore, this imaginary component increases with the squeezing strength, while at the same time the corresponding ionization times shift away from the extrema of the field towards earlier values. These observations indicate that the presence of squeezing continuously modifies the structure of the stationary solutions, in a manner consistent with previous studies showing that squeezing can influence the effective electron propagation dynamics~\cite{rivera-dean_microscopic_2025}.

We further evaluate how bona fide the obtained saddle-points are, with Fig.~\ref{Fig:SP:sols}~(e)-(h) displaying the eigenvalues of the real part of the Hessian for each case shown in panels (a)-(d).~In absolute value, the eigenvalues appear in pairs of opposite sign, confirming the presence of true saddle points. However, their magnitudes differ significantly, and only become more comparable as the squeezing strength increases.~This suggests, consistent with physical intuition, that applying contour deformations as discussed here might become more effective as $r$ grows.~Conversely, the limit $r \to 0$ corresponds to fields with vanishing mean intensity.~The behavior is further illustrated in Fig.~\ref{Fig:SP:sols}~(i), which shows the imaginary part of the ionization time as a function of the squeezing parameter.~As expected, the imaginary part increases as the squeezing strength decreases, indicating that ionization becomes progressively less probable.

\section{About the quantum optics and quantum information measures}
The main objective of this section is to provide some analytical background to some of the quantum optics and quantum information measures considered in the main text.

\subsection{Approximate measurements}\label{Sec:SM:Heralding}
The analysis presented thus far has implicitly assumed an ideal detector for heralding the coherent state superposition.~In practice, however, detectors possess a finite momentum resolution and may not perfectly distinguish outcomes $\boldsymbol{p}_f$ and $\boldsymbol{p}_f + \boldsymbol{\delta}$.~Within quantum measurement theory, such finite-resolution measurements are described as approximate measurements~\cite{breuer_quantum_2007}.

To model this effect, we introduce a conditional probability density $w(p_{f,\mu}|p_{f,\mu}')$, representing the probability that the detector reports outcome $p_{f,\mu}$ when the true value is $p_{f,\mu}'$.~This function satisfies the normalization condition $\int \dd p_{f,\mu} w(p_{f,\mu}|p_{f,\mu}') = 1$, and we model it by a Gaussian function 
\begin{equation}
	w(p_{f,\mu}|p'_{f,\mu})
		= \dfrac{1}{\sqrt{2\pi \sigma_\mu}}
			\exp[-\dfrac{(p_{f,\mu}-p_{f,\mu}')^2}{2\sigma_\mu^2}],
\end{equation}
where $\sigma_\mu$ characterizes the detector resolution along direction $\mu$. For simplicity, we restrict the discussion to a single momentum component $\mu$ to gain better understanding of how the electron propagation direction affects the final state.~Nevertheless, the extension to higher dimensions is straightforward and does not modify the structure of our argument, but only of the conditional probability $w(\cdot |\cdot)$.

In this context, the measurement is described by the POVM elements
\begin{equation}
	\hat{\Pi}(p_{f,\mu})
		= \int \dd p_{f,\mu}'w(p_{f,\mu}\vert p_{f,\mu}') \dyad{p_{f,\mu}'},
\end{equation}
which satisfy the completeness relation $\int \dd p_{f,\mu} \hat{\Pi}(p_{f,\mu}) = \hat{\mathbbm{1}}$.~Assuming a minimally disturbing measurement model in which the detector introduces only finite classical resolution but no additional unitary operations~\cite{breuer_quantum_2007}, the post-selected state conditioned on outcome $p_{f,\mu}$ is, up to normalization,
\begin{equation}
	\hat{\rho}(\boldsymbol{p}_{f,\mu})
		= 	\tr[\hat{\Pi}(p_{f,\mu})\dyad{\Psi(t)}]
		\approx \int \dd p_{f,\mu}' 
			w(p_{f,\mu}\vert p_{f,\mu}')
				\dyad{\Phi(\boldsymbol{p}_{f,\mu}',t)},
\end{equation}
where $\boldsymbol{p}_{f,\mu}' = (p_{f,\mu}',p_{f,\bar{\mu}})$.~The approximation arises from the strong-field approximations used in deriving Eq.~\eqref{Eq:SM:final:state}, and is independent of the measurement model itself.

\subsection{Bell nonlocality}\label{Sec:SM:Bell}
The concept of nonlocality arises in the context of measurements and the structure of joint probability distributions.~We consider two initially separate parties, Alice and Bob, who perform measurements labeled $x$ and $y$ on their respective subsytems of a bipartite physical system, upon which they obtain outcomes $a$ and $b$, respectively.~The statistics of the experiment are described by conditional probabilities $P(a,b|x,y)$, which define the probability of obtaining outcomes $(a,b)$ given measurement choices $(x,y)$. A behavior is said to admit a local hidden-variable model if it can be written as~\cite{brunner_bell_2014}
\begin{equation}\label{Eq:SM:LHM}
	P(a,b|x,y) = \int \dd \lambda \ q(\lambda)  P(a|x,\lambda)P(b|y,\lambda),
\end{equation}
where $\lambda$ denotes \emph{hidden variable} distributed accordingly to a probability density $q(\lambda)$.~The hidden variable represents shared classical randomness, which may influence the outcomes locally but is assumed to be independent of the measurement choices $x$ and $y$. Behaviors satisfying Eq.~\eqref{Eq:SM:LHM} are termed \emph{local}.

In quantum mechanics, if Alice and Bob share a bipartite system $\hat{\rho}$ and perform measurements described by POVM elements $\{\hat{M}_{a|x}\}$ and $\{\hat{M}_{b|x}\}$, the corresponding conditional probabilities are given by Born's rule~\cite{born_zur_1926,NielsenBookCh1}
\begin{equation}\label{Eq:SM:Born:rule}
	P(a,b|x,y)
		= \tr(\hat{\rho} \hat{M}_{a|x}\otimes \hat{M}_{b|y}).
\end{equation}
We denote by $\mathcal{L}$ the set of local behaviors satisfying Eq.~\eqref{Eq:SM:LHM}, and by $\mathcal{Q}$ the set of behavior achievable within quantum mechanics via Eq.~\eqref{Eq:SM:Born:rule}. Both sets are closed, bounded and convex~\cite{brunner_bell_2014} and, moreover, the inclusion $\mathcal{L}\subset \mathcal{Q}$ holds~\cite{pitowsky_range_1986}, meaning that all local behaviors can be realized within quantum theory.~The set of nonlocal behaviors is therefore defined as $\mathcal{NL} = \mathcal{Q} \backslash \mathcal{L}$, and it can be shown that entanglement is a necessary but not sufficient condition for a behavior to be nonlocal. This is where Bell inequalities arise.

A powerful way to distinguish local from nonlocal behaviors is through Bell inequalities, which formally delimit the extension of the local set in the $\mathsf{P}$-space. These are linear constraints of the form~\cite{bell_einstein_1964,brunner_bell_2014}
\begin{equation}\label{Eq:SM:BI}
	\sum_{a,b,x,y}
		s^{a,b}_{x,y}
			P(a,b|x,y)
				\leq S,
\end{equation}
while are satisfied by all behaviors in $\mathcal{L}$ but not by those in $\mathcal{L}$; a violation of Eq.~\eqref{Eq:SM:BI} certifies nonlocality.~In the scenario where Alice and Bob each choose between two dichotomic measurements, the Clauser-Horne-Shimony-Holt (CHSH) inequalities completely characterize the local set.~In our work, we consider this scenario, where the measurements performed by Alice and Bob correspond to~\cite{banaszek_nonlocality_1998,banaszek_testing_1999}
\begin{equation}
	\hat{\Pi}(\beta) 
		= \hat{D}(\beta)(-1)^{\hat{n}} \hat{D}^\dagger(\beta),
\end{equation}
where $\hat{D}(\beta)$ is the displacement operator and $(-1)^{\hat{n}}$ is the parity operator.~Measuring $\hat{\Pi}(\beta)$ corresponds to probing the Wigner function at the phase-space point $\beta$.~The measurement settings $x$ and $y$ introduced earlier, are therefore identified with the phase-space points $\beta^{(j)}_A$ and $\beta^{(j)}_B$ (with $j=1,2$) chosen by Alice and Bob.~The corresponding CHSH operator reads
\begin{equation}\label{Eq:SM:Bell:op}
	\langle\hat{\mathcal{B}}(\boldsymbol{\beta})\rangle
		= 
			\langle \hat{\Pi}(\beta^{(1)}_A)\otimes \hat{\Pi}(\beta^{(1)}_B) \rangle
				+ \langle \hat{\Pi}(\beta^{(1)}_A)\otimes \hat{\Pi}(\beta^{(2)}_B) \rangle
					+ \langle \hat{\Pi}(\beta^{(2)}_A)\otimes \hat{\Pi}(\beta^{(1)}_B) \rangle
						 -\langle \hat{\Pi}(\beta^{(2)}_A)\otimes \hat{\Pi}(\beta^{(2)}_B) \rangle,
\end{equation}
such that for all local behaviors $2<\lvert\langle \hat{B}(\boldsymbol{\beta})\rangle\rvert\leq 2$ while quantum mechanics allows violations up to the Tsirelson bound $2<\lvert \langle \hat{B}(\boldsymbol{\beta})\rangle \lvert \leq 2\sqrt{2}$.~Thus, observing $\lvert \langle \hat{B}(\boldsymbol{\beta})\rangle\rvert > 2$ certifies Bell nonlocality.

\begin{figure}[h!]
	\centering
	\includegraphics[width=1\textwidth]{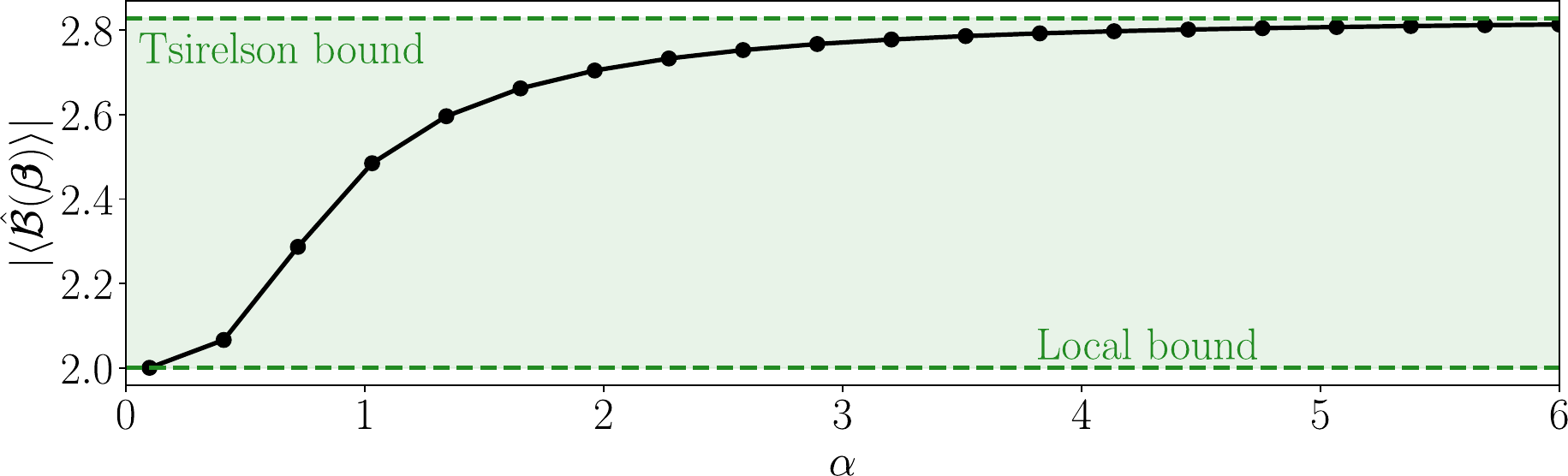}
	\caption{Maximum value of Eq.~\eqref{Eq:SM:Bell:op} when optimized with respect to $\boldsymbol{\beta}_A$ and $\boldsymbol{\beta}_B$ for the state in Eq.~\eqref{Eq:SM:ECS}.}
	\label{Fig:SM:CHSH:cat}
\end{figure}
Demonstrating the local bound is straightforward when considering deterministic local strategies.~In this case, the correlators factorize, such that
\begin{equation}
	\langle \hat{\mathcal{B}}\rangle
		= 	\langle \hat{\Pi}(\beta^{(1)}_A)\rangle\langle\hat{\Pi}(\beta^{(1)}_B) \rangle
		+ \langle \hat{\Pi}(\beta^{(1)}_A)\rangle\langle \hat{\Pi}(\beta^{(2)}_B) \rangle
		+ \langle \hat{\Pi}(\beta^{(2)}_A)\rangle\langle \hat{\Pi}(\beta^{(1)}_B) \rangle
		-\langle \hat{\Pi}(\beta^{(2)}_A)\rangle\langle \hat{\Pi}(\beta^{(2)}_B) \rangle.
\end{equation}
Since the displaced parity operator has eigenvalues $\pm 1$, it then follows that the maximum value one can find for the quantity above is $\lvert\!\langle \hat{\mathcal{B}}\rangle\!\rvert = 2$, obtained for instance by setting $\langle \hat{\Pi}(\beta^{(1)}_A)\rangle = \langle \hat{\Pi}(\beta^{(1)}_B)\rangle = \langle \hat{\Pi}(\beta^{(2)}_B)\rangle = 1$ and $\langle \hat{\Pi}(\beta^{(1)}_A)\rangle=-1$, or any possible combination.~We note that such factorization of the correlators occurs when considering separable states.~On the other hand, Fig.~\ref{Fig:SM:CHSH:cat} shows that by employing entangled states of the form
\begin{equation}\label{Eq:SM:ECS}
	\ket{\psi}
		= \dfrac{1}{\sqrt{\mathcal{N}}}
				\big[ \ket{\alpha,\alpha} - \ket{-\alpha,-\alpha}\big],
\end{equation} 
one can obtain violations of the Bell inequality, with the Tsirelson bound approaches in the limit $\alpha \to \infty$.~Given the structural similarity between these entangled coherent states and the states obtained in our analysis, we employ the same measurement scheme to investigate whether a Bell inequality violation can be achieved using the measurement configuration depicted in Fig.~\ref{Fig:CHSH}(a).

\section{NUMERICAL ANALYSIS}\label{Sec:SM:Numerics}
This section details the numerical procedures used to evaluate the quantum optical and quantum information measures presented in the main text.~We also discuss the choice of numerical parameters.

\subsection{Coherent state expansions}
The numerical implementation of the state in Eq.~\eqref{Eq:SM:final:state}, or analogously under the saddle-point approximation Eq.~\eqref{Eq:SM:final:state:SPA}, requires discretizing the integral over the coherent state amplitudes $\alpha$, which arises due to the continuous nature of the coherent state representation. Since coherent states are parameterized by a complex number $\alpha = x + iy$, a natural discretization consists of introducing a square lattice in phase-space
\begin{equation}\label{Eq:SM:lattice}
	\mathsf{L} = \{\ket{\alpha_{m,n}}: \alpha_{m,n} = \gamma(m+i n); m,n \in \mathbbm{Z}; \gamma \in \mathbbm{R}^+\}
\end{equation}
where $\gamma > 0$ denotes the lattice spacing. By setting $\gamma\to0$, we recover the continuous nature of the coherent state basis, rendering the state exact.~A fundamental requirement of such discretization is therefore that it preserves the representability of the state, i.e., that the discretized expansion retains the properties of the continuous one.~In the context of the lattice $\mathsf{L}$, this translates into studying its sparsity, namely whether one can define an upper bound on the lattice spacing $\gamma$ beyond which the discretized set no longer reproduces the continuous coherent state representation.

This question was addressed independently in Refs.~\cite{bargmann_completeness_1971,perelomov_completeness_1971}, which investigated the overcompleteness of coherent states and the conditions under which a discrete subset remains total.~More precisely, they determined when the set $\mathsf{L}$ spans a dense subspace of the Hilbert space, allowing any quantum state to be approximated arbitrarily well by linear combinations of states in $\mathsf{L}$.~The conclusion of these studies is that the lattice $\mathsf{L}$ is total whenever $\gamma \leq \sqrt{\pi}$, with the equality case marking the threshold of completeness and corresponds to the so-called von Neumann lattice.~In contrast, for $\gamma \geq \sqrt{\pi}$ the lattice becomes too sparse and fails to be total.

In our simulations, we choose $\gamma = 0.035\sqrt{\pi}$, ensuring that $\mathsf{L}$ is well within the completeness regime, while keeping it sufficiently large to maintain computational feasibility.~The bounds imposed on $m$ and $n$ in Eq.~\eqref{Eq:SM:lattice} are determined dynamically according to the support of the Husimi function of the state under consideration, ensuring that contributions outside the truncated region are numerically negligible.

\subsection{Fock state truncation and parameter scaling}
All quantum optical quantities were computed using \texttt{QuTiP}~\cite{johansson_qutip_2012,johansson_qutip_2013}, which represents quantum states in a truncated Fock basis, therefore requiring to introduce a dimensional cutoff $n_{\text{cutoff}}$.~In strong-field scenarios, typical parameters may reach $r\sim 10$ and $g(\omega_L)\sim10^{-8}$, leading to coherent state amplitudes as large as $\alpha \sim 10^{6}$.~Direct numerical simulation in this regime would require prohibitively large lattice sizes and Fock cutoffs.

However, the physically relevant parameter governing the electron dynamics if $\varepsilon_\alpha = 2 g(\omega_L)\alpha$, which determines the strength of the light-matter interaction.~Thus, in order to maintain the system within the strong-field regime while rendering treatment feasible, we adopt a rescaling strategy similar to that in previous HHG studies~\cite{wang_high-order_2025}, whereby $g(\omega_L)$ is increased and the corresponding coherent amplitudes (or effectively the squeezing parameter $r$) is reduced such that $\varepsilon_\alpha\sim10^{-2}$ a.u.~remains unchanged.

We emphasize, however, that modifying $\alpha$ alters the structure of the optical quantum state itself. While qualitative features remain, such as the position of maxima in $\abs{\Tilde{c}(\alpha)}$ which are determined by $\varepsilon_\alpha$, the separation between coherent state components in phase-space and the detailed interference features of the Wigner function depend on $\alpha$.~This effect can be seen by comparing Fig.~\ref{Fig:C_matrix} and Fig.~\ref{Fig:SM:C:before:after}, with the former computed with $g(\omega_L)= 10^{-6}$ a.u.~and $r = 10$. While the qualitative features are the same, larger coherent state amplitudes related to each peak would produce more widely separated intralobe structures in phase-space.

In particular, for the evaluation of the quantum optical and quantum information measures considered here, we set $g(\omega_L) = 5 \times 10^{-3}$ a.u.~that keep $\alpha \sim 10$, allowing accurate simulations with $n_{\text{cutoff}} = 200$.~Convergence was verified by showing that for higher values of $n_{\text{cutoff}} = 300$, the photon number statistics of the generated states remained the same.~In comparison, more realistic parameters with a smaller $g(\omega_L)$ are expected to result in Wigner functions exhibiting more widely separated intralobe structures and sharper interference regions, while preserving the qualitative interlobe interference pattern arising from the coherent superposition between both lobes at each side of the phase-space origin.

\subsection{Optimization of the CHSH inequality}\label{Sec:SM:Numerics:CHSH}
As discussed in Sec.~\ref{Sec:SM:Bell}, we consider a nonlocality probe based on displaced parity measurements, whereby Alice and Bob sample both positive and negative regions of the Wigner functions in order to maximize the CHSH score.~The measurement settings are parametrized by complex displacements $\beta$, with $X_1 = \text{Re}[\beta]$ and $X_2 = \text{Im}[\beta]$.~Since each party chooses two measurement settings, the optimization of Eq.~\eqref{Eq:SM:Bell:op} involves eight real variables in total. Performing a global optimization based on local optimizations in this parameter space is numerically demanding, particularly as the CHSH landscape may contain multiple local maxima, as they depend on the number of interference fringes in the generated states.

To reduce the computational cost, we restrict our analysis to displacements satisfying $X_1 = 0$, and optimize only over $X_2$, as here lie the maximum CHSH score for the ideal optical cat states considered in Fig.~\ref{Fig:SM:CHSH:cat}. This reduces the dimensionality of the optimization problem from eight to four real variables. Consequently, the CHSH values reported here should be interpreted as lower bounds on the true maximal violation attainable with unrestricted displacements.

The optimization procedure for each value of the final momentum $\boldsymbol{p}_f$ is as follows:
\begin{enumerate}
	\item \textbf{Initial sampling.}~For a given state $\hat{\rho}(\boldsymbol{p}_f)$, we generate 1000 random initial configurations of measurement settings $\{\boldsymbol{\beta}_0(\boldsymbol{p}_f)\}$, evaluate the corresponding CHSH scores, and retain the configuration yielding the largest absolute value. This configuration serves as the initial guess $\boldsymbol{\beta}_0(\boldsymbol{p}_f)$.
	
	\item \textbf{First local refinement.}~Starting from $\boldsymbol{\beta}_0(\boldsymbol{p}_f)$, we perform a local optimization using the Nelder-Mead algorithm, obtaining an improved set of settings denoted $\boldsymbol{\beta}_0^*(\boldsymbol{p}_f)$.
	\item \textbf{Momentum sweep.}~Steps 1 and 2 are repeated for all values in the momentum range $\{\boldsymbol{p}_f\}$, producing a set of locally optimized configurations $\{\boldsymbol{\beta}_0^*(\boldsymbol{p}_f)\}$.
	\item \textbf{Global reseeding across momentum.}~Among the configurations $\boldsymbol{\beta}_0^*(\boldsymbol{p}_f)\in \{\boldsymbol{\beta}_0^*(\boldsymbol{p}_f)\}$, we identify the one yielding the largest CHSH value within the selected momentum range.~This configuration is then used as an initial condition for a second optimization sweep across the full momentum range, producing a new set $\{\boldsymbol{\beta}_1^*(\boldsymbol{p}_f)\}$. For each momentum value, we retain the better configuration between the first and second sweeps
	 \begin{equation}
		\boldsymbol{\beta}_2^*(\boldsymbol{p}_f)
			=
				\arg 
					\max\big[
						\big\lvert\!\big\langle \hat{\mathcal{B}}\big(\boldsymbol{\beta}_0^*(\boldsymbol{p}_f)\big)\big\rangle\!\big\rvert,
						\big\lvert\!\big\langle \hat{\mathcal{B}}\big(\boldsymbol{\beta}_1^*(\boldsymbol{p}_f)\big)\big\rangle\!\big\rvert
						\big].	
	 \end{equation}
 
 	\item \textbf{Sequential continuation.}~Finally, we perform a third sweep in which the optimized configuration at momentum $\boldsymbol{p}_{f,i-1}$ is used as the initial guess for the neighboring point $\boldsymbol{p}_{f,i}$. This produces the new set $\{\boldsymbol{\beta}_3^*(\boldsymbol{p}_f)\}$ with the final measurement settings used in Fig.~\ref{Fig:CHSH}~(c) defined as
 	\begin{equation}
 		\boldsymbol{\beta}_4^*(\boldsymbol{p}_f)
 			=
 				\arg 
 					\max\big[
 					\big\lvert\!\big\langle \hat{\mathcal{B}}\big(\boldsymbol{\beta}_2^*(\boldsymbol{p}_f)\big)\big\rangle\!\big\rvert,
 					\big\lvert\!\big\langle \hat{\mathcal{B}}\big(\boldsymbol{\beta}_3^*(\boldsymbol{p}_f)\big)\big\rangle\!\big\rvert
 				\big].	
 	\end{equation}
\end{enumerate}

We emphasize that the reported CHSH values correspond to local maxima within the constrained optimization landscape. Since the procedure relies on local optimization methods, global optimality cannot be guaranteed.~Improved violations may in principle be obtained using more sophisticated global-search strategies or alternative initialization schemes.~Furthermore, allowing displacements with $X_1 \neq 0$ would enlarge the accessible parameter space and could lead to higher CHSH values, albeit at substantially increased computational cost.
\end{document}